\documentclass[
superscriptaddress,
aps,
prd,
preprint,
amsmath,
amssymb]{revtex4-2}

\usepackage{graphicx}
\usepackage{hyperref}
\usepackage{color}
\usepackage{cancel}
\usepackage{ulem}

\begin{document}

\preprint{RIKEN-iTHEMS-Report-25, FQSP-2025-4, YITP-25-133}

\title{Erratum: $S$-wave kaon-nucleon interactions from lattice QCD at the physical point [Phys.~Rev.~D \textbf{113}, 054506 (2026)]}

\author{Kotaro~Murakami}
\email{kotaro.murakami@yukawa.kyoto-u.ac.jp}
\affiliation{Department of Physics, Institute of Science Tokyo, 2-12-1 Ookayama, Megro, Tokyo 152-8551, Japan.}
\affiliation{RIKEN Center for Interdisciplinary Theoretical and
Mathematical Sciences(iTHEMS), RIKEN, Wako 351-0198, Japan}
\author{Sinya Aoki}
\affiliation{Fundamental Quantum Science Program (FQSP), TRIP Headquarters, RIKEN, Wako 351-0198, Japan}
\affiliation{Center for Gravitational Physics and Quantum Information, Yukawa Institute for Theoretical Physics, Kyoto University, Kitashirakawa Oiwakecho, Sakyo-ku, Kyoto 606-8502, Japan}
\author{Takumi~Doi}
\affiliation{RIKEN Center for Interdisciplinary Theoretical and
Mathematical Sciences(iTHEMS), RIKEN, Wako 351-0198, Japan}
\author{Yan~Lyu}
\affiliation{RIKEN Center for Interdisciplinary Theoretical and
Mathematical Sciences(iTHEMS), RIKEN, Wako 351-0198, Japan}
\author{Wren~Yamada}
\affiliation{RIKEN Center for Interdisciplinary Theoretical and
Mathematical Sciences(iTHEMS), RIKEN, Wako 351-0198, Japan}

\collaboration{HAL QCD Collaboration}\noaffiliation

\date{\today}

\maketitle

{\bf This version contains an erratum followed by the original manuscript, which is identical to \href{https://arxiv.org/abs/2509.00838v2}{arXiv:2509.00838v2}.}

After publication of the paper, we found that
the weights of the contractions involving the neutron at the sink in the $KN$ four-point correlation functions were erroneously smaller by a factor of two.
Specifically, the contribution to the correlation function in Eq.~(8) of the original paper arising from the second term of the sink operator in Eq.~(9) was underestimated by a factor of two. 
This affected all numerical results derived from the $KN$ four-point correlation functions in the original paper.
We have therefore repeated the analysis using the corrected four-point correlation functions with the same simulation setup as described in Sec.~III of the original paper.
Below, we present the corrected results, 
following the section structure of the original paper, and summarize how the conclusions are affected.

\section{Corrected results in Section~IV.A}
\label{sec:erratum_pot}

In this section, we show the corrected results in Sec.~IV.A (``$KN$ potentials'') of the original paper.

\begin{figure}[htbp]
  \centering
  \includegraphics[width=0.49\textwidth]{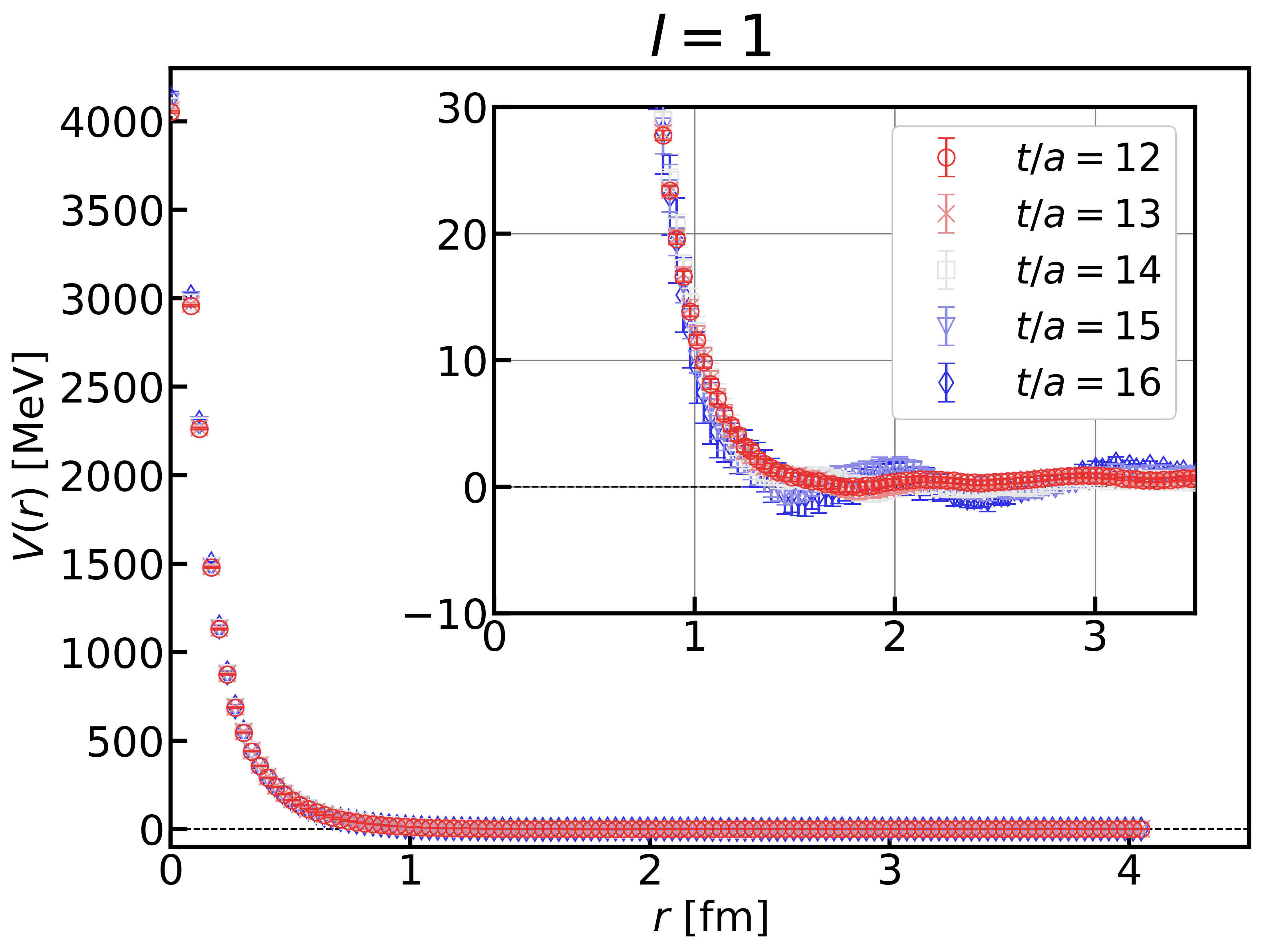}
  \includegraphics[width=0.49\textwidth]{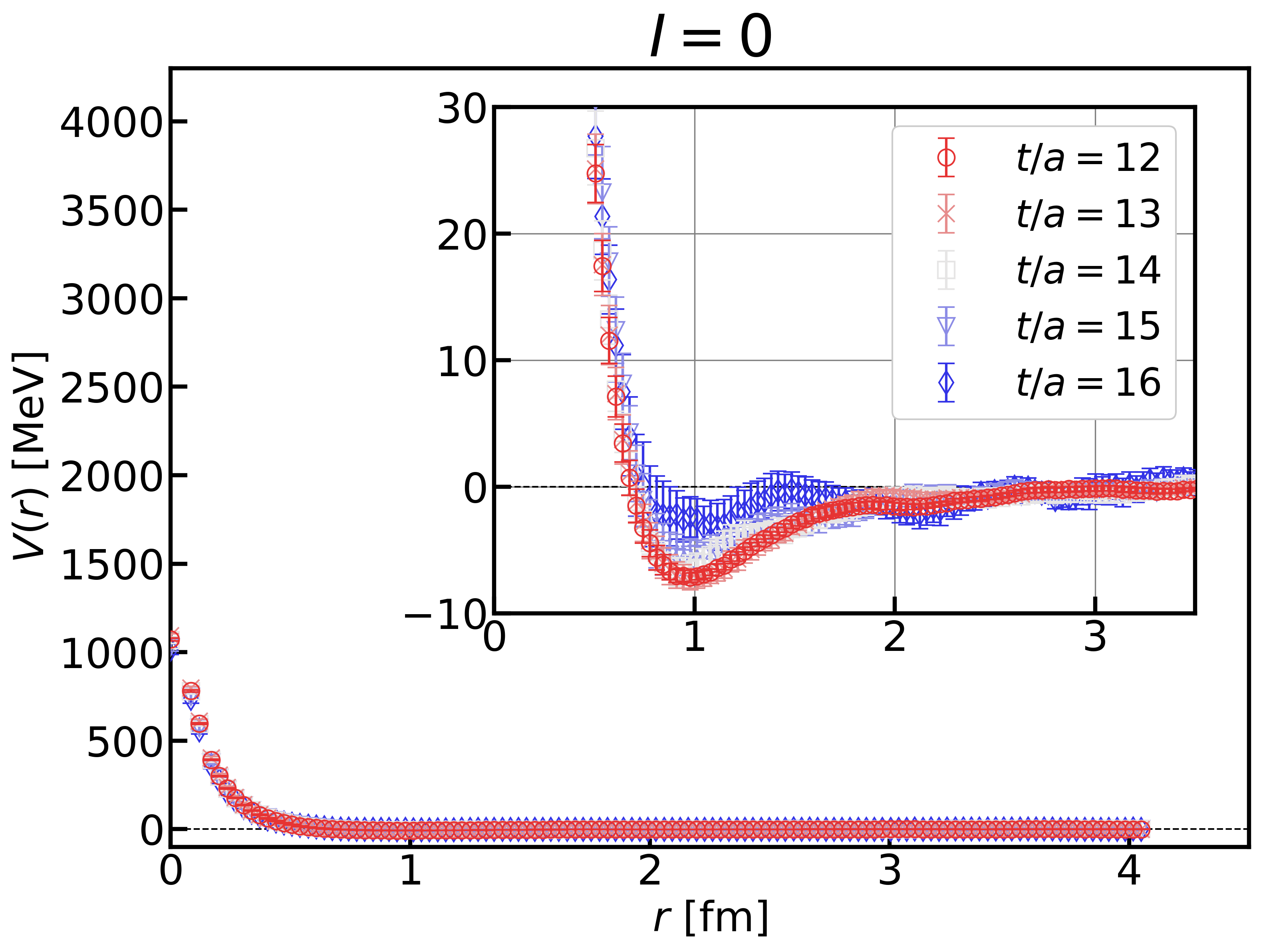}
\caption{Corrected leading-order potentials in $I=1$ (left panel) and $I=0$ channels (right panel), replacing Fig.~1 of the original paper.}
  \label{fig:erratum_pot_orig}
\end{figure}

\begin{figure}[htbp]
  \centering
  \includegraphics[width=0.49\textwidth]{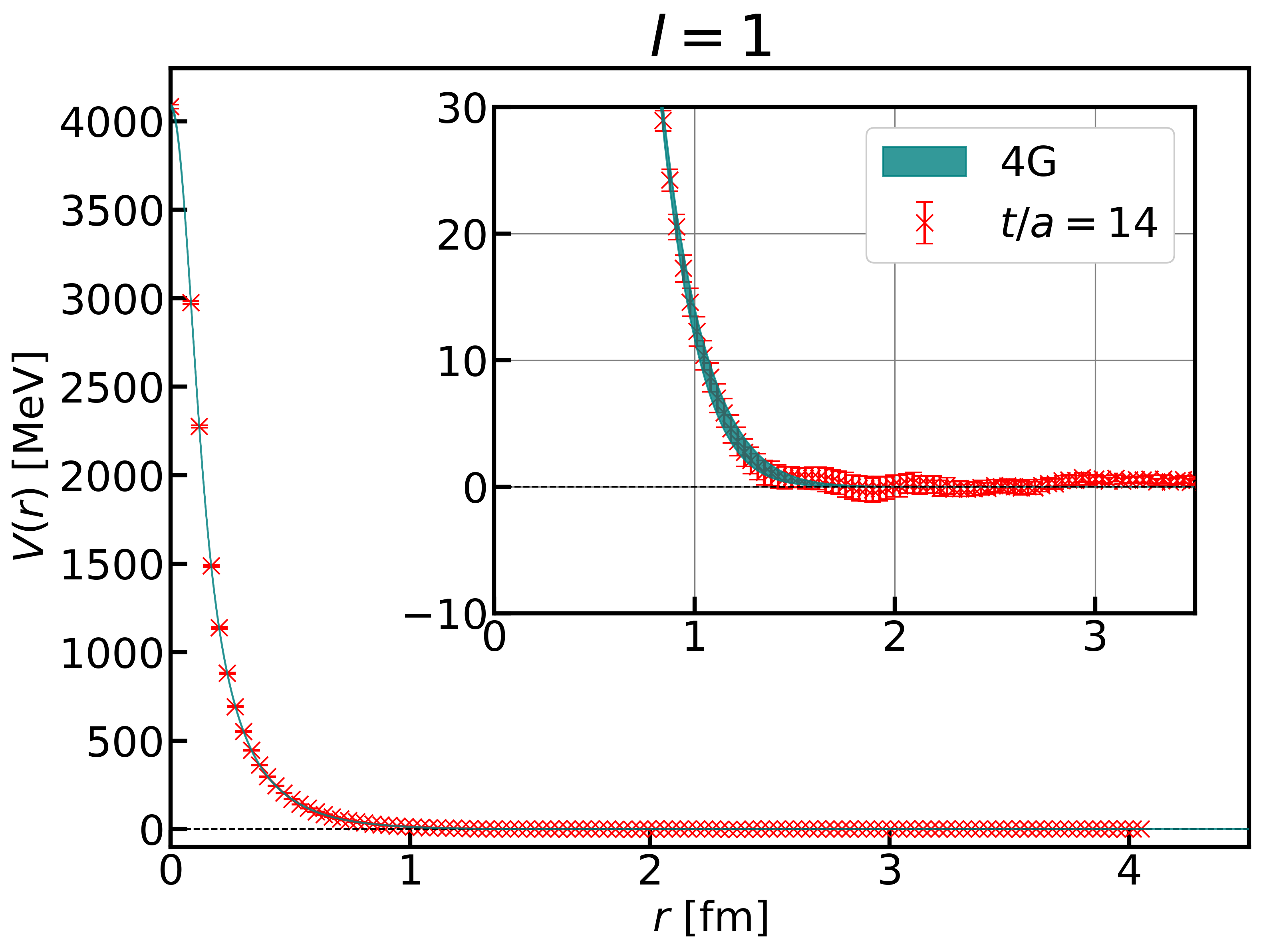}
  \includegraphics[width=0.49\textwidth]{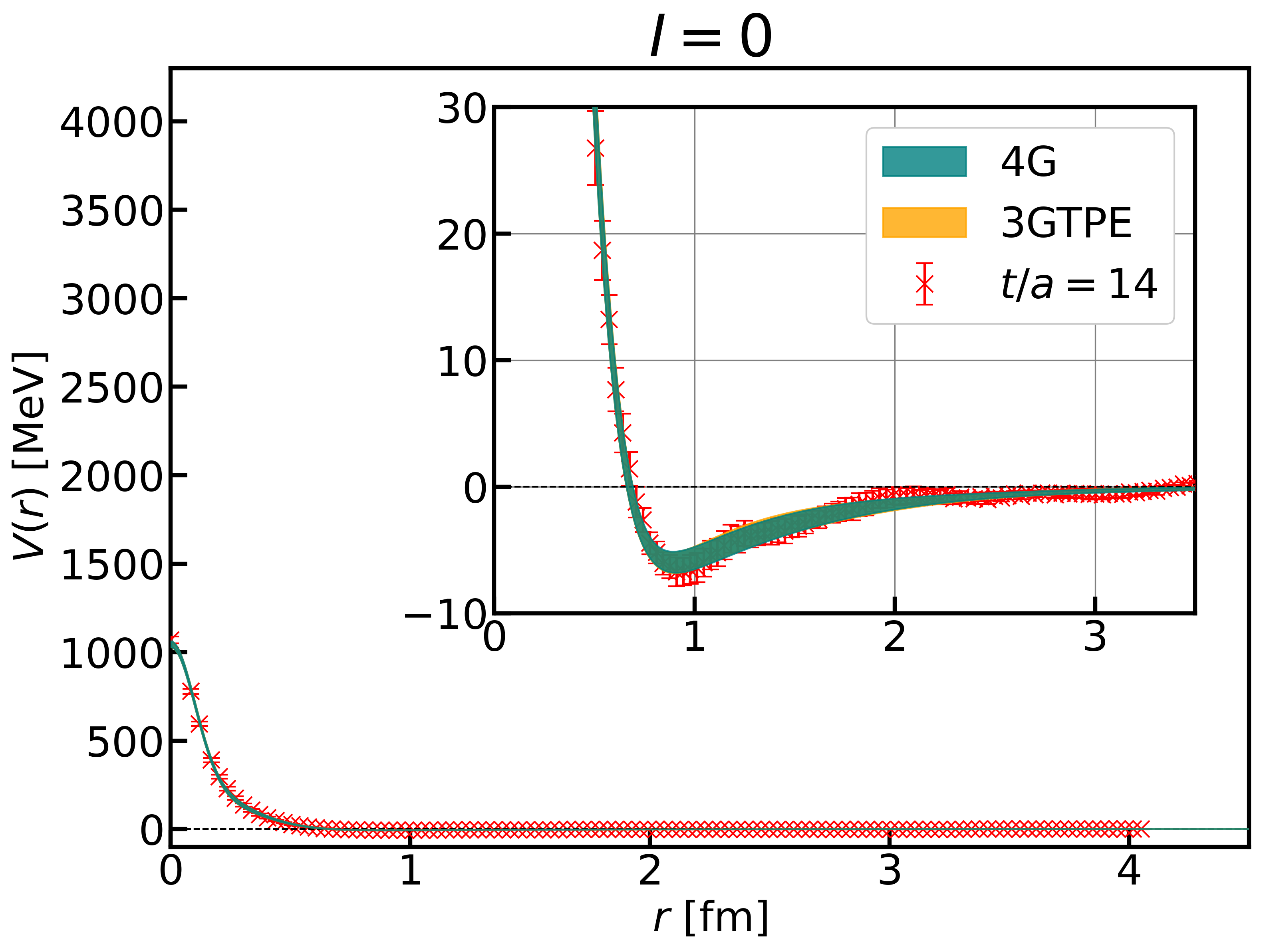}
\caption{Corrected fit results and corrected potential data for the $I=1$ (left panel) and $I=0$ channels (right panel), replacing Fig.~2 of the original paper.}
  \label{fig:erratum_pot_fit}
\end{figure}

\begin{table}[htbp]
  \centering
\caption{Corrected fit parameters of the $KN$ potential at $t/a=14$ for $V^{\textrm{4G}}_{I}(r)$, replacing Table~I of the original paper.
        The values of $\chi^2/\textrm{d.o.f.}=0.14(20)$ [$0.55(24)$]  for $I=1$ ($I=0$).}
  \label{tab:erratum_fitparams_4G}
  \begin{tabular}{c||c|c|c|c|c|c|c|c}
    \hline\hline
    $I$ & $a^{I}_{1}$ [MeV] & $b^{I}_{1}$ [fm] & $a^{I}_{2}$ [MeV] & $b^{I}_{2}$ [fm] & $a^{I}_{3}$ [MeV] & $b^{I}_{3}$ [fm] & $a^{I}_{4}$ [MeV] & $b^{I}_{4}$ [fm] \\
    \hline
    $1$ & $1995.8(102.0)$ & $0.113(1)$ & $1307.4(27.6)$ & $0.198(10)$ & $609.6(61.7)$ & $0.355(28)$ & $170.2(55.9)$ & $0.622(45)$ \\
    \hline
    $0$ & $781.3(33.8)$ & $0.140(4)$ & $275.1(34.5)$ & $0.361(13)$ & $-8.5(2.8)$ & $1.059(184)$ & $-2.9(2.3)$ & $2.098(575)$ \\
    \hline\hline
  \end{tabular}
\end{table}

\begin{table}[htbp]
  \centering
\caption{Corrected fit parameters of the $I=0$ $KN$ potential at $t/a=14$ for $V^{\textrm{3GTPE}}_{I}(r)$, replacing Table~II of the original paper.
        The value of $\chi^2/\textrm{d.o.f.}=0.57(25)$.}
  \label{tab:erratum_fitparams_3GTPE}
  \begin{tabular}{c|c|c|c|c|c|c}
    \hline\hline
    $c^{I=0}_{1}$ [MeV] & $d^{I=0}_{1}$ [fm] & $c^{I=0}_{2}$ [MeV] & $d^{I=0}_{2}$ [fm] & $c^{I=0}_{3}$ [MeV] & $d^{I=0}_{3}$ [fm] & $\alpha^{I=0}$ [MeV fm$^2$] \\
    \hline
    $783.9(35.4)$ & $0.140(5)$ & $263.6(34.6)$ & $0.363(15)$ & $-3.5(2.9)$ & $1.961(428)$ & $-16.8(7.9)$ \\
    \hline\hline
  \end{tabular}
\end{table}

Compared with the original results, the corrected potential for $I=1$ is more repulsive, 
whereas the corrected potential for $I=0$ exhibits a deeper attractive pocket.
The corrected results also have larger statistical fluctuations than the original results, especially at $t/a=16$.

In the original paper, 
the contribution from two-pion exchange (TPE) 
was estimated to be small based on the fit results.
In the corrected analysis, 
the TPE contribution is not well constrained 
in either isospin channel,
given the statistical and systematic uncertainties.
Therefore, no definitive conclusion can be drawn on the magnitude of the TPE coupling.

\section{Corrected results in Section~IV.B}
\label{sec:erratum_obs}

We next present the corrected results in Sec.~IV.B (``$KN$ scattering observables'') of the original paper.

\begin{figure}[htbp]
  \centering
  \includegraphics[width=0.49\textwidth]{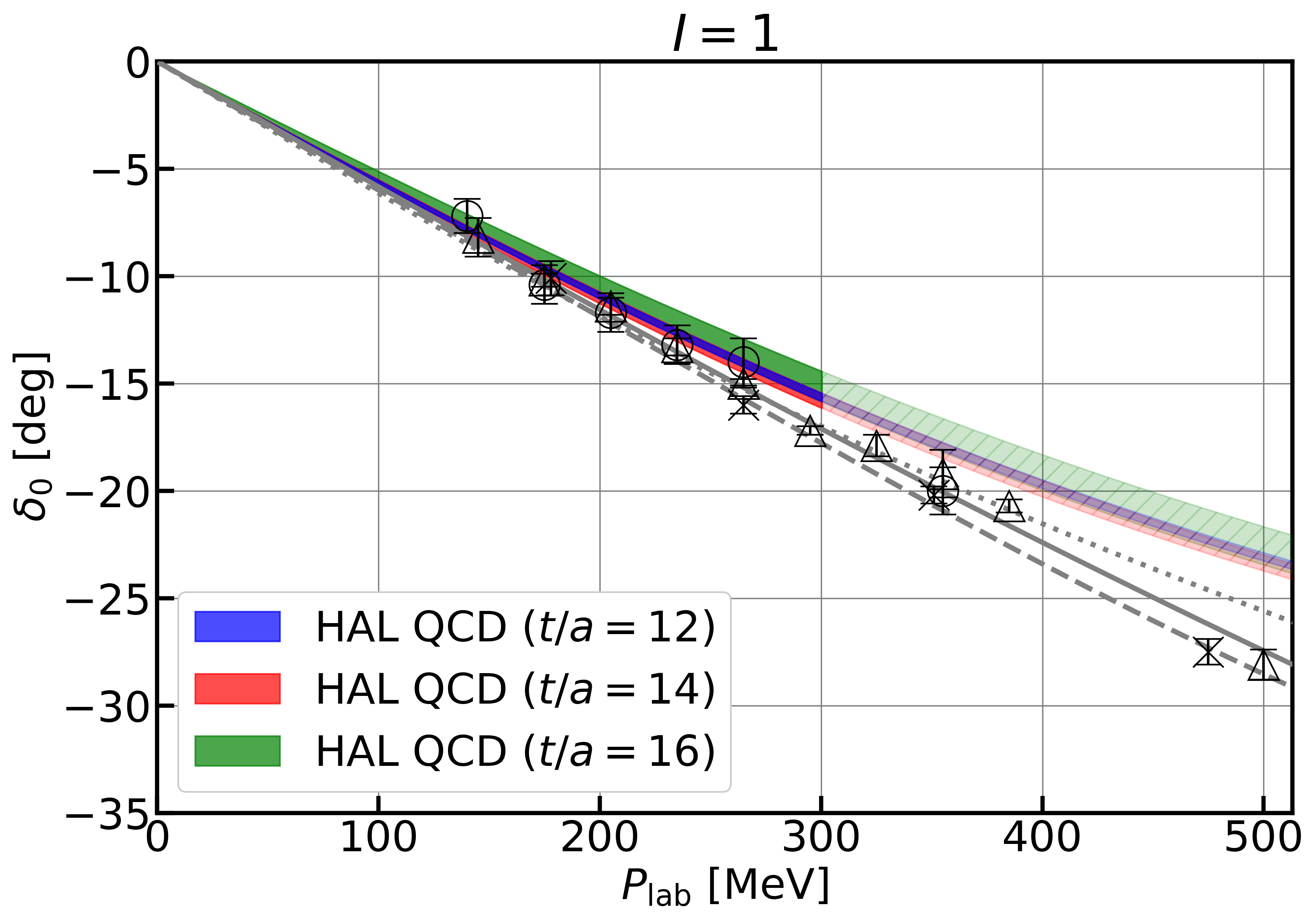}
  \includegraphics[width=0.49\textwidth]{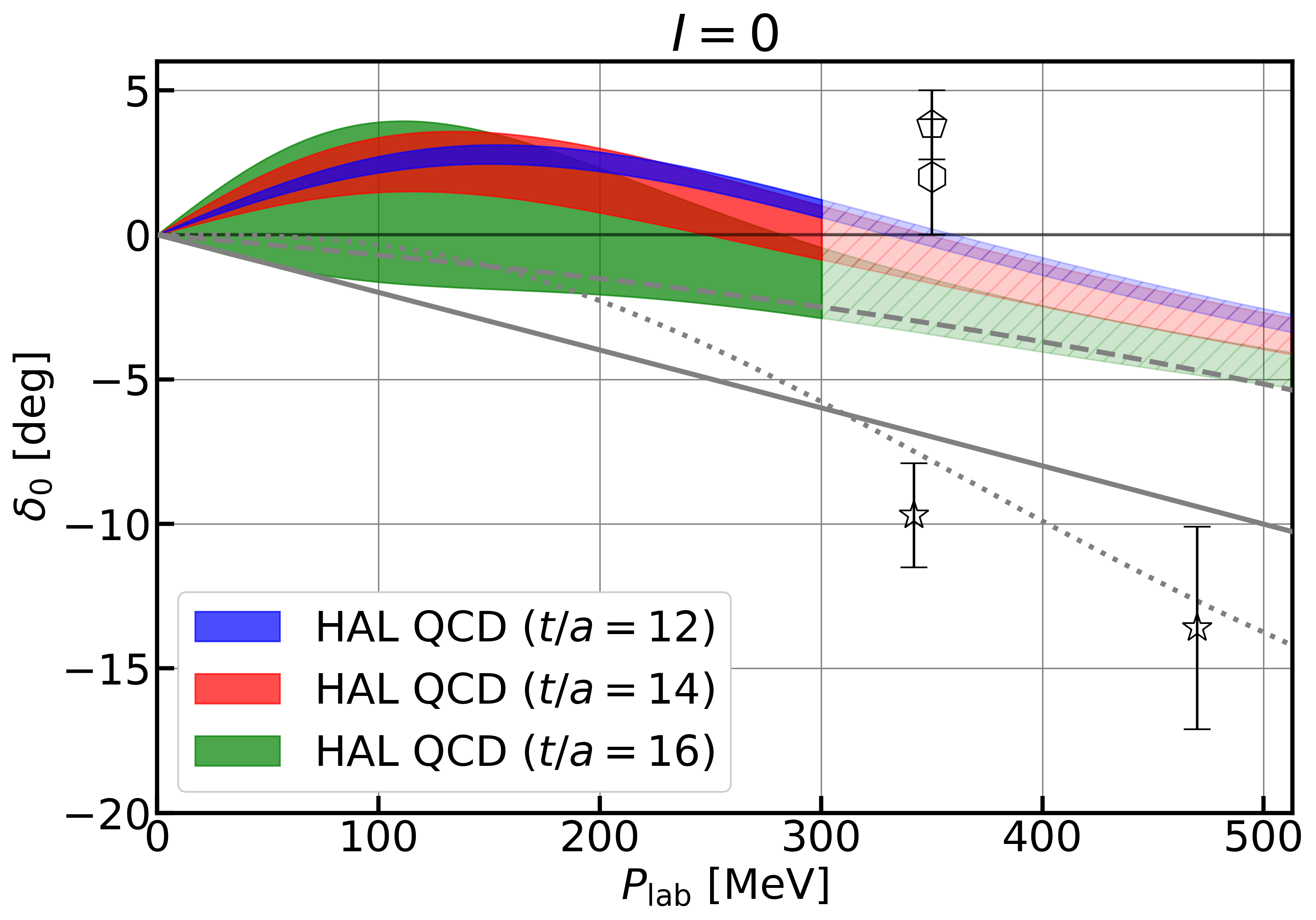}
\caption{Corrected phase shifts for the S-wave $KN$ scatterings in $I=1$ (left panel) and $I=0$ channels (right panel), replacing Fig.~3 of the original paper.}
  \label{fig:erratum_phase}
\end{figure}

\begin{figure}[htbp]
  \centering
  \includegraphics[width=0.49\textwidth]{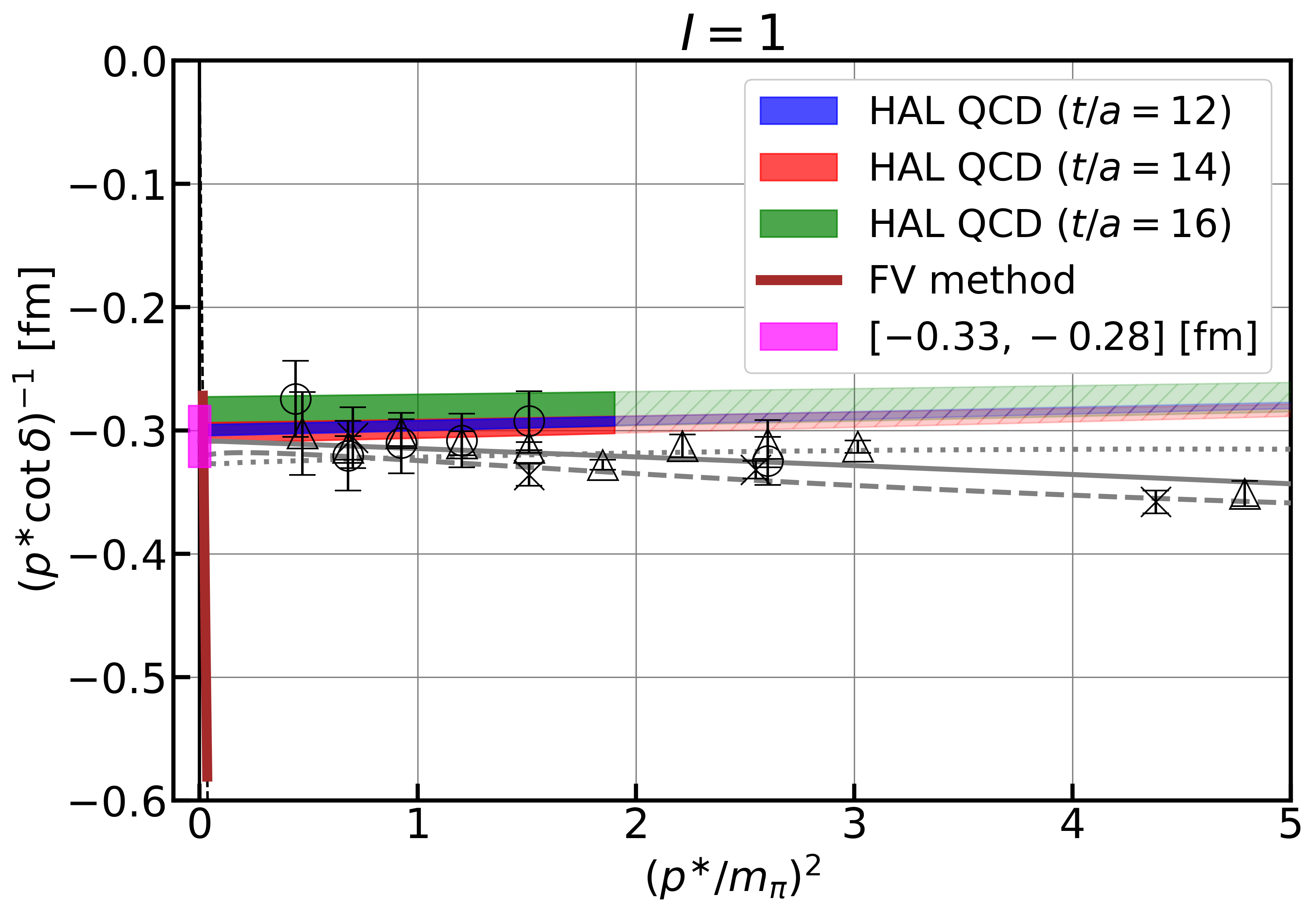}
  \includegraphics[width=0.49\textwidth]{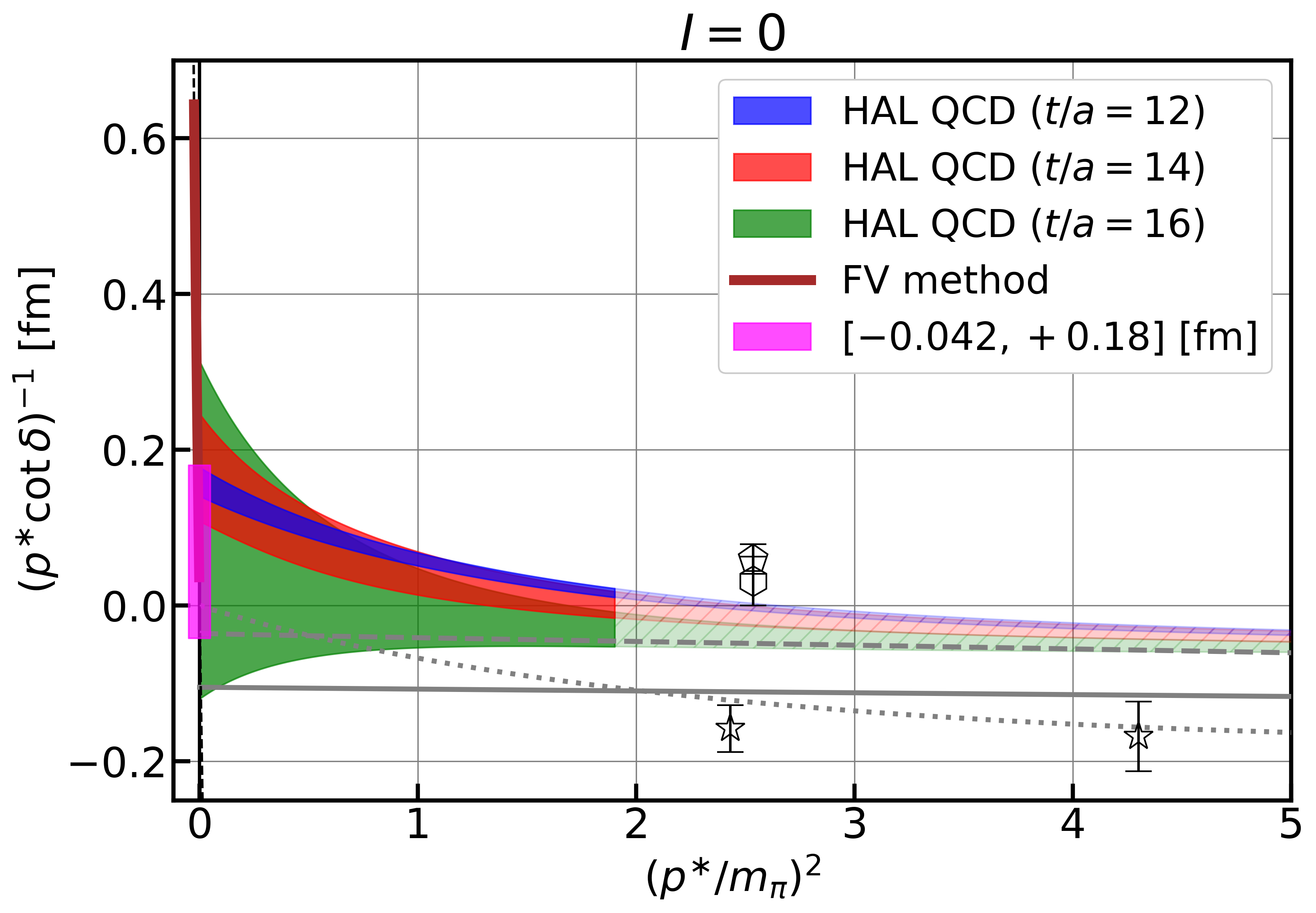}
\caption{Corrected $(p^{\ast}\cot{\delta_{0}(p^{\ast})})^{-1}$ for the S-wave $KN$ scatterings in $I=1$ (left panel) and $I=0$ channels (right panel), replacing Fig.~4 of the original paper.}
  \label{fig:erratum_kcotdinv}
\end{figure}

\begin{figure}[htbp]
  \centering
  \includegraphics[width=0.49\textwidth]{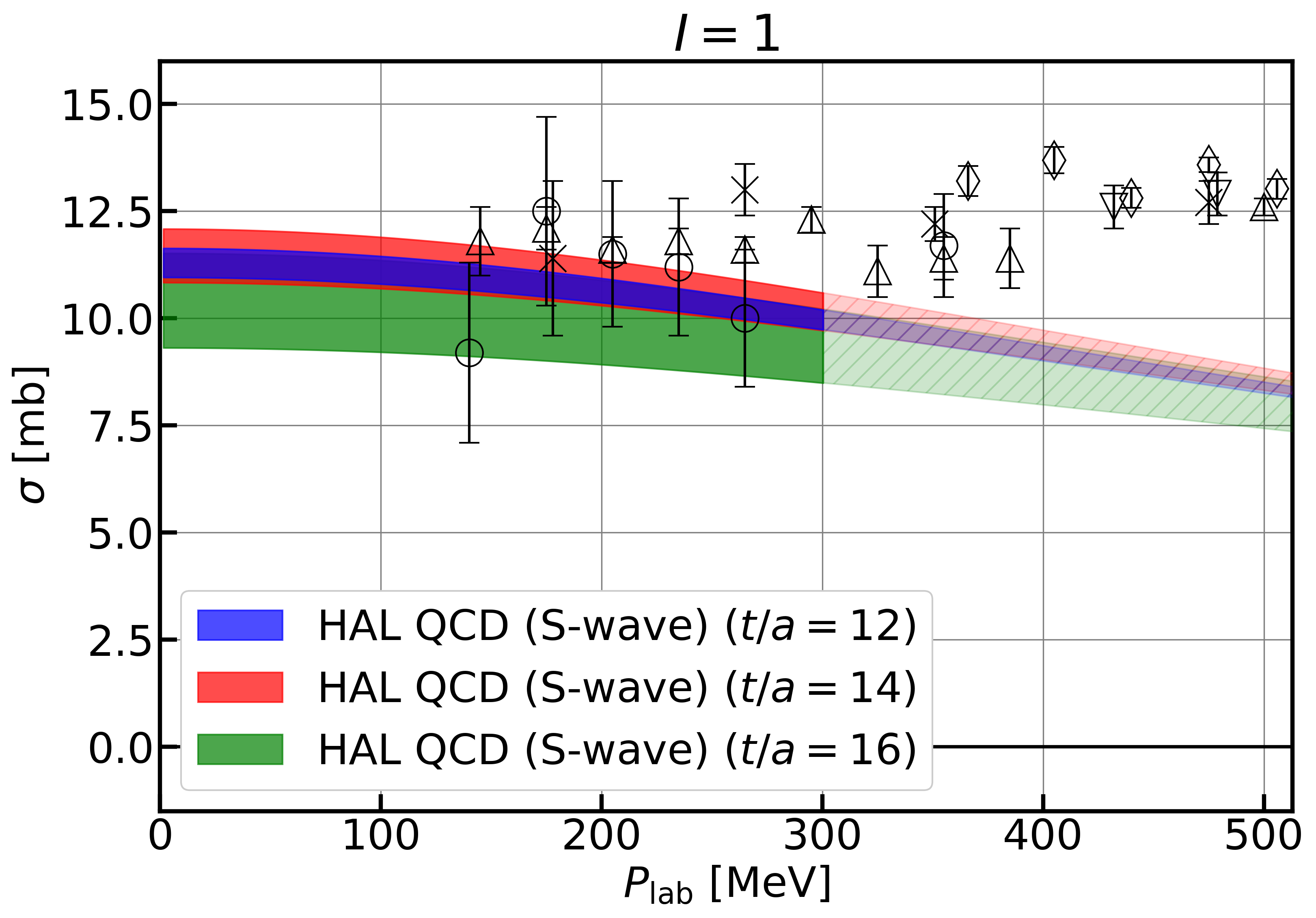}
  \includegraphics[width=0.49\textwidth]{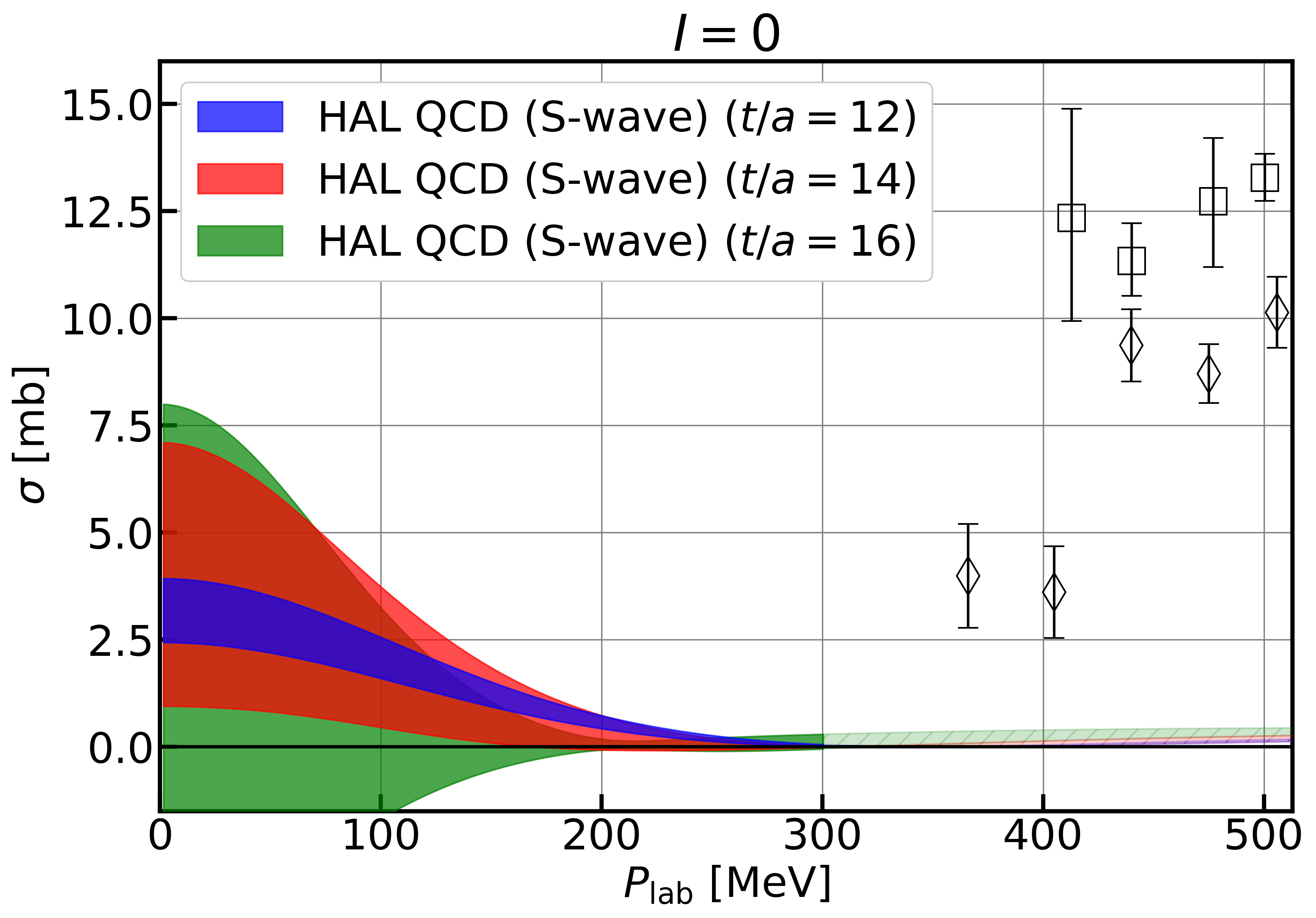}
\caption{Corrected S-wave total cross sections in $I=1$ (left panel) and $I=0$ channels (right panel), replacing Fig.~5 of the original paper.}
  \label{fig:erratum_crosssec}
\end{figure}

The corrected scattering lengths and effective range are
\begin{eqnarray}
\begin{aligned}
a^{I=1}_{0} &= -0.302 (8)(^{+14}_{-0})~\textrm{fm},\quad
r^{I=1}_{\textrm{eff}} = -0.149 (26)(^{+23}_{-24})~\textrm{fm}, \\
a^{I=0}_{0} &= +0.177 (69)(^{+0}_{-84})~\textrm{fm},
\end{aligned}
\end{eqnarray}
replacing the values given in Eq.~(13) of the original paper.
Note that, in accordance with the original paper, systematic errors are estimated from the differences between the central values at $t/a = 14$ and the results at $t/a = 12–16$.
Other uncertainties, including those associated with fit function dependence, are discussed separately in Sec. III below, following the structure of the original paper.

As in the original paper, the corrected phase shifts in both isospin channels start from $0$ degrees at the threshold and show neither a sudden rise nor a crossing of $90$ degrees. 
They therefore continue to indicate the absence of bound or resonant states corresponding to the $\Theta^{+}(1540)$ pentaquark in the S-wave $KN$ system, as concluded in the original paper.
In the $I=1$ channel, the corrected phase shifts agree with experiment at small momenta, thereby resolving the discrepancy reported in the original paper.
The corresponding S-wave cross section,
$11.5\pm 0.6$ mb at the threshold, is also consistent with the experimental data.
In the $I=0$ channel, the corrected phase shifts move toward positive values at small momenta, leading to a positive central value for the scattering length and to a larger S-wave contribution to the total cross section in this region.  
The corrected results appear to differ from the partial-wave analyses near threshold, though the difference may partly reflect the lack of low-energy experimental data available to constrain those analyses. 
Above $P_{\rm lab}=300$ MeV, the corrected S-wave total cross section remains too small to account for the experimental data,
and the results therefore still suggest that the experimental $I=0$ cross section is dominated by P-wave contributions, 
as noted in the original paper.
We note that the present analysis follows the original paper
by considering only the S-wave contributions to the total cross sections and by estimating the statistical errors 
using the same jackknife analysis. 
Higher partial wave contributions and 
the statistical uncertainties in the total cross sections 
will be investigated in more detail elsewhere.

\section{Corrected results in Section~IV.C}
\label{sec:erratum_FV}

We then present the corrected results in Sec.~IV.C (``Discussions on systematic uncertainties'') of the original paper.

\begin{figure}[htbp]
  \centering
  \includegraphics[width=0.49\textwidth]{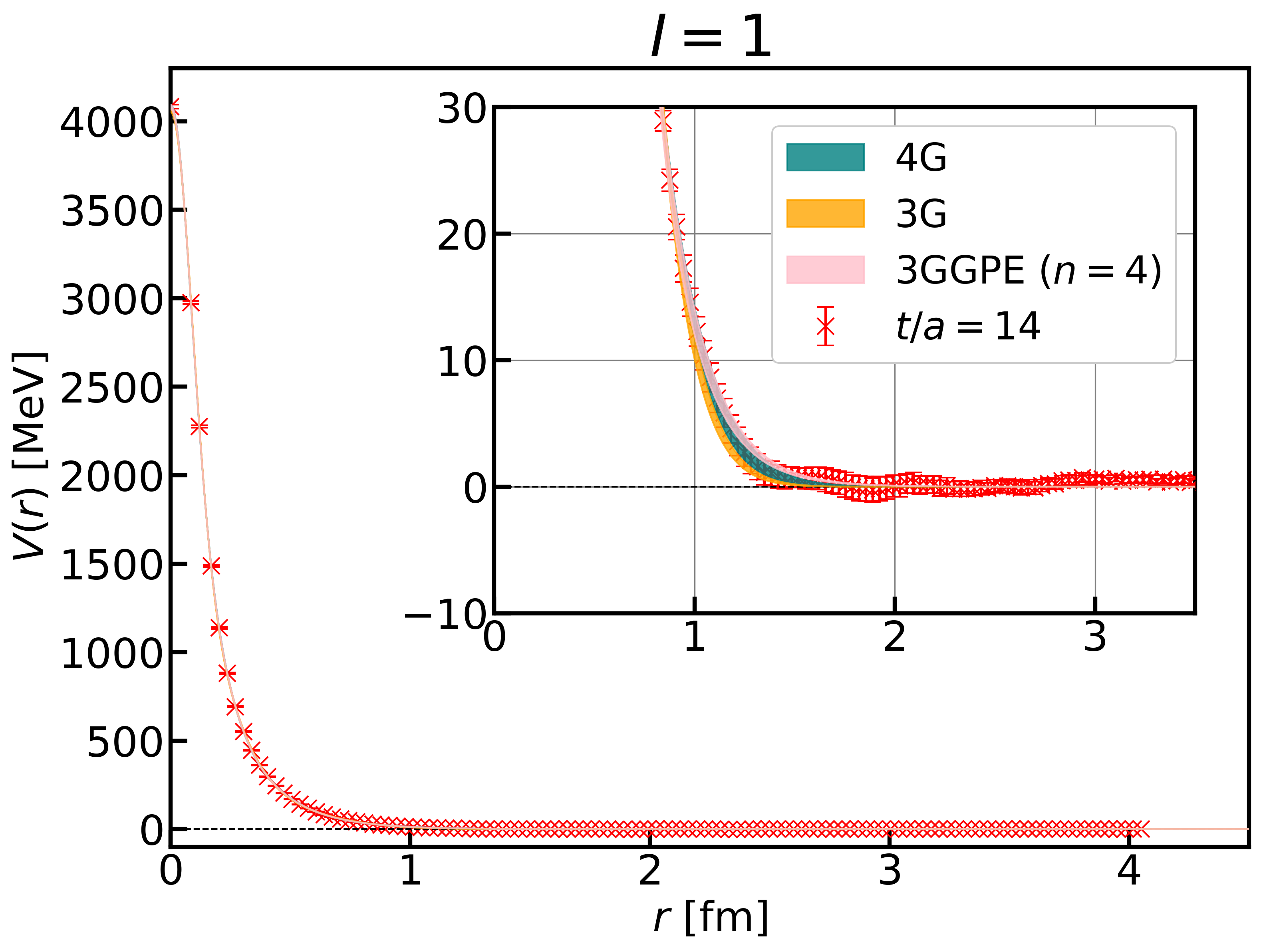}
  \includegraphics[width=0.49\textwidth]{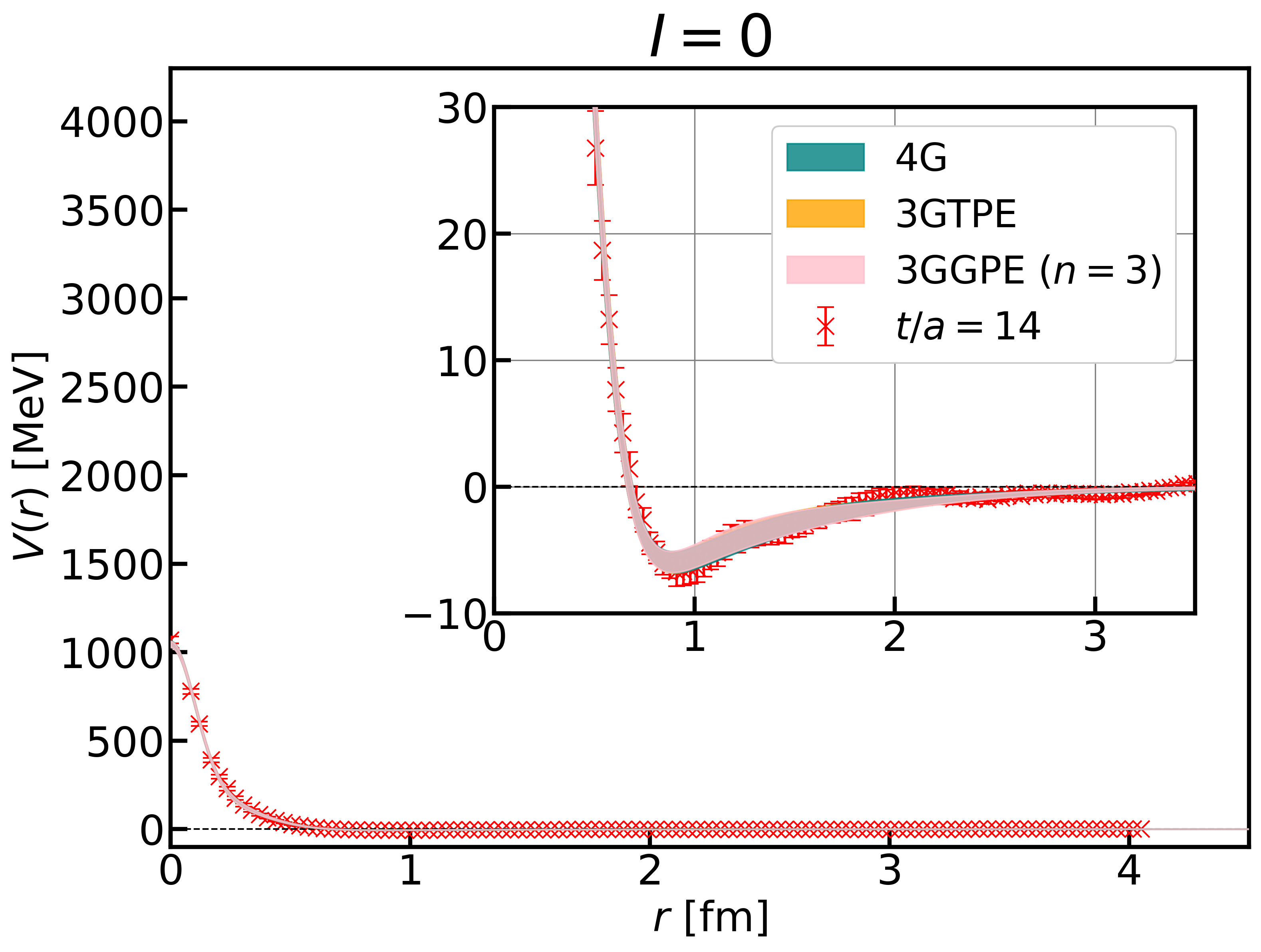}
\caption{Corrected fit results for different fit ansatz 
as well as corrected potential data
in $I=1$ (left panel) and $I=0$ channels (right panel), replacing Fig.~6 of the original paper.}
  \label{fig:erratum_potfit_comp}
\end{figure}

\begin{figure}[htbp]
  \centering
  \includegraphics[width=0.49\textwidth]{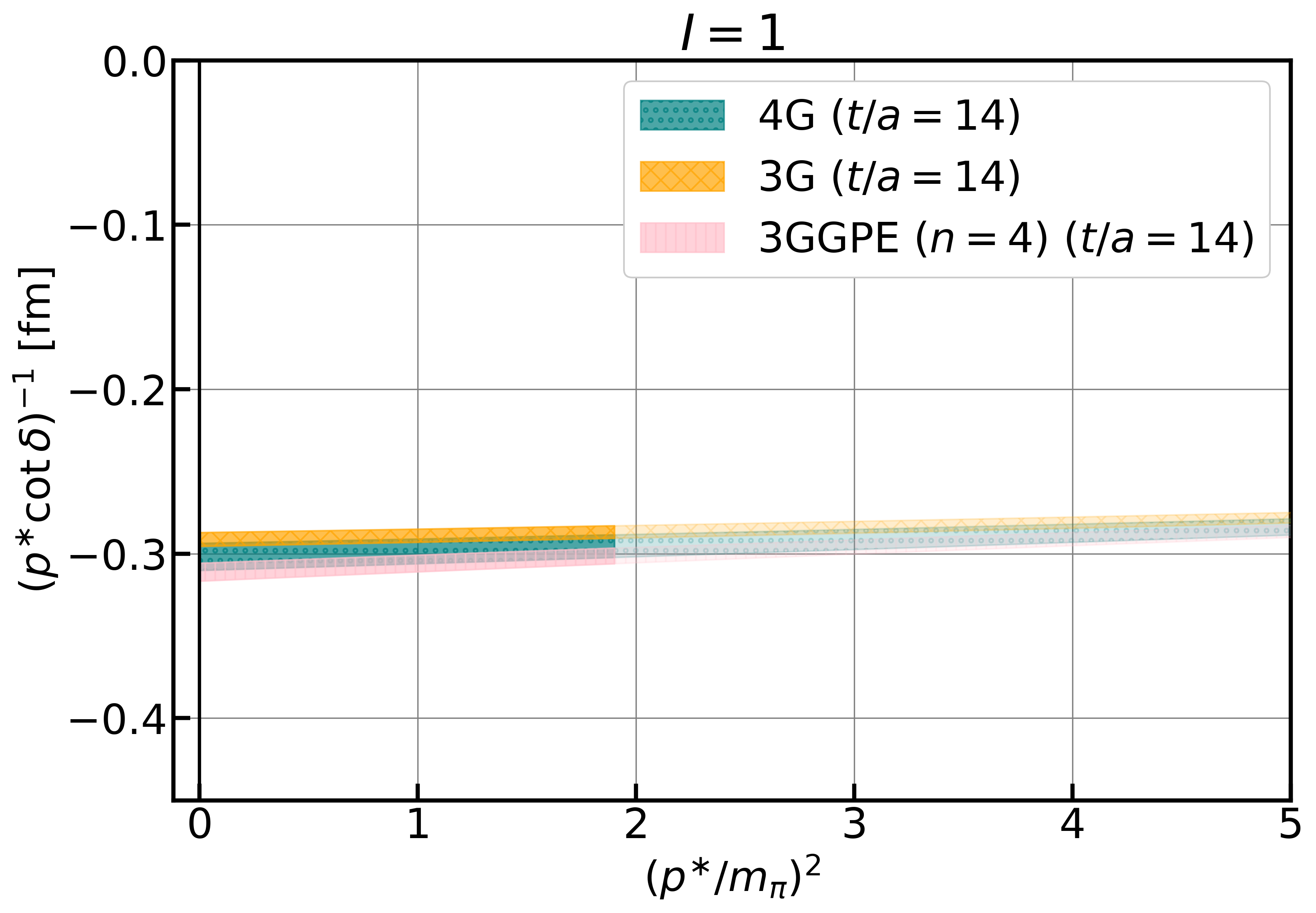}
  \includegraphics[width=0.49\textwidth]{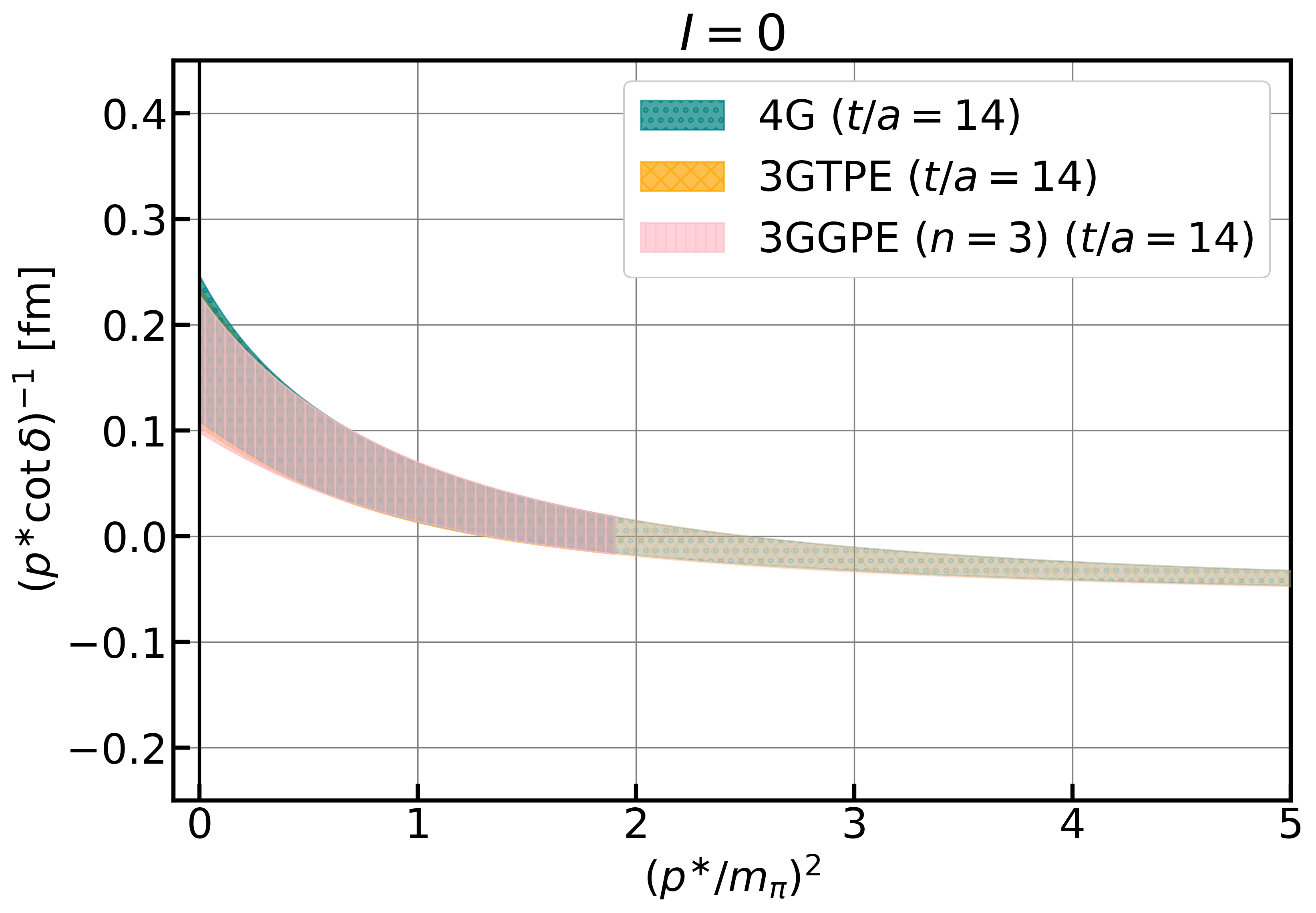}
\caption{Corrected $(p^{\ast}\cot{\delta_{0}(p^{\ast})})^{-1}$ for different fit ansatz in $I=1$ (left panel) and $I=0$ channels (right panel), replacing Fig.~7 of the original paper.
}
  \label{fig:erratum_kcotdinv_comp}
\end{figure}

The corrected scattering lengths evaluated from the ground-state energies on a finite volume via the $1/L$ expansion are
\begin{eqnarray}
\begin{aligned}
a^{I=1}_{0, \textrm{FV}} &= -0.434 (153)~\textrm{fm}, \quad a^{I=0}_{0, \textrm{FV}} &= +0.313 (299)~\textrm{fm},
\end{aligned}
\end{eqnarray}
replacing the values given below Eq.~(15) of the original paper.

In the corrected results, it is found that all the functions introduced in the original paper, $V^{\textrm{4G}}_{I}$,  $V^{\textrm{3G}}_{I}$,  $V^{\textrm{3GTPE}}_{I}$, and  $V^{\textrm{3GGPE}}_{I,n}$ with $n=3$ and $4$, 
provide reasonable fit results for both isospins at $t/a=14$.
The corrected phase shifts obtained using these fit functions are consistent with each other considering the statistical uncertainties.
A more detailed examination of the dependence of the phase shifts on the fit functions, together with an estimate of the associated systematic uncertainty, is left for future work.

For the systematic checks on remaining uncertainties in the original paper, the corrected results also lead to the same conclusions. 
The consistency between the scattering lengths from the potentials and those from the energy shifts holds also in the corrected results, as in the case of the original paper.
In addition, for the small difference in the kaon and nucleon masses between our setup and nature, and the lattice discretization effect at short distances, the estimates using the same methods as in the original paper continue to indicate no significant effect after the correction.

\section{Corrected results in Appendices A and B}
\label{sec:erratum_appAB}

In this section, we show the corrected results in Appendices~A (``Plots of the comparison of the LO $KN$ potentials between different pion masses'') and~B (``Energy shifts on finite volume from $KN$ temporal correlation functions with optimized operators'') of the original paper.

\begin{figure}[htbp]
  \centering
  \includegraphics[width=0.49\textwidth]{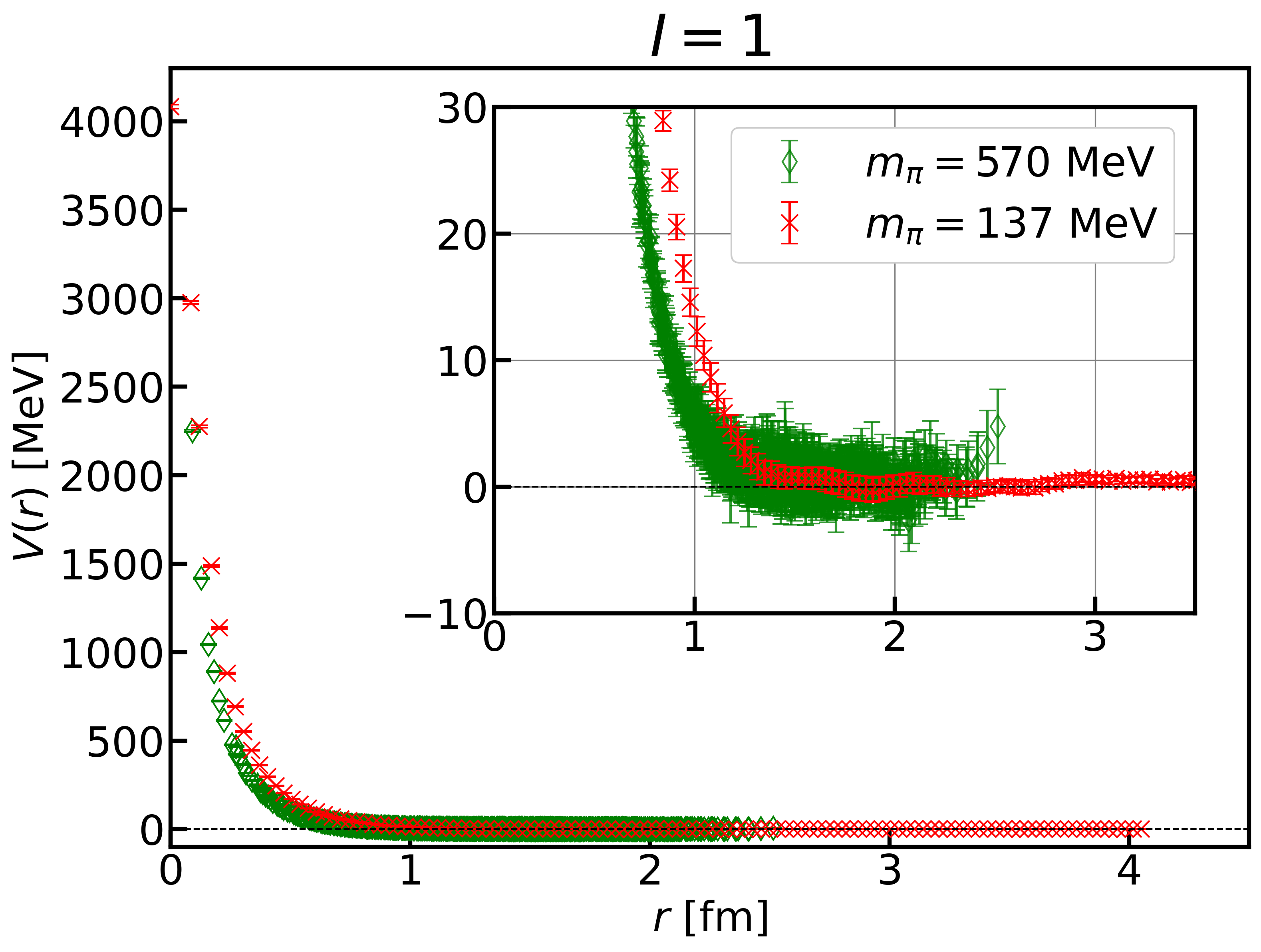}
  \includegraphics[width=0.49\textwidth]{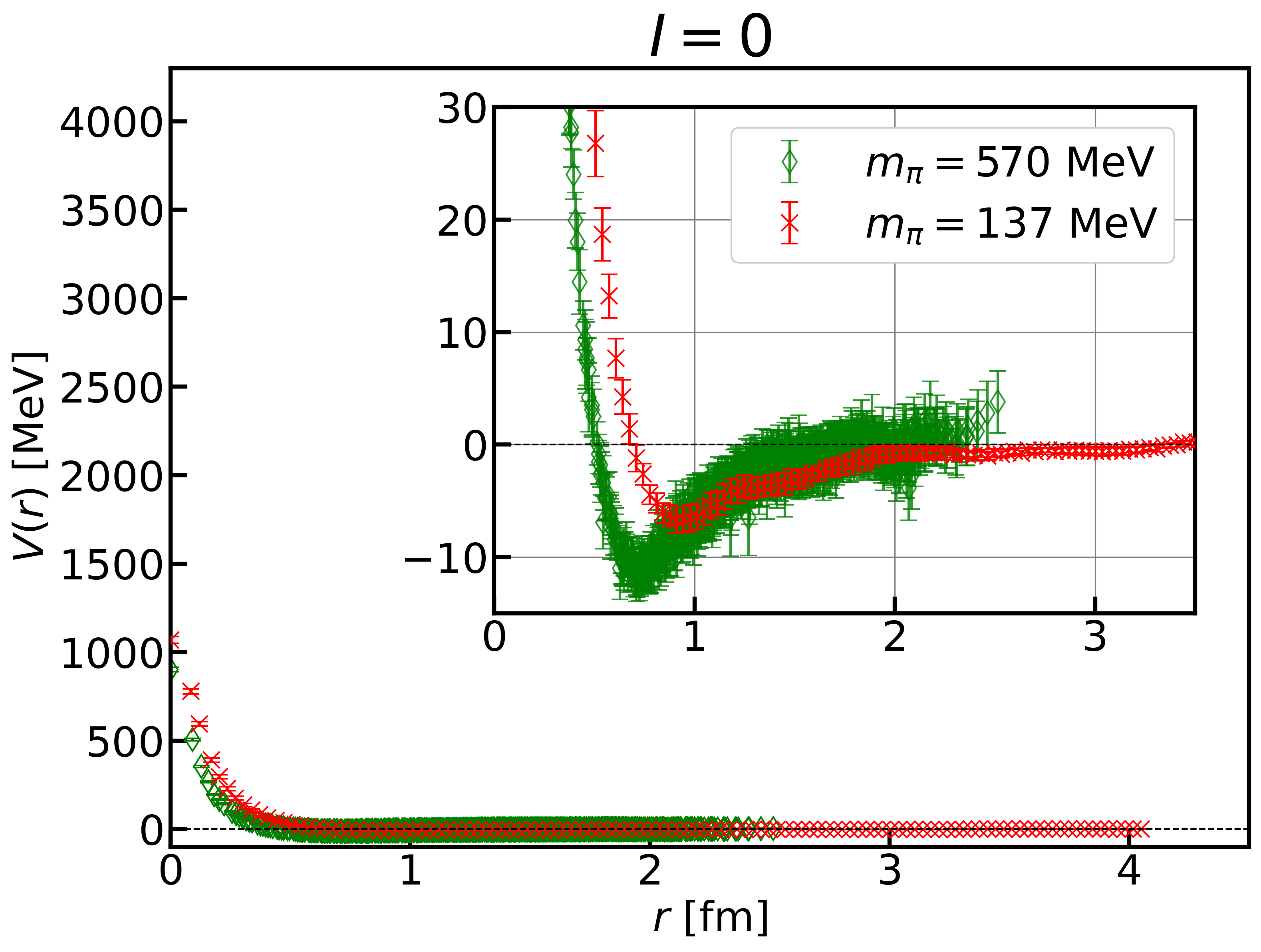}
\caption{Corrected comparison between the LO potentials at $m_{\pi}\approx 137~\textrm{MeV}$ with those obtained in the previous study at $m_{\pi}\approx 570~\textrm{MeV}$, for $I=1$ (left panel) and $I=0$ (right panel), replacing Fig.~8 of the original paper.}
  \label{fig:erratum_pot_comp_570MeV}
\end{figure}

In corrected Fig.~\ref{fig:erratum_pot_comp_570MeV},
we compare the LO $KN$ potentials at two different quark masses. 
Note that the correction was required only for the new results 
at $m_{\pi}\approx 137~\textrm{MeV}$,
whereas the results at $m_{\pi}\approx 570~\textrm{MeV}$
obtained in our previous study remain unchanged. 
In the corrected results for $I=1$, the quark mass dependence
is more clearly observed, as the corrected potential at $m_{\pi} \approx 137~\textrm{MeV}$ is more repulsive.

\begin{figure}[htbp]
  \centering
  \includegraphics[width=0.49\textwidth]{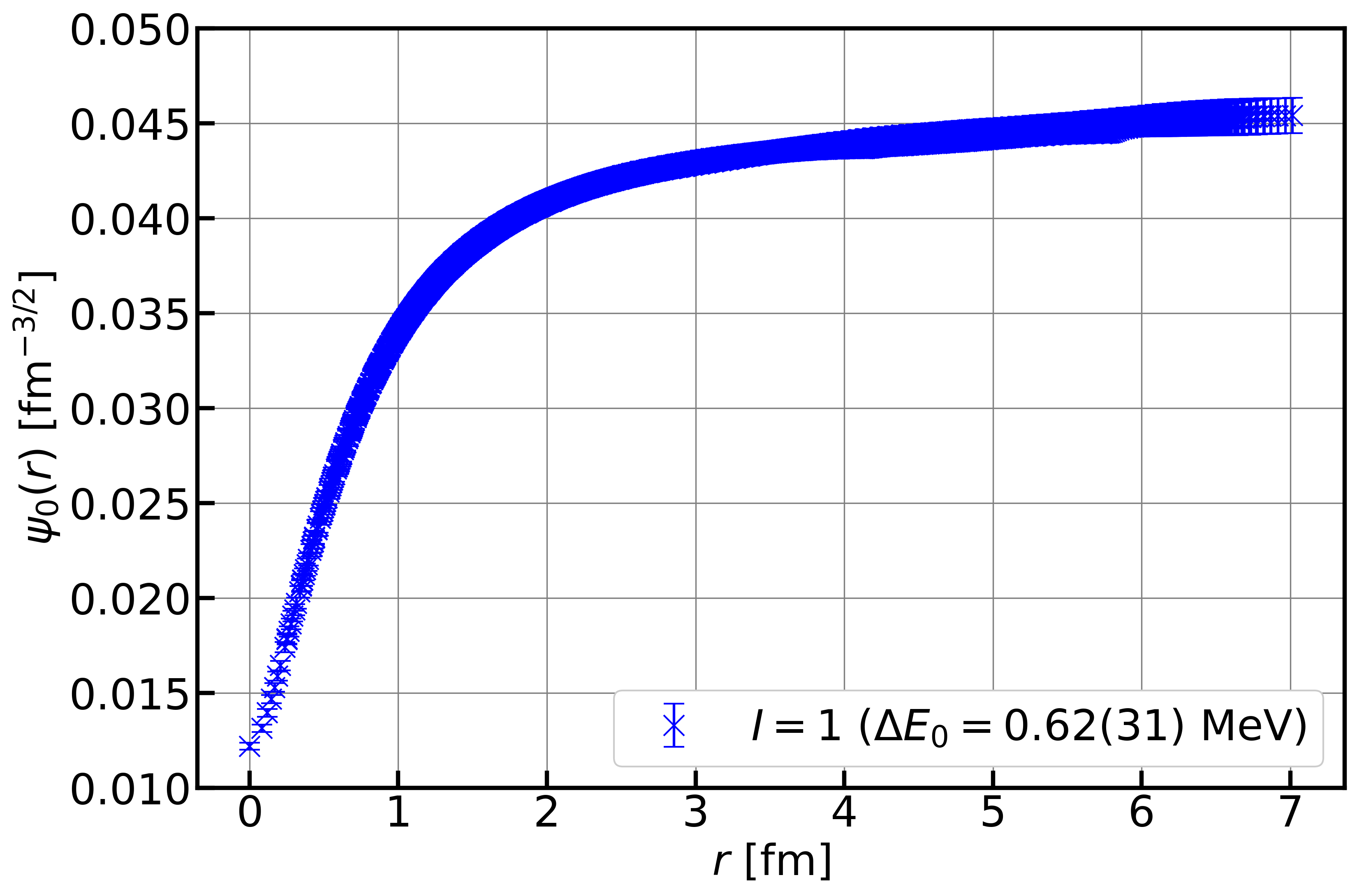}
  \includegraphics[width=0.49\textwidth]{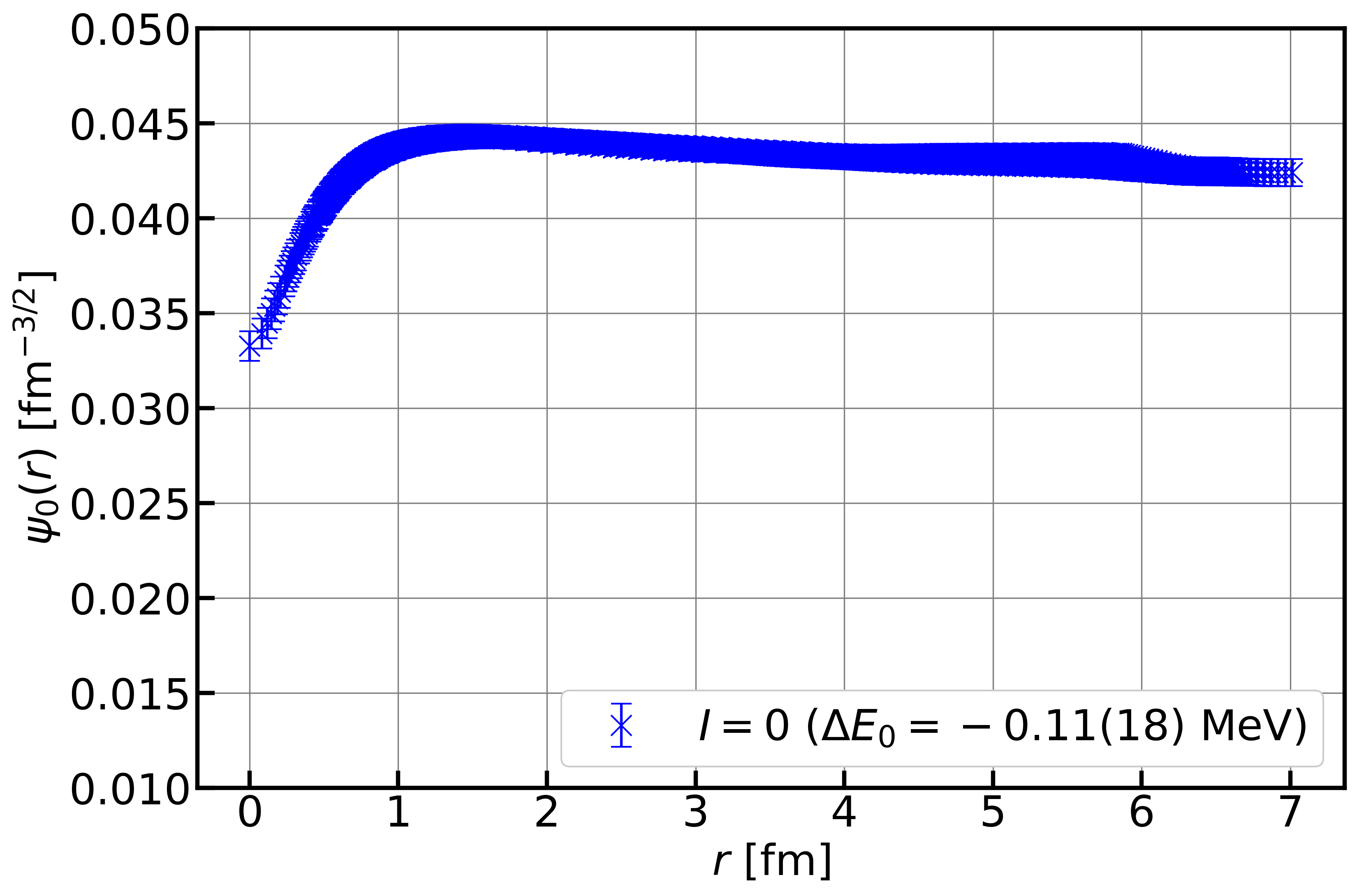}
\caption{Corrected wave function of the ground state on the lattice in the finite volume with the potential in $I=1$ (left panel) and $I=0$ channels (right panel), replacing Fig.~9 of the original paper.}
  \label{fig:erratum_wavefunc_FV}
\end{figure}

\begin{figure}[htbp]
  \centering
  \includegraphics[width=0.49\textwidth]{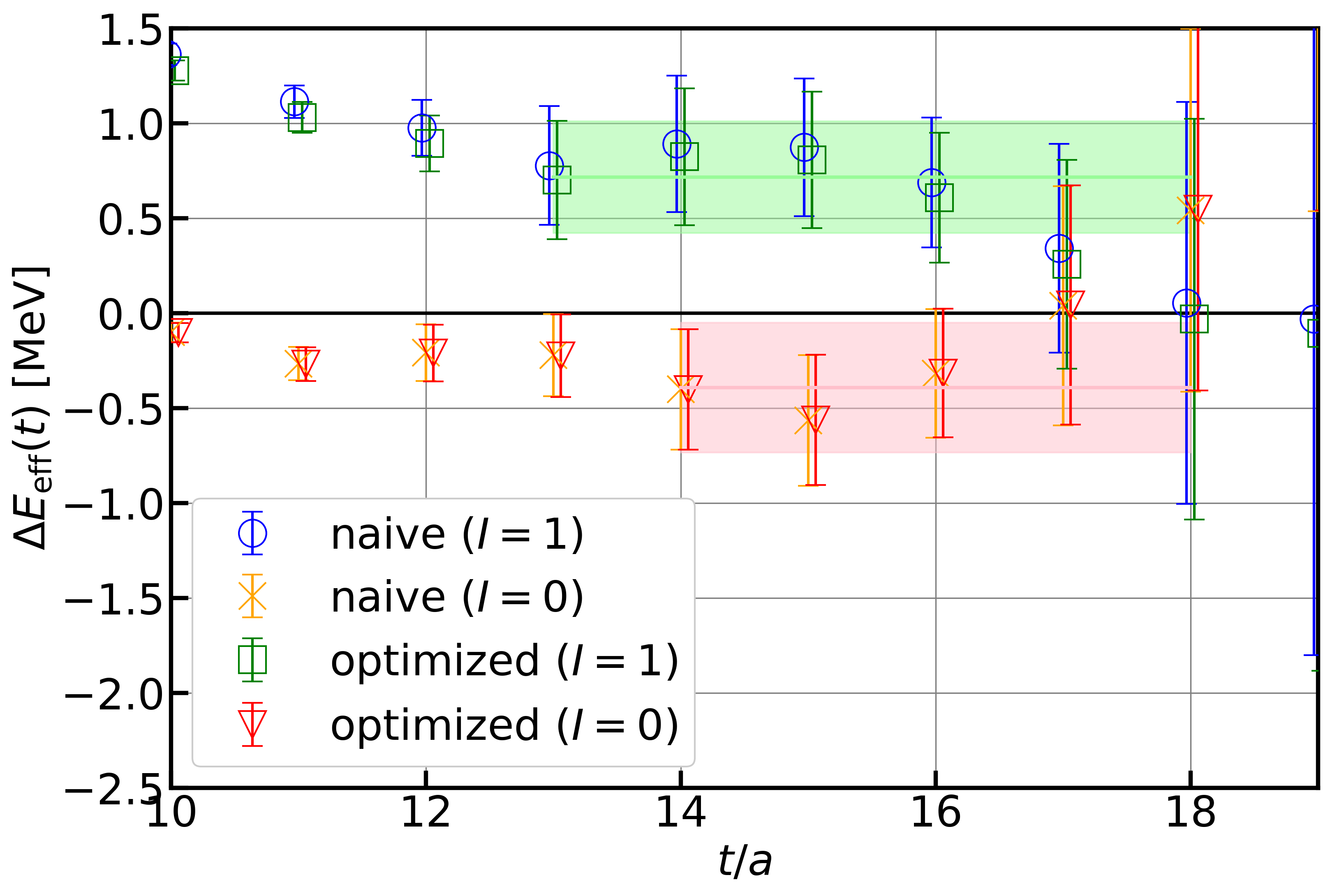}
\caption{Corrected effective energy for the two-point correlation functions with the optimized and naive sink operators, replacing Fig.~10 of the original paper.
}
  \label{fig:erratum_eff_energy}
\end{figure}

The corrected energy shifts of the $KN$ ground states on the finite volume are
\begin{eqnarray}
\begin{aligned}
E^{I=1}_{\textrm{gs}}-m_{N}-m_{K} &= +0.716(295)~\textrm{MeV},\\
E^{I=0}_{\textrm{gs}}-m_{N}-m_{K} &= -0.392(342)~\textrm{MeV},
\end{aligned}
\end{eqnarray}
replacing the values given in Eq.~(B2) of the original paper.
Compared with the original results, the magnitudes of the energy shifts are larger for both isospins.
These energy shifts are consistent with
the eigenenergies, $\Delta E_0$,
shown in the legends of the corrected Fig. 9.

\section{Summary of changes to the central conclusions}
\label{sec:erratum_conclusion}

Finally, changes to and unchanged aspects of the central conclusions of the original paper are summarized.
The conclusion regarding the absence of a bound or resonant state corresponding to the pentaquark $\Theta^{+} (1540)$ in the S-wave kaon-nucleon system remains unchanged.
In the $I=1$ channel, the discrepancy between our results and the experimental data reported in the original paper is now resolved by the correction.
In the $I=0$ channel, the phase shifts move toward positive values at small momenta, leading to an enhanced S-wave cross section near threshold. 
Nevertheless, the corrected results still suggest P-wave dominance in the scattering amplitude,
as in the original paper.
Finally, unlike in the original paper, 
the corrected fit results do not allow a robust conclusion about the magnitude of the TPE contribution.



\begin{thebibliography}{0}%
\makeatletter
\providecommand \@ifxundefined [1]{%
 \@ifx{#1\undefined}
}%
\providecommand \@ifnum [1]{%
 \ifnum #1\expandafter \@firstoftwo
 \else \expandafter \@secondoftwo
 \fi
}%
\providecommand \@ifx [1]{%
 \ifx #1\expandafter \@firstoftwo
 \else \expandafter \@secondoftwo
 \fi
}%
\providecommand \natexlab [1]{#1}%
\providecommand \enquote  [1]{``#1''}%
\providecommand \bibnamefont  [1]{#1}%
\providecommand \bibfnamefont [1]{#1}%
\providecommand \citenamefont [1]{#1}%
\providecommand \href@noop [0]{\@secondoftwo}%
\providecommand \href [0]{\begingroup \@sanitize@url \@href}%
\providecommand \@href[1]{\@@startlink{#1}\@@href}%
\providecommand \@@href[1]{\endgroup#1\@@endlink}%
\providecommand \@sanitize@url [0]{\catcode `\\12\catcode `\$12\catcode
  `\&12\catcode `\#12\catcode `\^12\catcode `\_12\catcode `\%12\relax}%
\providecommand \@@startlink[1]{}%
\providecommand \@@endlink[0]{}%
\providecommand \url  [0]{\begingroup\@sanitize@url \@url }%
\providecommand \@url [1]{\endgroup\@href {#1}{\urlprefix }}%
\providecommand \urlprefix  [0]{URL }%
\providecommand \Eprint [0]{\href }%
\providecommand \doibase [0]{https://doi.org/}%
\providecommand \selectlanguage [0]{\@gobble}%
\providecommand \bibinfo  [0]{\@secondoftwo}%
\providecommand \bibfield  [0]{\@secondoftwo}%
\providecommand \translation [1]{[#1]}%
\providecommand \BibitemOpen [0]{}%
\providecommand \bibitemStop [0]{}%
\providecommand \bibitemNoStop [0]{.\EOS\space}%
\providecommand \EOS [0]{\spacefactor3000\relax}%
\providecommand \BibitemShut  [1]{\csname bibitem#1\endcsname}%
\let\auto@bib@innerbib\@empty
\end{thebibliography}%


\providecommand{\noopsort}[1]{}\providecommand{\singleletter}[1]{#1}%
\begin{thebibliography}{52}%
\makeatletter
\providecommand \@ifxundefined [1]{%
 \@ifx{#1\undefined}
}%
\providecommand \@ifnum [1]{%
 \ifnum #1\expandafter \@firstoftwo
 \else \expandafter \@secondoftwo
 \fi
}%
\providecommand \@ifx [1]{%
 \ifx #1\expandafter \@firstoftwo
 \else \expandafter \@secondoftwo
 \fi
}%
\providecommand \natexlab [1]{#1}%
\providecommand \enquote  [1]{``#1''}%
\providecommand \bibnamefont  [1]{#1}%
\providecommand \bibfnamefont [1]{#1}%
\providecommand \citenamefont [1]{#1}%
\providecommand \href@noop [0]{\@secondoftwo}%
\providecommand \href [0]{\begingroup \@sanitize@url \@href}%
\providecommand \@href[1]{\@@startlink{#1}\@@href}%
\providecommand \@@href[1]{\endgroup#1\@@endlink}%
\providecommand \@sanitize@url [0]{\catcode `\\12\catcode `\$12\catcode
  `\&12\catcode `\#12\catcode `\^12\catcode `\_12\catcode `\%12\relax}%
\providecommand \@@startlink[1]{}%
\providecommand \@@endlink[0]{}%
\providecommand \url  [0]{\begingroup\@sanitize@url \@url }%
\providecommand \@url [1]{\endgroup\@href {#1}{\urlprefix }}%
\providecommand \urlprefix  [0]{URL }%
\providecommand \Eprint [0]{\href }%
\providecommand \doibase [0]{https://doi.org/}%
\providecommand \selectlanguage [0]{\@gobble}%
\providecommand \bibinfo  [0]{\@secondoftwo}%
\providecommand \bibfield  [0]{\@secondoftwo}%
\providecommand \translation [1]{[#1]}%
\providecommand \BibitemOpen [0]{}%
\providecommand \bibitemStop [0]{}%
\providecommand \bibitemNoStop [0]{.\EOS\space}%
\providecommand \EOS [0]{\spacefactor3000\relax}%
\providecommand \BibitemShut  [1]{\csname bibitem#1\endcsname}%
\let\auto@bib@innerbib\@empty
\bibitem [{\citenamefont {Ishii}\ \emph {et~al.}(2007)\citenamefont {Ishii},
  \citenamefont {Aoki},\ and\ \citenamefont {Hatsuda}}]{Ishii:2006ec}%
  \BibitemOpen
  \bibfield  {author} {\bibinfo {author} {\bibfnamefont {N.}~\bibnamefont
  {Ishii}}, \bibinfo {author} {\bibfnamefont {S.}~\bibnamefont {Aoki}},\ and\
  \bibinfo {author} {\bibfnamefont {T.}~\bibnamefont {Hatsuda}},\ }\bibfield
  {title} {\bibinfo {title} {{The Nuclear Force from Lattice QCD}},\ }\href
  {https://doi.org/10.1103/PhysRevLett.99.022001} {\bibfield  {journal}
  {\bibinfo  {journal} {Phys. Rev. Lett.}\ }\textbf {\bibinfo {volume} {99}},\
  \bibinfo {pages} {022001} (\bibinfo {year} {2007})},\ \Eprint
  {https://arxiv.org/abs/nucl-th/0611096} {arXiv:nucl-th/0611096} \BibitemShut
  {NoStop}%
\bibitem [{\citenamefont {Aoki}\ \emph {et~al.}(2010)\citenamefont {Aoki},
  \citenamefont {Hatsuda},\ and\ \citenamefont {Ishii}}]{Aoki:2009ji}%
  \BibitemOpen
  \bibfield  {author} {\bibinfo {author} {\bibfnamefont {S.}~\bibnamefont
  {Aoki}}, \bibinfo {author} {\bibfnamefont {T.}~\bibnamefont {Hatsuda}},\ and\
  \bibinfo {author} {\bibfnamefont {N.}~\bibnamefont {Ishii}},\ }\bibfield
  {title} {\bibinfo {title} {{Theoretical Foundation of the Nuclear Force in
  QCD and its applications to Central and Tensor Forces in Quenched Lattice QCD
  Simulations}},\ }\href {https://doi.org/10.1143/PTP.123.89} {\bibfield
  {journal} {\bibinfo  {journal} {Prog. Theor. Phys.}\ }\textbf {\bibinfo
  {volume} {123}},\ \bibinfo {pages} {89} (\bibinfo {year} {2010})},\ \Eprint
  {https://arxiv.org/abs/0909.5585} {arXiv:0909.5585 [hep-lat]} \BibitemShut
  {NoStop}%
\bibitem [{\citenamefont {Ishii}\ \emph {et~al.}(2012)\citenamefont {Ishii},
  \citenamefont {Aoki}, \citenamefont {Doi}, \citenamefont {Hatsuda},
  \citenamefont {Ikeda}, \citenamefont {Inoue}, \citenamefont {Murano},
  \citenamefont {Nemura},\ and\ \citenamefont {Sasaki}}]{Ishii:2012ssm}%
  \BibitemOpen
  \bibfield  {author} {\bibinfo {author} {\bibfnamefont {N.}~\bibnamefont
  {Ishii}}, \bibinfo {author} {\bibfnamefont {S.}~\bibnamefont {Aoki}},
  \bibinfo {author} {\bibfnamefont {T.}~\bibnamefont {Doi}}, \bibinfo {author}
  {\bibfnamefont {T.}~\bibnamefont {Hatsuda}}, \bibinfo {author} {\bibfnamefont
  {Y.}~\bibnamefont {Ikeda}}, \bibinfo {author} {\bibfnamefont
  {T.}~\bibnamefont {Inoue}}, \bibinfo {author} {\bibfnamefont
  {K.}~\bibnamefont {Murano}}, \bibinfo {author} {\bibfnamefont
  {H.}~\bibnamefont {Nemura}},\ and\ \bibinfo {author} {\bibfnamefont
  {K.}~\bibnamefont {Sasaki}} (\bibinfo {collaboration} {HAL QCD}),\ }\bibfield
   {title} {\bibinfo {title} {{Hadron\textendash{}hadron interactions from
  imaginary-time Nambu\textendash{}Bethe\textendash{}Salpeter wave function on
  the lattice}},\ }\href {https://doi.org/10.1016/j.physletb.2012.04.076}
  {\bibfield  {journal} {\bibinfo  {journal} {Phys. Lett. B}\ }\textbf
  {\bibinfo {volume} {712}},\ \bibinfo {pages} {437} (\bibinfo {year}
  {2012})},\ \Eprint {https://arxiv.org/abs/1203.3642} {arXiv:1203.3642
  [hep-lat]} \BibitemShut {NoStop}%
\bibitem [{\citenamefont {Doi}\ \emph {et~al.}(2018)\citenamefont {Doi} \emph
  {et~al.}}]{Doi:2017zov}%
  \BibitemOpen
  \bibfield  {author} {\bibinfo {author} {\bibfnamefont {T.}~\bibnamefont
  {Doi}} \emph {et~al.},\ }\bibfield  {title} {\bibinfo {title} {{Baryon
  interactions from lattice QCD with physical quark masses -- Nuclear forces
  and $\Xi\Xi$ forces --}},\ }\href
  {https://doi.org/10.1051/epjconf/201817505009} {\bibfield  {journal}
  {\bibinfo  {journal} {EPJ Web Conf.}\ }\textbf {\bibinfo {volume} {175}},\
  \bibinfo {pages} {05009} (\bibinfo {year} {2018})},\ \Eprint
  {https://arxiv.org/abs/1711.01952} {arXiv:1711.01952 [hep-lat]} \BibitemShut
  {NoStop}%
\bibitem [{\citenamefont {Gongyo}\ \emph {et~al.}(2018)\citenamefont {Gongyo}
  \emph {et~al.}}]{Gongyo:2017fjb}%
  \BibitemOpen
  \bibfield  {author} {\bibinfo {author} {\bibfnamefont {S.}~\bibnamefont
  {Gongyo}} \emph {et~al.},\ }\bibfield  {title} {\bibinfo {title} {{Most
  Strange Dibaryon from Lattice QCD}},\ }\href
  {https://doi.org/10.1103/PhysRevLett.120.212001} {\bibfield  {journal}
  {\bibinfo  {journal} {Phys. Rev. Lett.}\ }\textbf {\bibinfo {volume} {120}},\
  \bibinfo {pages} {212001} (\bibinfo {year} {2018})},\ \Eprint
  {https://arxiv.org/abs/1709.00654} {arXiv:1709.00654 [hep-lat]} \BibitemShut
  {NoStop}%
\bibitem [{\citenamefont {Iritani}\ \emph
  {et~al.}(2019{\natexlab{a}})\citenamefont {Iritani} \emph
  {et~al.}}]{HALQCD:2018qyu}%
  \BibitemOpen
  \bibfield  {author} {\bibinfo {author} {\bibfnamefont {T.}~\bibnamefont
  {Iritani}} \emph {et~al.} (\bibinfo {collaboration} {HAL QCD}),\ }\bibfield
  {title} {\bibinfo {title} {{$N\Omega$ dibaryon from lattice QCD near the
  physical point}},\ }\href {https://doi.org/10.1016/j.physletb.2019.03.050}
  {\bibfield  {journal} {\bibinfo  {journal} {Phys. Lett. B}\ }\textbf
  {\bibinfo {volume} {792}},\ \bibinfo {pages} {284} (\bibinfo {year}
  {2019}{\natexlab{a}})},\ \Eprint {https://arxiv.org/abs/1810.03416}
  {arXiv:1810.03416 [hep-lat]} \BibitemShut {NoStop}%
\bibitem [{\citenamefont {Sasaki}\ \emph {et~al.}(2020)\citenamefont {Sasaki}
  \emph {et~al.}}]{HALQCD:2019wsz}%
  \BibitemOpen
  \bibfield  {author} {\bibinfo {author} {\bibfnamefont {K.}~\bibnamefont
  {Sasaki}} \emph {et~al.} (\bibinfo {collaboration} {HAL QCD}),\ }\bibfield
  {title} {\bibinfo {title} {{$\Lambda\Lambda$ and N$\Xi$ interactions from
  lattice QCD near the physical point}},\ }\href
  {https://doi.org/10.1016/j.nuclphysa.2020.121737} {\bibfield  {journal}
  {\bibinfo  {journal} {Nucl. Phys. A}\ }\textbf {\bibinfo {volume} {998}},\
  \bibinfo {pages} {121737} (\bibinfo {year} {2020})},\ \Eprint
  {https://arxiv.org/abs/1912.08630} {arXiv:1912.08630 [hep-lat]} \BibitemShut
  {NoStop}%
\bibitem [{\citenamefont {Lyu}\ \emph {et~al.}(2021)\citenamefont {Lyu},
  \citenamefont {Tong}, \citenamefont {Sugiura}, \citenamefont {Aoki},
  \citenamefont {Doi}, \citenamefont {Hatsuda}, \citenamefont {Meng},\ and\
  \citenamefont {Miyamoto}}]{Lyu:2021qsh}%
  \BibitemOpen
  \bibfield  {author} {\bibinfo {author} {\bibfnamefont {Y.}~\bibnamefont
  {Lyu}}, \bibinfo {author} {\bibfnamefont {H.}~\bibnamefont {Tong}}, \bibinfo
  {author} {\bibfnamefont {T.}~\bibnamefont {Sugiura}}, \bibinfo {author}
  {\bibfnamefont {S.}~\bibnamefont {Aoki}}, \bibinfo {author} {\bibfnamefont
  {T.}~\bibnamefont {Doi}}, \bibinfo {author} {\bibfnamefont {T.}~\bibnamefont
  {Hatsuda}}, \bibinfo {author} {\bibfnamefont {J.}~\bibnamefont {Meng}},\ and\
  \bibinfo {author} {\bibfnamefont {T.}~\bibnamefont {Miyamoto}},\ }\bibfield
  {title} {\bibinfo {title} {{Dibaryon with Highest Charm Number near Unitarity
  from Lattice QCD}},\ }\href {https://doi.org/10.1103/PhysRevLett.127.072003}
  {\bibfield  {journal} {\bibinfo  {journal} {Phys. Rev. Lett.}\ }\textbf
  {\bibinfo {volume} {127}},\ \bibinfo {pages} {072003} (\bibinfo {year}
  {2021})},\ \Eprint {https://arxiv.org/abs/2102.00181} {arXiv:2102.00181
  [hep-lat]} \BibitemShut {NoStop}%
\bibitem [{\citenamefont {Lyu}\ \emph {et~al.}(2022{\natexlab{a}})\citenamefont
  {Lyu}, \citenamefont {Doi}, \citenamefont {Hatsuda}, \citenamefont {Ikeda},
  \citenamefont {Meng}, \citenamefont {Sasaki},\ and\ \citenamefont
  {Sugiura}}]{Lyu:2022imf}%
  \BibitemOpen
  \bibfield  {author} {\bibinfo {author} {\bibfnamefont {Y.}~\bibnamefont
  {Lyu}}, \bibinfo {author} {\bibfnamefont {T.}~\bibnamefont {Doi}}, \bibinfo
  {author} {\bibfnamefont {T.}~\bibnamefont {Hatsuda}}, \bibinfo {author}
  {\bibfnamefont {Y.}~\bibnamefont {Ikeda}}, \bibinfo {author} {\bibfnamefont
  {J.}~\bibnamefont {Meng}}, \bibinfo {author} {\bibfnamefont {K.}~\bibnamefont
  {Sasaki}},\ and\ \bibinfo {author} {\bibfnamefont {T.}~\bibnamefont
  {Sugiura}},\ }\bibfield  {title} {\bibinfo {title} {{Attractive
  N-\ensuremath{\phi} interaction and two-pion tail from lattice QCD near
  physical point}},\ }\href {https://doi.org/10.1103/PhysRevD.106.074507}
  {\bibfield  {journal} {\bibinfo  {journal} {Phys. Rev. D}\ }\textbf {\bibinfo
  {volume} {106}},\ \bibinfo {pages} {074507} (\bibinfo {year}
  {2022}{\natexlab{a}})},\ \Eprint {https://arxiv.org/abs/2205.10544}
  {arXiv:2205.10544 [hep-lat]} \BibitemShut {NoStop}%
\bibitem [{\citenamefont {Lyu}\ \emph {et~al.}(2023)\citenamefont {Lyu},
  \citenamefont {Aoki}, \citenamefont {Doi}, \citenamefont {Hatsuda},
  \citenamefont {Ikeda},\ and\ \citenamefont {Meng}}]{Lyu:2023xro}%
  \BibitemOpen
  \bibfield  {author} {\bibinfo {author} {\bibfnamefont {Y.}~\bibnamefont
  {Lyu}}, \bibinfo {author} {\bibfnamefont {S.}~\bibnamefont {Aoki}}, \bibinfo
  {author} {\bibfnamefont {T.}~\bibnamefont {Doi}}, \bibinfo {author}
  {\bibfnamefont {T.}~\bibnamefont {Hatsuda}}, \bibinfo {author} {\bibfnamefont
  {Y.}~\bibnamefont {Ikeda}},\ and\ \bibinfo {author} {\bibfnamefont
  {J.}~\bibnamefont {Meng}},\ }\bibfield  {title} {\bibinfo {title} {{Doubly
  Charmed Tetraquark Tcc+ from Lattice QCD near Physical Point}},\ }\href
  {https://doi.org/10.1103/PhysRevLett.131.161901} {\bibfield  {journal}
  {\bibinfo  {journal} {Phys. Rev. Lett.}\ }\textbf {\bibinfo {volume} {131}},\
  \bibinfo {pages} {161901} (\bibinfo {year} {2023})},\ \Eprint
  {https://arxiv.org/abs/2302.04505} {arXiv:2302.04505 [hep-lat]} \BibitemShut
  {NoStop}%
\bibitem [{\citenamefont {Lyu}\ \emph {et~al.}(2025)\citenamefont {Lyu},
  \citenamefont {Doi}, \citenamefont {Hatsuda},\ and\ \citenamefont
  {Sugiura}}]{Lyu:2024ttm}%
  \BibitemOpen
  \bibfield  {author} {\bibinfo {author} {\bibfnamefont {Y.}~\bibnamefont
  {Lyu}}, \bibinfo {author} {\bibfnamefont {T.}~\bibnamefont {Doi}}, \bibinfo
  {author} {\bibfnamefont {T.}~\bibnamefont {Hatsuda}},\ and\ \bibinfo {author}
  {\bibfnamefont {T.}~\bibnamefont {Sugiura}},\ }\bibfield  {title} {\bibinfo
  {title} {{Nucleon-charmonium interactions from lattice QCD}},\ }\href
  {https://doi.org/10.1016/j.physletb.2024.139178} {\bibfield  {journal}
  {\bibinfo  {journal} {Phys. Lett. B}\ }\textbf {\bibinfo {volume} {860}},\
  \bibinfo {pages} {139178} (\bibinfo {year} {2025})},\ \Eprint
  {https://arxiv.org/abs/2410.22755} {arXiv:2410.22755 [hep-lat]} \BibitemShut
  {NoStop}%
\bibitem [{\citenamefont {Aoyama}\ \emph {et~al.}(2024)\citenamefont {Aoyama},
  \citenamefont {Doi}, \citenamefont {Doi}, \citenamefont {Itou}, \citenamefont
  {Lyu}, \citenamefont {Murakami},\ and\ \citenamefont
  {Sugiura}}]{Aoyama:2024cko}%
  \BibitemOpen
  \bibfield  {author} {\bibinfo {author} {\bibfnamefont {T.}~\bibnamefont
  {Aoyama}}, \bibinfo {author} {\bibfnamefont {T.~M.}\ \bibnamefont {Doi}},
  \bibinfo {author} {\bibfnamefont {T.}~\bibnamefont {Doi}}, \bibinfo {author}
  {\bibfnamefont {E.}~\bibnamefont {Itou}}, \bibinfo {author} {\bibfnamefont
  {Y.}~\bibnamefont {Lyu}}, \bibinfo {author} {\bibfnamefont {K.}~\bibnamefont
  {Murakami}},\ and\ \bibinfo {author} {\bibfnamefont {T.}~\bibnamefont
  {Sugiura}} (\bibinfo {collaboration} {HAL QCD}),\ }\bibfield  {title}
  {\bibinfo {title} {{Scale setting and hadronic properties in the light quark
  sector with (2+1)-flavor Wilson fermions at the physical point}},\ }\href
  {https://doi.org/10.1103/PhysRevD.110.094502} {\bibfield  {journal} {\bibinfo
   {journal} {Phys. Rev. D}\ }\textbf {\bibinfo {volume} {110}},\ \bibinfo
  {pages} {094502} (\bibinfo {year} {2024})},\ \Eprint
  {https://arxiv.org/abs/2406.16665} {arXiv:2406.16665 [hep-lat]} \BibitemShut
  {NoStop}%
\bibitem [{\citenamefont {Aoki}\ and\ \citenamefont
  {Jido}(2017)}]{Aoki:2017hel}%
  \BibitemOpen
  \bibfield  {author} {\bibinfo {author} {\bibfnamefont {K.}~\bibnamefont
  {Aoki}}\ and\ \bibinfo {author} {\bibfnamefont {D.}~\bibnamefont {Jido}},\
  }\bibfield  {title} {\bibinfo {title} {{$K^+$-nucleus elastic scattering
  revisited from perspective of partial restoration of chiral symmetry}},\
  }\href {https://doi.org/10.1093/ptep/ptx133} {\bibfield  {journal} {\bibinfo
  {journal} {PTEP}\ }\textbf {\bibinfo {volume} {2017}},\ \bibinfo {pages}
  {103D01} (\bibinfo {year} {2017})},\ \bibinfo {note} {[Erratum: PTEP 2019,
  069201 (2019)]},\ \Eprint {https://arxiv.org/abs/1705.07548}
  {arXiv:1705.07548 [nucl-th]} \BibitemShut {NoStop}%
\bibitem [{\citenamefont {Iizawa}\ \emph {et~al.}(2024)\citenamefont {Iizawa},
  \citenamefont {Jido},\ and\ \citenamefont {H\"ubsch}}]{Iizawa:2023xsi}%
  \BibitemOpen
  \bibfield  {author} {\bibinfo {author} {\bibfnamefont {Y.}~\bibnamefont
  {Iizawa}}, \bibinfo {author} {\bibfnamefont {D.}~\bibnamefont {Jido}},\ and\
  \bibinfo {author} {\bibfnamefont {S.}~\bibnamefont {H\"ubsch}},\ }\bibfield
  {title} {\bibinfo {title} {{K+N Elastic Scatterings for Estimation of the
  In-Medium Quark Condensate with Strange Quarks}},\ }\href
  {https://doi.org/10.1093/ptep/ptae050} {\bibfield  {journal} {\bibinfo
  {journal} {PTEP}\ }\textbf {\bibinfo {volume} {2024}},\ \bibinfo {pages}
  {053D01} (\bibinfo {year} {2024})},\ \Eprint
  {https://arxiv.org/abs/2308.09397} {arXiv:2308.09397 [hep-ph]} \BibitemShut
  {NoStop}%
\bibitem [{\citenamefont {Hosaka}\ \emph {et~al.}(2017)\citenamefont {Hosaka},
  \citenamefont {Hyodo}, \citenamefont {Sudoh}, \citenamefont {Yamaguchi},\
  and\ \citenamefont {Yasui}}]{Hosaka:2016ypm}%
  \BibitemOpen
  \bibfield  {author} {\bibinfo {author} {\bibfnamefont {A.}~\bibnamefont
  {Hosaka}}, \bibinfo {author} {\bibfnamefont {T.}~\bibnamefont {Hyodo}},
  \bibinfo {author} {\bibfnamefont {K.}~\bibnamefont {Sudoh}}, \bibinfo
  {author} {\bibfnamefont {Y.}~\bibnamefont {Yamaguchi}},\ and\ \bibinfo
  {author} {\bibfnamefont {S.}~\bibnamefont {Yasui}},\ }\bibfield  {title}
  {\bibinfo {title} {{Heavy Hadrons in Nuclear Matter}},\ }\href
  {https://doi.org/10.1016/j.ppnp.2017.04.003} {\bibfield  {journal} {\bibinfo
  {journal} {Prog. Part. Nucl. Phys.}\ }\textbf {\bibinfo {volume} {96}},\
  \bibinfo {pages} {88} (\bibinfo {year} {2017})},\ \Eprint
  {https://arxiv.org/abs/1606.08685} {arXiv:1606.08685 [hep-ph]} \BibitemShut
  {NoStop}%
\bibitem [{\citenamefont {Dover}\ and\ \citenamefont
  {Walker}(1982)}]{Dover:1982zh}%
  \BibitemOpen
  \bibfield  {author} {\bibinfo {author} {\bibfnamefont {C.~B.}\ \bibnamefont
  {Dover}}\ and\ \bibinfo {author} {\bibfnamefont {G.~E.}\ \bibnamefont
  {Walker}},\ }\bibfield  {title} {\bibinfo {title} {{THE INTERACTION OF KAONS
  WITH NUCLEONS AND NUCLEI}},\ }\href
  {https://doi.org/10.1016/0370-1573(82)90043-6} {\bibfield  {journal}
  {\bibinfo  {journal} {Phys. Rept.}\ }\textbf {\bibinfo {volume} {89}},\
  \bibinfo {pages} {1} (\bibinfo {year} {1982})}\BibitemShut {NoStop}%
\bibitem [{\citenamefont {Nakano}\ \emph {et~al.}(2003)\citenamefont {Nakano}
  \emph {et~al.}}]{LEPS:2003wug}%
  \BibitemOpen
  \bibfield  {author} {\bibinfo {author} {\bibfnamefont {T.}~\bibnamefont
  {Nakano}} \emph {et~al.} (\bibinfo {collaboration} {LEPS}),\ }\bibfield
  {title} {\bibinfo {title} {{Evidence for a narrow S = +1 baryon resonance in
  photoproduction from the neutron}},\ }\href
  {https://doi.org/10.1103/PhysRevLett.91.012002} {\bibfield  {journal}
  {\bibinfo  {journal} {Phys. Rev. Lett.}\ }\textbf {\bibinfo {volume} {91}},\
  \bibinfo {pages} {012002} (\bibinfo {year} {2003})},\ \Eprint
  {https://arxiv.org/abs/hep-ex/0301020} {arXiv:hep-ex/0301020} \BibitemShut
  {NoStop}%
\bibitem [{\citenamefont {Danilov}\ and\ \citenamefont
  {Mizuk}(2008)}]{Danilov:2008uxa}%
  \BibitemOpen
  \bibfield  {author} {\bibinfo {author} {\bibfnamefont {M.}~\bibnamefont
  {Danilov}}\ and\ \bibinfo {author} {\bibfnamefont {R.}~\bibnamefont
  {Mizuk}},\ }\bibfield  {title} {\bibinfo {title} {{Experimental review on
  pentaquarks}},\ }\href {https://doi.org/10.1134/S1063778808040029} {\bibfield
   {journal} {\bibinfo  {journal} {Phys. Atom. Nucl.}\ }\textbf {\bibinfo
  {volume} {71}},\ \bibinfo {pages} {605} (\bibinfo {year} {2008})},\ \Eprint
  {https://arxiv.org/abs/0704.3531} {arXiv:0704.3531 [hep-ex]} \BibitemShut
  {NoStop}%
\bibitem [{\citenamefont {Liu}\ and\ \citenamefont
  {Mathur}(2006)}]{Liu:2005yc}%
  \BibitemOpen
  \bibfield  {author} {\bibinfo {author} {\bibfnamefont {K.-F.}\ \bibnamefont
  {Liu}}\ and\ \bibinfo {author} {\bibfnamefont {N.}~\bibnamefont {Mathur}},\
  }\bibfield  {title} {\bibinfo {title} {{A Review of pentaquark calculations
  on the lattice}},\ }\href {https://doi.org/10.1142/S0217751X06032162}
  {\bibfield  {journal} {\bibinfo  {journal} {Int. J. Mod. Phys. A}\ }\textbf
  {\bibinfo {volume} {21}},\ \bibinfo {pages} {851} (\bibinfo {year} {2006})},\
  \Eprint {https://arxiv.org/abs/hep-lat/0510036} {arXiv:hep-lat/0510036}
  \BibitemShut {NoStop}%
\bibitem [{\citenamefont {Luscher}(1991)}]{Luscher:1990ux}%
  \BibitemOpen
  \bibfield  {author} {\bibinfo {author} {\bibfnamefont {M.}~\bibnamefont
  {Luscher}},\ }\bibfield  {title} {\bibinfo {title} {{Two particle states on a
  torus and their relation to the scattering matrix}},\ }\href
  {https://doi.org/10.1016/0550-3213(91)90366-6} {\bibfield  {journal}
  {\bibinfo  {journal} {Nucl. Phys. B}\ }\textbf {\bibinfo {volume} {354}},\
  \bibinfo {pages} {531} (\bibinfo {year} {1991})}\BibitemShut {NoStop}%
\bibitem [{\citenamefont {Fukugita}\ \emph {et~al.}(1995)\citenamefont
  {Fukugita}, \citenamefont {Kuramashi}, \citenamefont {Okawa}, \citenamefont
  {Mino},\ and\ \citenamefont {Ukawa}}]{Fukugita:1994ve}%
  \BibitemOpen
  \bibfield  {author} {\bibinfo {author} {\bibfnamefont {M.}~\bibnamefont
  {Fukugita}}, \bibinfo {author} {\bibfnamefont {Y.}~\bibnamefont {Kuramashi}},
  \bibinfo {author} {\bibfnamefont {M.}~\bibnamefont {Okawa}}, \bibinfo
  {author} {\bibfnamefont {H.}~\bibnamefont {Mino}},\ and\ \bibinfo {author}
  {\bibfnamefont {A.}~\bibnamefont {Ukawa}},\ }\bibfield  {title} {\bibinfo
  {title} {{Hadron scattering lengths in lattice QCD}},\ }\href
  {https://doi.org/10.1103/PhysRevD.52.3003} {\bibfield  {journal} {\bibinfo
  {journal} {Phys. Rev. D}\ }\textbf {\bibinfo {volume} {52}},\ \bibinfo
  {pages} {3003} (\bibinfo {year} {1995})}\BibitemShut {NoStop}%
\bibitem [{\citenamefont {Meng}\ \emph {et~al.}(2004)\citenamefont {Meng},
  \citenamefont {Miao}, \citenamefont {Du},\ and\ \citenamefont
  {Liu}}]{Meng:2003gm}%
  \BibitemOpen
  \bibfield  {author} {\bibinfo {author} {\bibfnamefont {G.-w.}\ \bibnamefont
  {Meng}}, \bibinfo {author} {\bibfnamefont {C.}~\bibnamefont {Miao}}, \bibinfo
  {author} {\bibfnamefont {X.-n.}\ \bibnamefont {Du}},\ and\ \bibinfo {author}
  {\bibfnamefont {C.}~\bibnamefont {Liu}},\ }\bibfield  {title} {\bibinfo
  {title} {{Lattice study on kaon nucleon scattering length in the I = 1
  channel}},\ }\href {https://doi.org/10.1142/S0217751X04019627} {\bibfield
  {journal} {\bibinfo  {journal} {Int. J. Mod. Phys. A}\ }\textbf {\bibinfo
  {volume} {19}},\ \bibinfo {pages} {4401} (\bibinfo {year}
  {2004})}\BibitemShut {NoStop}%
\bibitem [{\citenamefont {Torok}\ \emph {et~al.}(2010)\citenamefont {Torok},
  \citenamefont {Beane}, \citenamefont {Detmold}, \citenamefont {Luu},
  \citenamefont {Orginos}, \citenamefont {Parreno}, \citenamefont {Savage},\
  and\ \citenamefont {Walker-Loud}}]{Torok:2009dg}%
  \BibitemOpen
  \bibfield  {author} {\bibinfo {author} {\bibfnamefont {A.}~\bibnamefont
  {Torok}}, \bibinfo {author} {\bibfnamefont {S.~R.}\ \bibnamefont {Beane}},
  \bibinfo {author} {\bibfnamefont {W.}~\bibnamefont {Detmold}}, \bibinfo
  {author} {\bibfnamefont {T.~C.}\ \bibnamefont {Luu}}, \bibinfo {author}
  {\bibfnamefont {K.}~\bibnamefont {Orginos}}, \bibinfo {author} {\bibfnamefont
  {A.}~\bibnamefont {Parreno}}, \bibinfo {author} {\bibfnamefont {M.~J.}\
  \bibnamefont {Savage}},\ and\ \bibinfo {author} {\bibfnamefont
  {A.}~\bibnamefont {Walker-Loud}},\ }\bibfield  {title} {\bibinfo {title}
  {{Meson-Baryon Scattering Lengths from Mixed-Action Lattice QCD}},\ }\href
  {https://doi.org/10.1103/PhysRevD.81.074506} {\bibfield  {journal} {\bibinfo
  {journal} {Phys. Rev. D}\ }\textbf {\bibinfo {volume} {81}},\ \bibinfo
  {pages} {074506} (\bibinfo {year} {2010})}\BibitemShut {NoStop}%
\bibitem [{\citenamefont {Detmold}\ and\ \citenamefont
  {Nicholson}(2013)}]{Detmold:2013gua}%
  \BibitemOpen
  \bibfield  {author} {\bibinfo {author} {\bibfnamefont {W.}~\bibnamefont
  {Detmold}}\ and\ \bibinfo {author} {\bibfnamefont {A.~N.}\ \bibnamefont
  {Nicholson}},\ }\bibfield  {title} {\bibinfo {title} {{Baryon masses at
  nonzero isospin/kaon density}},\ }\href
  {https://doi.org/10.1103/PhysRevD.88.074501} {\bibfield  {journal} {\bibinfo
  {journal} {Phys. Rev. D}\ }\textbf {\bibinfo {volume} {88}},\ \bibinfo
  {pages} {074501} (\bibinfo {year} {2013})}\BibitemShut {NoStop}%
\bibitem [{\citenamefont {Detmold}\ and\ \citenamefont
  {Nicholson}(2016)}]{Detmold:2015qwf}%
  \BibitemOpen
  \bibfield  {author} {\bibinfo {author} {\bibfnamefont {W.}~\bibnamefont
  {Detmold}}\ and\ \bibinfo {author} {\bibfnamefont {A.}~\bibnamefont
  {Nicholson}},\ }\bibfield  {title} {\bibinfo {title} {{Low energy scattering
  phase shifts for meson-baryon systems}},\ }\href
  {https://doi.org/10.1103/PhysRevD.93.114511} {\bibfield  {journal} {\bibinfo
  {journal} {Phys. Rev. D}\ }\textbf {\bibinfo {volume} {93}},\ \bibinfo
  {pages} {114511} (\bibinfo {year} {2016})}\BibitemShut {NoStop}%
\bibitem [{\citenamefont {Ikeda}\ \emph {et~al.}(2010)\citenamefont {Ikeda},
  \citenamefont {Aoki}, \citenamefont {Doi}, \citenamefont {Hatsuda},
  \citenamefont {Inoue}, \citenamefont {Ishii}, \citenamefont {Murano},
  \citenamefont {Nemura},\ and\ \citenamefont {Sasaki}}]{Ikeda:2010sg}%
  \BibitemOpen
  \bibfield  {author} {\bibinfo {author} {\bibfnamefont {Y.}~\bibnamefont
  {Ikeda}}, \bibinfo {author} {\bibfnamefont {S.}~\bibnamefont {Aoki}},
  \bibinfo {author} {\bibfnamefont {T.}~\bibnamefont {Doi}}, \bibinfo {author}
  {\bibfnamefont {T.}~\bibnamefont {Hatsuda}}, \bibinfo {author} {\bibfnamefont
  {T.}~\bibnamefont {Inoue}}, \bibinfo {author} {\bibfnamefont
  {N.}~\bibnamefont {Ishii}}, \bibinfo {author} {\bibfnamefont
  {K.}~\bibnamefont {Murano}}, \bibinfo {author} {\bibfnamefont
  {H.}~\bibnamefont {Nemura}},\ and\ \bibinfo {author} {\bibfnamefont
  {K.}~\bibnamefont {Sasaki}},\ }\bibfield  {title} {\bibinfo {title}
  {{Kaon-Nucleon potential from lattice QCD}},\ }\href
  {https://doi.org/10.1051/epjconf/20100303007} {\bibfield  {journal} {\bibinfo
   {journal} {EPJ Web Conf.}\ }\textbf {\bibinfo {volume} {3}},\ \bibinfo
  {pages} {03007} (\bibinfo {year} {2010})},\ \Eprint
  {https://arxiv.org/abs/1002.2309} {arXiv:1002.2309 [hep-lat]} \BibitemShut
  {NoStop}%
\bibitem [{\citenamefont {Ikeda}(2011)}]{Ikeda:2011qm}%
  \BibitemOpen
  \bibfield  {author} {\bibinfo {author} {\bibfnamefont {Y.}~\bibnamefont
  {Ikeda}} (\bibinfo {collaboration} {HAL QCD}),\ }\bibfield  {title} {\bibinfo
  {title} {{S-wave meson-baryon potentials with strangeness from Lattice
  QCD}},\ }\href {https://doi.org/10.22323/1.139.0159} {\bibfield  {journal}
  {\bibinfo  {journal} {PoS}\ }\textbf {\bibinfo {volume} {LATTICE2011}},\
  \bibinfo {pages} {159} (\bibinfo {year} {2011})},\ \Eprint
  {https://arxiv.org/abs/1111.2663} {arXiv:1111.2663 [hep-lat]} \BibitemShut
  {NoStop}%
\bibitem [{\citenamefont {Murakami}\ \emph {et~al.}(2020)\citenamefont
  {Murakami}, \citenamefont {Akahoshi},\ and\ \citenamefont
  {Aoki}}]{Murakami:2020yzt}%
  \BibitemOpen
  \bibfield  {author} {\bibinfo {author} {\bibfnamefont {K.}~\bibnamefont
  {Murakami}}, \bibinfo {author} {\bibfnamefont {Y.}~\bibnamefont {Akahoshi}},\
  and\ \bibinfo {author} {\bibfnamefont {S.}~\bibnamefont {Aoki}} (\bibinfo
  {collaboration} {LATTICE-HALQCD}),\ }\bibfield  {title} {\bibinfo {title}
  {{S-wave kaon\textendash{}nucleon potentials with all-to-all propagators in
  the HAL QCD method}},\ }\href {https://doi.org/10.1093/ptep/ptaa118}
  {\bibfield  {journal} {\bibinfo  {journal} {PTEP}\ }\textbf {\bibinfo
  {volume} {2020}},\ \bibinfo {pages} {093B03} (\bibinfo {year} {2020})},\
  \Eprint {https://arxiv.org/abs/2006.01383} {arXiv:2006.01383 [hep-lat]}
  \BibitemShut {NoStop}%
\bibitem [{\citenamefont {Okubo}\ and\ \citenamefont
  {Marshak}(1958)}]{OKUBO1958166}%
  \BibitemOpen
  \bibfield  {author} {\bibinfo {author} {\bibfnamefont {S.}~\bibnamefont
  {Okubo}}\ and\ \bibinfo {author} {\bibfnamefont {R.}~\bibnamefont
  {Marshak}},\ }\bibfield  {title} {\bibinfo {title} {Velocity dependence of
  the two-nucleon interaction},\ }\href
  {https://doi.org/https://doi.org/10.1016/0003-4916(58)90031-9} {\bibfield
  {journal} {\bibinfo  {journal} {Annals of Physics}\ }\textbf {\bibinfo
  {volume} {4}},\ \bibinfo {pages} {166 } (\bibinfo {year} {1958})}\BibitemShut
  {NoStop}%
\bibitem [{\citenamefont {Miyamoto}\ \emph {et~al.}(2020)\citenamefont
  {Miyamoto}, \citenamefont {Akahoshi}, \citenamefont {Aoki}, \citenamefont
  {Aoyama}, \citenamefont {Doi}, \citenamefont {Gongyo},\ and\ \citenamefont
  {Sasaki}}]{Miyamoto:2019jjc}%
  \BibitemOpen
  \bibfield  {author} {\bibinfo {author} {\bibfnamefont {T.}~\bibnamefont
  {Miyamoto}}, \bibinfo {author} {\bibfnamefont {Y.}~\bibnamefont {Akahoshi}},
  \bibinfo {author} {\bibfnamefont {S.}~\bibnamefont {Aoki}}, \bibinfo {author}
  {\bibfnamefont {T.}~\bibnamefont {Aoyama}}, \bibinfo {author} {\bibfnamefont
  {T.}~\bibnamefont {Doi}}, \bibinfo {author} {\bibfnamefont {S.}~\bibnamefont
  {Gongyo}},\ and\ \bibinfo {author} {\bibfnamefont {K.}~\bibnamefont
  {Sasaki}},\ }\bibfield  {title} {\bibinfo {title} {{Partial wave
  decomposition on the lattice and its applications to the HAL QCD method}},\
  }\href {https://doi.org/10.1103/PhysRevD.101.074514} {\bibfield  {journal}
  {\bibinfo  {journal} {Phys. Rev. D}\ }\textbf {\bibinfo {volume} {101}},\
  \bibinfo {pages} {074514} (\bibinfo {year} {2020})},\ \Eprint
  {https://arxiv.org/abs/1906.01987} {arXiv:1906.01987 [hep-lat]} \BibitemShut
  {NoStop}%
\bibitem [{\citenamefont {Wiringa}\ \emph {et~al.}(1984)\citenamefont
  {Wiringa}, \citenamefont {Smith},\ and\ \citenamefont
  {Ainsworth}}]{Wiringa:1984tg}%
  \BibitemOpen
  \bibfield  {author} {\bibinfo {author} {\bibfnamefont {R.~B.}\ \bibnamefont
  {Wiringa}}, \bibinfo {author} {\bibfnamefont {R.~A.}\ \bibnamefont {Smith}},\
  and\ \bibinfo {author} {\bibfnamefont {T.~L.}\ \bibnamefont {Ainsworth}},\
  }\bibfield  {title} {\bibinfo {title} {{Nucleon Nucleon Potentials with and
  Without Delta (1232) Degrees of Freedom}},\ }\href
  {https://doi.org/10.1103/PhysRevC.29.1207} {\bibfield  {journal} {\bibinfo
  {journal} {Phys. Rev. C}\ }\textbf {\bibinfo {volume} {29}},\ \bibinfo
  {pages} {1207} (\bibinfo {year} {1984})}\BibitemShut {NoStop}%
\bibitem [{\citenamefont {Aaron}\ \emph {et~al.}(1971)\citenamefont {Aaron},
  \citenamefont {Silbar},\ and\ \citenamefont {Amado}}]{Aaron:1971ka}%
  \BibitemOpen
  \bibfield  {author} {\bibinfo {author} {\bibfnamefont {R.}~\bibnamefont
  {Aaron}}, \bibinfo {author} {\bibfnamefont {R.~R.}\ \bibnamefont {Silbar}},\
  and\ \bibinfo {author} {\bibfnamefont {R.~D.}\ \bibnamefont {Amado}},\
  }\bibfield  {title} {\bibinfo {title} {{Theoretical evidence for i=0 z*'s}},\
  }\href {https://doi.org/10.1103/PhysRevLett.26.407} {\bibfield  {journal}
  {\bibinfo  {journal} {Phys. Rev. Lett.}\ }\textbf {\bibinfo {volume} {26}},\
  \bibinfo {pages} {407} (\bibinfo {year} {1971})}\BibitemShut {NoStop}%
\bibitem [{\citenamefont {Goldhaber}\ \emph {et~al.}(1962)\citenamefont
  {Goldhaber}, \citenamefont {Chinowsky}, \citenamefont {Goldhaber},
  \citenamefont {Lee}, \citenamefont {O'Halloran}, \citenamefont {Stubbs},
  \citenamefont {Pjerrou}, \citenamefont {Stork},\ and\ \citenamefont
  {Ticho}}]{Goldhaber:1962zz}%
  \BibitemOpen
  \bibfield  {author} {\bibinfo {author} {\bibfnamefont {S.}~\bibnamefont
  {Goldhaber}}, \bibinfo {author} {\bibfnamefont {W.}~\bibnamefont
  {Chinowsky}}, \bibinfo {author} {\bibfnamefont {G.}~\bibnamefont
  {Goldhaber}}, \bibinfo {author} {\bibfnamefont {W.}~\bibnamefont {Lee}},
  \bibinfo {author} {\bibfnamefont {T.}~\bibnamefont {O'Halloran}}, \bibinfo
  {author} {\bibfnamefont {T.~F.}\ \bibnamefont {Stubbs}}, \bibinfo {author}
  {\bibfnamefont {G.~M.}\ \bibnamefont {Pjerrou}}, \bibinfo {author}
  {\bibfnamefont {D.~H.}\ \bibnamefont {Stork}},\ and\ \bibinfo {author}
  {\bibfnamefont {H.~K.}\ \bibnamefont {Ticho}},\ }\bibfield  {title} {\bibinfo
  {title} {{K+-p Interaction From 140 to 642 MeV/c}},\ }\href
  {https://doi.org/10.1103/PhysRevLett.9.135} {\bibfield  {journal} {\bibinfo
  {journal} {Phys. Rev. Lett.}\ }\textbf {\bibinfo {volume} {9}},\ \bibinfo
  {pages} {135} (\bibinfo {year} {1962})}\BibitemShut {NoStop}%
\bibitem [{\citenamefont {Burnstein}\ \emph {et~al.}(1974)\citenamefont
  {Burnstein}, \citenamefont {LeFebvre}, \citenamefont {Petersen},
  \citenamefont {Rubin}, \citenamefont {Day}, \citenamefont {Fram},
  \citenamefont {Glasser}, \citenamefont {McClellan}, \citenamefont
  {Sechi-Zorn},\ and\ \citenamefont {Snow}}]{Burnstein:1974ax}%
  \BibitemOpen
  \bibfield  {author} {\bibinfo {author} {\bibfnamefont {R.~A.}\ \bibnamefont
  {Burnstein}}, \bibinfo {author} {\bibfnamefont {J.~J.}\ \bibnamefont
  {LeFebvre}}, \bibinfo {author} {\bibfnamefont {D.~V.}\ \bibnamefont
  {Petersen}}, \bibinfo {author} {\bibfnamefont {H.~A.}\ \bibnamefont {Rubin}},
  \bibinfo {author} {\bibfnamefont {T.~B.}\ \bibnamefont {Day}}, \bibinfo
  {author} {\bibfnamefont {J.~R.}\ \bibnamefont {Fram}}, \bibinfo {author}
  {\bibfnamefont {R.~G.}\ \bibnamefont {Glasser}}, \bibinfo {author}
  {\bibfnamefont {G.}~\bibnamefont {McClellan}}, \bibinfo {author}
  {\bibfnamefont {B.}~\bibnamefont {Sechi-Zorn}},\ and\ \bibinfo {author}
  {\bibfnamefont {G.~A.}\ \bibnamefont {Snow}},\ }\bibfield  {title} {\bibinfo
  {title} {{K+- proton scattering from 200 to 600 MeV/c}},\ }\href
  {https://doi.org/10.1103/PhysRevD.10.2767} {\bibfield  {journal} {\bibinfo
  {journal} {Phys. Rev. D}\ }\textbf {\bibinfo {volume} {10}},\ \bibinfo
  {pages} {2767} (\bibinfo {year} {1974})}\BibitemShut {NoStop}%
\bibitem [{\citenamefont {Cameron}\ \emph {et~al.}(1974)\citenamefont {Cameron}
  \emph {et~al.}}]{Cameron:1974xx}%
  \BibitemOpen
  \bibfield  {author} {\bibinfo {author} {\bibfnamefont {W.}~\bibnamefont
  {Cameron}} \emph {et~al.},\ }\bibfield  {title} {\bibinfo {title} {{K+ p
  Elastic Scattering from 130-MeV/c to 755-MeV/c}},\ }\href
  {https://doi.org/10.1016/0550-3213(74)90117-5} {\bibfield  {journal}
  {\bibinfo  {journal} {Nucl. Phys. B}\ }\textbf {\bibinfo {volume} {78}},\
  \bibinfo {pages} {93} (\bibinfo {year} {1974})}\BibitemShut {NoStop}%
\bibitem [{\citenamefont {Stenger}\ \emph {et~al.}(1964)\citenamefont
  {Stenger}, \citenamefont {Slater}, \citenamefont {Stork}, \citenamefont
  {Ticho}, \citenamefont {Goldhaber},\ and\ \citenamefont
  {Goldhaber}}]{PhysRev.134.B1111}%
  \BibitemOpen
  \bibfield  {author} {\bibinfo {author} {\bibfnamefont {V.~J.}\ \bibnamefont
  {Stenger}}, \bibinfo {author} {\bibfnamefont {W.~E.}\ \bibnamefont {Slater}},
  \bibinfo {author} {\bibfnamefont {D.~H.}\ \bibnamefont {Stork}}, \bibinfo
  {author} {\bibfnamefont {H.~K.}\ \bibnamefont {Ticho}}, \bibinfo {author}
  {\bibfnamefont {G.}~\bibnamefont {Goldhaber}},\ and\ \bibinfo {author}
  {\bibfnamefont {S.}~\bibnamefont {Goldhaber}},\ }\bibfield  {title} {\bibinfo
  {title} {$k\ensuremath{-}n$ interaction in the $i=0$ state at low energies},\
  }\href {https://doi.org/10.1103/PhysRev.134.B1111} {\bibfield  {journal}
  {\bibinfo  {journal} {Phys. Rev.}\ }\textbf {\bibinfo {volume} {134}},\
  \bibinfo {pages} {B1111} (\bibinfo {year} {1964})}\BibitemShut {NoStop}%
\bibitem [{\citenamefont {Glasser}\ \emph {et~al.}(1977)\citenamefont
  {Glasser}, \citenamefont {Snow}, \citenamefont {Trevvett}, \citenamefont
  {Burnstein}, \citenamefont {Fu}, \citenamefont {Petri}, \citenamefont
  {Rosenblatt},\ and\ \citenamefont {Rubin}}]{PhysRevD.15.1200}%
  \BibitemOpen
  \bibfield  {author} {\bibinfo {author} {\bibfnamefont {R.~G.}\ \bibnamefont
  {Glasser}}, \bibinfo {author} {\bibfnamefont {G.~A.}\ \bibnamefont {Snow}},
  \bibinfo {author} {\bibfnamefont {D.}~\bibnamefont {Trevvett}}, \bibinfo
  {author} {\bibfnamefont {R.~A.}\ \bibnamefont {Burnstein}}, \bibinfo {author}
  {\bibfnamefont {C.}~\bibnamefont {Fu}}, \bibinfo {author} {\bibfnamefont
  {R.}~\bibnamefont {Petri}}, \bibinfo {author} {\bibfnamefont
  {G.}~\bibnamefont {Rosenblatt}},\ and\ \bibinfo {author} {\bibfnamefont
  {H.~A.}\ \bibnamefont {Rubin}},\ }\bibfield  {title} {\bibinfo {title}
  {Low-momentum ${K}^{+}d$ scattering},\ }\href
  {https://doi.org/10.1103/PhysRevD.15.1200} {\bibfield  {journal} {\bibinfo
  {journal} {Phys. Rev. D}\ }\textbf {\bibinfo {volume} {15}},\ \bibinfo
  {pages} {1200} (\bibinfo {year} {1977})}\BibitemShut {NoStop}%
\bibitem [{\citenamefont {Martin}(1975)}]{Martin:1975gs}%
  \BibitemOpen
  \bibfield  {author} {\bibinfo {author} {\bibfnamefont {B.~R.}\ \bibnamefont
  {Martin}},\ }\bibfield  {title} {\bibinfo {title} {{Kaon-Nucleon Partial Wave
  Amplitudes Below 1.5-GeV/c for I=0 and 1}},\ }\href
  {https://doi.org/10.1016/0550-3213(75)90104-2} {\bibfield  {journal}
  {\bibinfo  {journal} {Nucl. Phys. B}\ }\textbf {\bibinfo {volume} {94}},\
  \bibinfo {pages} {413} (\bibinfo {year} {1975})}\BibitemShut {NoStop}%
\bibitem [{\citenamefont {Hyslop}\ \emph {et~al.}(1992)\citenamefont {Hyslop},
  \citenamefont {Arndt}, \citenamefont {Roper},\ and\ \citenamefont
  {Workman}}]{Hyslop:1992cs}%
  \BibitemOpen
  \bibfield  {author} {\bibinfo {author} {\bibfnamefont {J.~S.}\ \bibnamefont
  {Hyslop}}, \bibinfo {author} {\bibfnamefont {R.~A.}\ \bibnamefont {Arndt}},
  \bibinfo {author} {\bibfnamefont {L.~D.}\ \bibnamefont {Roper}},\ and\
  \bibinfo {author} {\bibfnamefont {R.~L.}\ \bibnamefont {Workman}},\
  }\bibfield  {title} {\bibinfo {title} {{Partial wave analysis of K+ nucleon
  scattering}},\ }\href {https://doi.org/10.1103/PhysRevD.46.961} {\bibfield
  {journal} {\bibinfo  {journal} {Phys. Rev. D}\ }\textbf {\bibinfo {volume}
  {46}},\ \bibinfo {pages} {961} (\bibinfo {year} {1992})},\ \bibinfo {note}
  {phase shift data is taken from https://gwdac.phys.gwu.edu/}\BibitemShut
  {NoStop}%
\bibitem [{\citenamefont {Gibbs}\ and\ \citenamefont
  {Arceo}(2007)}]{Gibbs:2006ab}%
  \BibitemOpen
  \bibfield  {author} {\bibinfo {author} {\bibfnamefont {W.~R.}\ \bibnamefont
  {Gibbs}}\ and\ \bibinfo {author} {\bibfnamefont {R.}~\bibnamefont {Arceo}},\
  }\bibfield  {title} {\bibinfo {title} {{Partial-wave analysis of K+ nucleon
  scattering}},\ }\href {https://doi.org/10.1103/PhysRevC.75.035204} {\bibfield
   {journal} {\bibinfo  {journal} {Phys. Rev. C}\ }\textbf {\bibinfo {volume}
  {75}},\ \bibinfo {pages} {035204} (\bibinfo {year} {2007})},\ \Eprint
  {https://arxiv.org/abs/nucl-th/0611095} {arXiv:nucl-th/0611095} \BibitemShut
  {NoStop}%
\bibitem [{\citenamefont {Bowen}\ \emph {et~al.}(1970)\citenamefont {Bowen},
  \citenamefont {Caldwell}, \citenamefont {Dikmen}, \citenamefont {Jenkins},
  \citenamefont {Kalbach}, \citenamefont {Petersen},\ and\ \citenamefont
  {Pifer}}]{Bowen:1970azd}%
  \BibitemOpen
  \bibfield  {author} {\bibinfo {author} {\bibfnamefont {T.}~\bibnamefont
  {Bowen}}, \bibinfo {author} {\bibfnamefont {P.~K.}\ \bibnamefont {Caldwell}},
  \bibinfo {author} {\bibfnamefont {F.~N.}\ \bibnamefont {Dikmen}}, \bibinfo
  {author} {\bibfnamefont {E.~W.}\ \bibnamefont {Jenkins}}, \bibinfo {author}
  {\bibfnamefont {R.~M.}\ \bibnamefont {Kalbach}}, \bibinfo {author}
  {\bibfnamefont {D.~V.}\ \bibnamefont {Petersen}},\ and\ \bibinfo {author}
  {\bibfnamefont {A.~E.}\ \bibnamefont {Pifer}},\ }\bibfield  {title} {\bibinfo
  {title} {{Kaon-nucleon total cross-sections from 0.36 to 0.72 gev/c}},\
  }\href {https://doi.org/10.1103/PhysRevD.2.2599} {\bibfield  {journal}
  {\bibinfo  {journal} {Phys. Rev. D}\ }\textbf {\bibinfo {volume} {2}},\
  \bibinfo {pages} {2599} (\bibinfo {year} {1970})}\BibitemShut {NoStop}%
\bibitem [{\citenamefont {Carroll}\ \emph {et~al.}(1973)\citenamefont
  {Carroll}, \citenamefont {Kycia}, \citenamefont {Li}, \citenamefont
  {Michael}, \citenamefont {Mockett}, \citenamefont {Rahm},\ and\ \citenamefont
  {Rubinstein}}]{Carroll:1973ux}%
  \BibitemOpen
  \bibfield  {author} {\bibinfo {author} {\bibfnamefont {A.~S.}\ \bibnamefont
  {Carroll}}, \bibinfo {author} {\bibfnamefont {T.~F.}\ \bibnamefont {Kycia}},
  \bibinfo {author} {\bibfnamefont {K.~K.}\ \bibnamefont {Li}}, \bibinfo
  {author} {\bibfnamefont {D.~N.}\ \bibnamefont {Michael}}, \bibinfo {author}
  {\bibfnamefont {P.~M.}\ \bibnamefont {Mockett}}, \bibinfo {author}
  {\bibfnamefont {D.~C.}\ \bibnamefont {Rahm}},\ and\ \bibinfo {author}
  {\bibfnamefont {R.}~\bibnamefont {Rubinstein}},\ }\bibfield  {title}
  {\bibinfo {title} {{Structure in the k+ nucleon, i=0 total cross-section
  below 1.1 gev/c}},\ }\href {https://doi.org/10.1016/0370-2693(73)90662-X}
  {\bibfield  {journal} {\bibinfo  {journal} {Phys. Lett. B}\ }\textbf
  {\bibinfo {volume} {45}},\ \bibinfo {pages} {531} (\bibinfo {year}
  {1973})}\BibitemShut {NoStop}%
\bibitem [{\citenamefont {Adams}\ \emph {et~al.}(1973)\citenamefont {Adams}
  \emph {et~al.}}]{Adams:1973elr}%
  \BibitemOpen
  \bibfield  {author} {\bibinfo {author} {\bibfnamefont {C.~J.}\ \bibnamefont
  {Adams}} \emph {et~al.},\ }\bibfield  {title} {\bibinfo {title} {{K+ p
  elastic scattering between 432 and 939 mev/c and phase shift analysis}},\
  }\href {https://doi.org/10.1016/0550-3213(73)90005-9} {\bibfield  {journal}
  {\bibinfo  {journal} {Nucl. Phys. B}\ }\textbf {\bibinfo {volume} {66}},\
  \bibinfo {pages} {36} (\bibinfo {year} {1973})}\BibitemShut {NoStop}%
\bibitem [{\citenamefont {Aoki}\ and\ \citenamefont
  {Jido}(2019)}]{Aoki:2018wug}%
  \BibitemOpen
  \bibfield  {author} {\bibinfo {author} {\bibfnamefont {K.}~\bibnamefont
  {Aoki}}\ and\ \bibinfo {author} {\bibfnamefont {D.}~\bibnamefont {Jido}},\
  }\bibfield  {title} {\bibinfo {title} {{KN scattering amplitude revisited in
  a chiral unitary approach and a possible broad resonance in S = +1
  channel}},\ }\href {https://doi.org/10.1093/ptep/pty130} {\bibfield
  {journal} {\bibinfo  {journal} {PTEP}\ }\textbf {\bibinfo {volume} {2019}},\
  \bibinfo {pages} {013D01} (\bibinfo {year} {2019})},\ \Eprint
  {https://arxiv.org/abs/1806.00925} {arXiv:1806.00925 [nucl-th]} \BibitemShut
  {NoStop}%
\bibitem [{\citenamefont {Slater}\ \emph {et~al.}(1961)\citenamefont {Slater},
  \citenamefont {Stork}, \citenamefont {Ticho}, \citenamefont {Lee},
  \citenamefont {Chinowsky}, \citenamefont {Goldhaber}, \citenamefont
  {Goldhaber},\ and\ \citenamefont {O'Halloran}}]{Slater:1961zz}%
  \BibitemOpen
  \bibfield  {author} {\bibinfo {author} {\bibfnamefont {W.}~\bibnamefont
  {Slater}}, \bibinfo {author} {\bibfnamefont {D.~H.}\ \bibnamefont {Stork}},
  \bibinfo {author} {\bibfnamefont {H.~K.}\ \bibnamefont {Ticho}}, \bibinfo
  {author} {\bibfnamefont {W.}~\bibnamefont {Lee}}, \bibinfo {author}
  {\bibfnamefont {W.}~\bibnamefont {Chinowsky}}, \bibinfo {author}
  {\bibfnamefont {G.}~\bibnamefont {Goldhaber}}, \bibinfo {author}
  {\bibfnamefont {S.}~\bibnamefont {Goldhaber}},\ and\ \bibinfo {author}
  {\bibfnamefont {T.}~\bibnamefont {O'Halloran}},\ }\bibfield  {title}
  {\bibinfo {title} {{K++d Charge -Exchange Reaction from 52 to 456 Mev}},\
  }\href {https://doi.org/10.1103/PhysRevLett.7.378} {\bibfield  {journal}
  {\bibinfo  {journal} {Phys. Rev. Lett.}\ }\textbf {\bibinfo {volume} {7}},\
  \bibinfo {pages} {378} (\bibinfo {year} {1961})}\BibitemShut {NoStop}%
\bibitem [{\citenamefont {Giacomelli}\ \emph {et~al.}(1974)\citenamefont
  {Giacomelli} \emph {et~al.}}]{Giacomelli:1974az}%
  \BibitemOpen
  \bibfield  {author} {\bibinfo {author} {\bibfnamefont {G.}~\bibnamefont
  {Giacomelli}} \emph {et~al.},\ }\bibfield  {title} {\bibinfo {title}
  {{Phase-shift analysis of k+ n ---\ensuremath{>} k n scattering in the i=0
  state up to 1.5 gev/c}},\ }\href
  {https://doi.org/10.1016/0550-3213(74)90261-2} {\bibfield  {journal}
  {\bibinfo  {journal} {Nucl. Phys. B}\ }\textbf {\bibinfo {volume} {71}},\
  \bibinfo {pages} {138} (\bibinfo {year} {1974})}\BibitemShut {NoStop}%
\bibitem [{\citenamefont {Sakitt}\ \emph {et~al.}(1975)\citenamefont {Sakitt},
  \citenamefont {Skelly},\ and\ \citenamefont {Thompson}}]{PhysRevD.12.3386}%
  \BibitemOpen
  \bibfield  {author} {\bibinfo {author} {\bibfnamefont {M.}~\bibnamefont
  {Sakitt}}, \bibinfo {author} {\bibfnamefont {J.}~\bibnamefont {Skelly}},\
  and\ \bibinfo {author} {\bibfnamefont {J.~A.}\ \bibnamefont {Thompson}},\
  }\bibfield  {title} {\bibinfo {title} {Study of ${K}^{+}d$ elastic scattering
  in the region of 600 to 900 mev/c},\ }\href
  {https://doi.org/10.1103/PhysRevD.12.3386} {\bibfield  {journal} {\bibinfo
  {journal} {Phys. Rev. D}\ }\textbf {\bibinfo {volume} {12}},\ \bibinfo
  {pages} {3386} (\bibinfo {year} {1975})}\BibitemShut {NoStop}%
\bibitem [{\citenamefont {Sakitt}\ \emph {et~al.}(1977)\citenamefont {Sakitt},
  \citenamefont {Skelly},\ and\ \citenamefont {Thompson}}]{PhysRevD.15.1846}%
  \BibitemOpen
  \bibfield  {author} {\bibinfo {author} {\bibfnamefont {M.}~\bibnamefont
  {Sakitt}}, \bibinfo {author} {\bibfnamefont {J.}~\bibnamefont {Skelly}},\
  and\ \bibinfo {author} {\bibfnamefont {J.}~\bibnamefont {Thompson}},\
  }\bibfield  {title} {\bibinfo {title} {Differential cross-section
  measurements of ${K}^{+}n\ensuremath{\rightarrow}{K}^{0}p$ at ${K}^{+}$
  momenta of 0.7, 0.8, and 0.9 gev/c},\ }\href
  {https://doi.org/10.1103/PhysRevD.15.1846} {\bibfield  {journal} {\bibinfo
  {journal} {Phys. Rev. D}\ }\textbf {\bibinfo {volume} {15}},\ \bibinfo
  {pages} {1846} (\bibinfo {year} {1977})}\BibitemShut {NoStop}%
\bibitem [{\citenamefont {Luscher}(1986)}]{Luscher:1986pf}%
  \BibitemOpen
  \bibfield  {author} {\bibinfo {author} {\bibfnamefont {M.}~\bibnamefont
  {Luscher}},\ }\bibfield  {title} {\bibinfo {title} {{Volume Dependence of the
  Energy Spectrum in Massive Quantum Field Theories. 2. Scattering States}},\
  }\href {https://doi.org/10.1007/BF01211097} {\bibfield  {journal} {\bibinfo
  {journal} {Commun. Math. Phys.}\ }\textbf {\bibinfo {volume} {105}},\
  \bibinfo {pages} {153} (\bibinfo {year} {1986})}\BibitemShut {NoStop}%
\bibitem [{\citenamefont {Green}\ \emph {et~al.}(2021)\citenamefont {Green},
  \citenamefont {Hanlon}, \citenamefont {Junnarkar},\ and\ \citenamefont
  {Wittig}}]{Green:2021qol}%
  \BibitemOpen
  \bibfield  {author} {\bibinfo {author} {\bibfnamefont {J.~R.}\ \bibnamefont
  {Green}}, \bibinfo {author} {\bibfnamefont {A.~D.}\ \bibnamefont {Hanlon}},
  \bibinfo {author} {\bibfnamefont {P.~M.}\ \bibnamefont {Junnarkar}},\ and\
  \bibinfo {author} {\bibfnamefont {H.}~\bibnamefont {Wittig}},\ }\bibfield
  {title} {\bibinfo {title} {{Weakly bound $H$ dibaryon from
  SU(3)-flavor-symmetric QCD}},\ }\href
  {https://doi.org/10.1103/PhysRevLett.127.242003} {\bibfield  {journal}
  {\bibinfo  {journal} {Phys. Rev. Lett.}\ }\textbf {\bibinfo {volume} {127}},\
  \bibinfo {pages} {242003} (\bibinfo {year} {2021})},\ \Eprint
  {https://arxiv.org/abs/2103.01054} {arXiv:2103.01054 [hep-lat]} \BibitemShut
  {NoStop}%
\bibitem [{\citenamefont {Iritani}\ \emph
  {et~al.}(2019{\natexlab{b}})\citenamefont {Iritani}, \citenamefont {Aoki},
  \citenamefont {Doi}, \citenamefont {Hatsuda}, \citenamefont {Ikeda},
  \citenamefont {Inoue}, \citenamefont {Ishii}, \citenamefont {Nemura},\ and\
  \citenamefont {Sasaki}}]{Iritani:2018vfn}%
  \BibitemOpen
  \bibfield  {author} {\bibinfo {author} {\bibfnamefont {T.}~\bibnamefont
  {Iritani}}, \bibinfo {author} {\bibfnamefont {S.}~\bibnamefont {Aoki}},
  \bibinfo {author} {\bibfnamefont {T.}~\bibnamefont {Doi}}, \bibinfo {author}
  {\bibfnamefont {T.}~\bibnamefont {Hatsuda}}, \bibinfo {author} {\bibfnamefont
  {Y.}~\bibnamefont {Ikeda}}, \bibinfo {author} {\bibfnamefont
  {T.}~\bibnamefont {Inoue}}, \bibinfo {author} {\bibfnamefont
  {N.}~\bibnamefont {Ishii}}, \bibinfo {author} {\bibfnamefont
  {H.}~\bibnamefont {Nemura}},\ and\ \bibinfo {author} {\bibfnamefont
  {K.}~\bibnamefont {Sasaki}} (\bibinfo {collaboration} {HAL QCD}),\ }\bibfield
   {title} {\bibinfo {title} {{Consistency between
  L{\"u}scher{\textquoteright}s finite volume method and HAL QCD method for
  two-baryon systems in lattice QCD}},\ }\href
  {https://doi.org/10.1007/JHEP03(2019)007} {\bibfield  {journal} {\bibinfo
  {journal} {JHEP}\ }\textbf {\bibinfo {volume} {03}},\ \bibinfo {pages}
  {007}},\ \Eprint {https://arxiv.org/abs/1812.08539} {arXiv:1812.08539
  [hep-lat]} \BibitemShut {NoStop}%
\bibitem [{\citenamefont {Lyu}\ \emph {et~al.}(2022{\natexlab{b}})\citenamefont
  {Lyu}, \citenamefont {Tong}, \citenamefont {Sugiura}, \citenamefont {Aoki},
  \citenamefont {Doi}, \citenamefont {Hatsuda}, \citenamefont {Meng},\ and\
  \citenamefont {Miyamoto}}]{Lyu:2022tsd}%
  \BibitemOpen
  \bibfield  {author} {\bibinfo {author} {\bibfnamefont {Y.}~\bibnamefont
  {Lyu}}, \bibinfo {author} {\bibfnamefont {H.}~\bibnamefont {Tong}}, \bibinfo
  {author} {\bibfnamefont {T.}~\bibnamefont {Sugiura}}, \bibinfo {author}
  {\bibfnamefont {S.}~\bibnamefont {Aoki}}, \bibinfo {author} {\bibfnamefont
  {T.}~\bibnamefont {Doi}}, \bibinfo {author} {\bibfnamefont {T.}~\bibnamefont
  {Hatsuda}}, \bibinfo {author} {\bibfnamefont {J.}~\bibnamefont {Meng}},\ and\
  \bibinfo {author} {\bibfnamefont {T.}~\bibnamefont {Miyamoto}},\ }\bibfield
  {title} {\bibinfo {title} {{Optimized two-baryon operators in lattice QCD}},\
  }\href {https://doi.org/10.1103/PhysRevD.105.074512} {\bibfield  {journal}
  {\bibinfo  {journal} {Phys. Rev. D}\ }\textbf {\bibinfo {volume} {105}},\
  \bibinfo {pages} {074512} (\bibinfo {year} {2022}{\natexlab{b}})},\ \Eprint
  {https://arxiv.org/abs/2201.02782} {arXiv:2201.02782 [hep-lat]} \BibitemShut
  {NoStop}%
\end{thebibliography}%
\end{document}


\preprint{RIKEN-iTHEMS-Report-25, FQSP-2025-4, YITP-25-133}

\title{$S$-wave kaon-nucleon interactions from lattice QCD at the physical point}

\author{Kotaro~Murakami}
\email{kotaro.murakami@yukawa.kyoto-u.ac.jp}
\affiliation{Department of Physics, Institute of Science Tokyo, 2-12-1 Ookayama, Megro, Tokyo 152-8551, Japan.}
\affiliation{RIKEN Center for Interdisciplinary Theoretical and
Mathematical Sciences(iTHEMS), RIKEN, Wako 351-0198, Japan}
\author{Sinya Aoki}
\affiliation{Fundamental Quantum Science Program (FQSP), TRIP Headquarters, RIKEN, Wako 351-0198, Japan}
\affiliation{Center for Gravitational Physics and Quantum Information, Yukawa Institute for Theoretical Physics, Kyoto University, Kitashirakawa Oiwakecho, Sakyo-ku, Kyoto 606-8502, Japan}
\author{Takumi~Doi}
\affiliation{RIKEN Center for Interdisciplinary Theoretical and
Mathematical Sciences(iTHEMS), RIKEN, Wako 351-0198, Japan}
\author{Yan~Lyu}
\affiliation{RIKEN Center for Interdisciplinary Theoretical and
Mathematical Sciences(iTHEMS), RIKEN, Wako 351-0198, Japan}
\author{Wren~Yamada}
\affiliation{RIKEN Center for Interdisciplinary Theoretical and
Mathematical Sciences(iTHEMS), RIKEN, Wako 351-0198, Japan}


\collaboration{HAL QCD Collaboration}\noaffiliation



\date{\today}

\begin{abstract}
We investigate S-wave kaon-nucleon ($KN$) interactions with strangeness $S=+1$ in lattice QCD using the time-dependent HAL QCD method. 
Employing the $(2+1)$-flavor gauge configuration with $m_{\pi}\approx 137~\textrm{MeV}$ and $m_{K}\approx 502~\textrm{MeV}$, we calculate the $KN$ potentials at the leading order in the derivative expansion.
The potentials in both isospin channels ($I=1$ and $I=0$) exhibit repulsion at short distances, while only the $I=0$ potential has a small attractive pocket at intermediate distances. 
The phase shifts computed from these potentials show no signals corresponding to resonances or bound states in both isospin channels, suggesting the absence of the $\Theta^{+}(1540)$ pentaquark in the S-wave $KN$ systems. 
The scattering lengths result in $a^{I=1}_{0} = -0.226(5)(^{+5}_{-0})~\textrm{fm}$ and  $a^{I=0}_{0} = +0.028(61)(^{+3}_{-26})~\textrm{fm}$. 
Our results of the S-wave cross sections for $I=1$ are consistent with some of the experimental data within $2-3$ $\sigma$, while they deviate from others.
The results for $I=0$, combined with recent studies on chiral perturbation theory, suggest that the scattering amplitudes in this channel are dominated by P-wave components rather than S-wave. 
\end{abstract}

\maketitle

\tableofcontents

\section{\label{sec:intro} Introduction}

Understanding the nature of hadrons from quantum chromodynamics (QCD) is a central objective in hadron physics. 
In particular, interactions between hadrons play an essential role in various phenomena such as the evolution of stars and the structure of nuclei. 
For studies on hadrons from the first principles, numerical simulation of lattice QCD is a powerful tool.
The HAL QCD method~\cite{Ishii:2006ec,Aoki:2009ji,Ishii:2012ssm} allows us to extract hadron interactions in lattice QCD without requiring ground-state saturation thanks to the time-dependent approach. 
This method has been applied to a wide range of hadron interactions, including those near the physical point, $m_{\pi} \approx 146~\textrm{MeV}$~\cite{Doi:2017zov,Gongyo:2017fjb,HALQCD:2018qyu,HALQCD:2019wsz,Lyu:2021qsh,Lyu:2022imf,Lyu:2023xro,Lyu:2024ttm}. 
In order to further perform simulations on the physical point,
the gauge configurations at $m_{\pi} \approx 137~\textrm{MeV}$ have been generated~\cite{Aoyama:2024cko}.

In this paper, we focus on the kaon-nucleon ($KN$) interaction with strangeness $S=+1$.
Interactions between a pseudo-scalar meson and a nucleon, including $KN$, are of interest in many aspects. 
One example is that they provide an essential input to determine the quantitative nature of the partial restoration of chiral symmetry in nuclear matter~\cite{Aoki:2017hel,Iizawa:2023xsi}. 
In addition, meson-nucleon interactions play a vital role in the search for exotic states such as genuine pentaquarks and mesic nuclei~\cite{Hosaka:2016ypm}.
The $KN$ interactions have been studied through scattering experiments with the $K^{+}$ beam and nuclear target for a long time. 
A comprehensive overview can be found in Ref.~\cite{Dover:1982zh}. 
If restricted to the low-momentum elastic region, most experimental measurements were conducted in the 1960s and 1970s. 
However, there are no experimental data near the $KN$ threshold due to the difficulty of producing a kaon beam with low momentum. 
As a result, the $KN$ scattering observables in this region are obtained only by extrapolation using partial-wave analysis or chiral perturbation theory. 
This leads to a large ambiguity, especially for isospin $I=0$. 
Indeed, Ref.~\cite{Iizawa:2023xsi} has pointed out that a much more precise determination of the $KN$ scattering amplitudes at low momenta is required to evaluate accurately the strange quark condensate in a nuclear medium.

In this work, we investigate the S-wave $KN$ interactions using the HAL QCD method on the physical point.
Lattice QCD calculation of the $KN$ interactions offers technical advantages. 
Since the $KN$ systems do not have a quark pair creation/annihilation process, 
the conventional computational framework for calculating point-to-all (wall-to-all) quark propagators is sufficient to determine four-point correlation functions, which enables simulations with high statistics at a reasonable computational cost, even under realistic simulation setups. 
Moreover, only a single channel analysis is required around the threshold in the $KN$ systems. 
With these features, the time-dependent HAL QCD method can provide reliable $KN$ scattering phase shifts with high precision. 
Therefore, the first-principles lattice QCD determination is the most desirable theoretical approach on the issue of the $KN$ interactions.

We also note that this study can probe the possible existence of the $\Theta^{+}(1540)$ pentaquark through the $KN$ system in the S-wave channel.
The existence of this pentaquark was first claimed by the LEPS Collaboration~\cite{LEPS:2003wug}, but was later largely disfavored by subsequent experiments.
For an experimental review, see~\cite{Danilov:2008uxa}. 
In addition, a dozen or so lattice QCD calculations of the spectrum from two-point correlation functions have been carried out (see Ref.~\cite{Liu:2005yc} and references therein). 
However, these calculations were performed only with unphysical quark masses with quenched approximation.
For a definitive conclusion, 
it is crucial to search for the pentaquark by full QCD lattice simulations with physical quark masses through the $KN$ scattering process. 
Thus, we address this issue by analyzing the S-wave phase shifts derived from the HAL QCD potentials.

For the previous lattice QCD studies on the $KN$ scatterings, several works have been carried out using the finite-volume method~\cite{Luscher:1990ux} with heavy pion masses~\cite{Fukugita:1994ve, Meng:2003gm, Torok:2009dg, Detmold:2013gua, Detmold:2015qwf}. 
For the HAL QCD approach, Refs.~\cite{Ikeda:2010sg,Ikeda:2011qm} are the pioneering studies. 
In addition, there is a recent study using kaon and nucleon sources projected onto zero momentum~\cite{Murakami:2020yzt}.
Remark that our present study represents the first lattice QCD calculation of the $KN$ interaction on the physical point, which enables direct comparison with the experimental data. 

This paper is organized as follows. In Sec.~\ref{sec:HAL}, we briefly review the HAL QCD method for octet meson–baryon systems. 
In Sec.~\ref{sec:setups}, we introduce the four-point correlation functions used to extract the S-wave $KN$ potential and outline our numerical setup.  
In Sec.~\ref{sec:results}, we present our numerical results for the potentials and the scattering observables, and discuss possible systematic uncertainties of the results. 
Sec.~\ref{sec:conclusion} is devoted to the conclusion of this paper.

\section{\label{sec:HAL} HAL QCD method for systems of octet meson and baryon}

Let us start with the four-point correlation functions for systems of octet (pseudo-scalar) meson and baryon defined by
\begin{eqnarray}
F_{\alpha}(\vb{r},t)  
=\langle M(\vb{r+x},t+t_{0})B_{\alpha}(\vb{x},t+t_{0}) \bar{J}_{MB}(t_0) \rangle,
\end{eqnarray}
where $M(\vb{x},t)$ and $B_{\alpha}(\vb{x},t)$ are sink operators located at $(\vb{x},t)$ for octet (PS) meson and baryon, respectively, and $\bar{J}_{MB}(t_0)$ is the source operator of the meson-baryon states at the time $t_0$. 
We then introduce the R-correlator given by
\begin{eqnarray}
R_{\alpha}(\vb{r},t)  
=\frac{F_{\alpha}(\vb{r},t)}{C_{M}(t)C_{B}(t)},\label{eq:def_Rcorr}
\end{eqnarray}
where $C_{M}(t)$ and  $C_{B}(t)$ are two-point correlation functions of the meson and baryon, respectively. 
At sufficiently large time $t$, the R-correlator can be expressed by a linear combination of equal-time Nambu-Bethe-Salpeter (NBS) wave functions, and we obtain the time-dependent equation with interaction potential, $U(\vb{r},\vb{r}')$ as
\begin{eqnarray}
\int d^3r' \ U_{\alpha\beta}(\vb{r},\vb{r}')R_{\beta}(\vb{r'},t) 
\simeq
\left(\frac{\nabla^2}{2\mu}-\pdv{t} + \frac{1+3\delta^2}{8\mu}\pdv[2]{t} \right)R_{\alpha}(\vb{r},t) + \order{\Delta W^3}, \label{eq:scheq_rela_approx}
\end{eqnarray}
where $\mu$ is the reduced mass, $\delta=(m_M-m_B)/(m_M+m_B)$, and $\Delta W$ is the typical energy of the meson-baryon system from the threshold, $\Delta W=W-m_M-m_B$. 
We employ the leading-order (LO) approximation in the derivative expansion as
\begin{eqnarray}\label{eq:LO_approx}
U_{\alpha\beta}(\vb{r},\vb{r}') \simeq V^{\textrm{LO}}(r) \delta_{\alpha\beta}\delta^{(3)}(\vb{r}-\vb{r}'),
\end{eqnarray}
which leads to the following equation:
\begin{eqnarray}\label{eq:LOpotential_gen}
V^{\textrm{LO}}(r) \simeq
\frac{1}{R_{\alpha}(\vb{r},t)}\left(\frac{\nabla^2}{2\mu}-\pdv{t} + \frac{1+3\delta^2}{8\mu}\pdv[2]{t} \right)R_{\alpha}(\vb{r},t)
 + \order{\Delta W^3}.\end{eqnarray}
We thus obtain the LO potential from a single R-correlator via this equation.
Note that, according to the Okubo-Marshak decomposition~\cite{OKUBO1958166} of the generic meson-baryon potential, the LO term has a simple structure of baryon spin indices $\alpha$ and $\beta$ as in Eq.~\eqref{eq:LO_approx}, while more complicated structures start appearing at the next-to-leading order. 
Eqs.~\eqref{eq:scheq_rela_approx} and~\eqref{eq:LOpotential_gen} contain higher order terms $\order{\Delta W^3}$, which require higher-order time derivatives but are numerically confirmed to be negligible in the current study.

\section{\label{sec:setups} Setups}
\subsection{\label{subsec:def_of_corr} Definition of the correlation functions}
We use the following interpolating operators for kaons:
\begin{eqnarray}
\begin{aligned}
K^{+}(x) &= i\bar{s}(x)\gamma_{5}u(x), \quad K^{-}(x) = i\bar{u}(x)\gamma_{5}s(x), \\
K^{0}(x) &=  i\bar{s}(x)\gamma_{5}d(x), \quad \bar{K}^{0}(x) =  i\bar{d}(x)\gamma_{5}s(x),\label{eq:def_kaonop}
\end{aligned}
\end{eqnarray}
and for nucleons:
\begin{eqnarray}
\begin{aligned}
N_{\alpha}(x) &= \epsilon_{abc}q_{a,\alpha}(x)(u^{{\textrm T}}_{b}(x) C\gamma_{5}d_{c}(x)), \\
\bar{N}_{\alpha}(x) &= -\epsilon_{abc}\bar{q}_{a,\alpha}(x)(\bar{u}_{b}(x) C\gamma_{5}\bar{d}^{{\textrm T}}_{c}(x)), \label{eq:def_nucleonop}
\end{aligned}
\end{eqnarray}
where $N=(p,n)$ and $q=(u,d)$, and $\alpha$ is the spinor indices. 
Note that all the quark operators in these equations are local.

The $KN$ four-point correlation function for isospin $I(=0,1)$ is given by
\begin{eqnarray}\label{eq:4ptfunc}
F^{I}_{\alpha}({\bf r},t) =  \langle (KN)^{(I)}_{\alpha}({\bf r}, t+t_{0}) \bar{K}^{0}(t_0)\bar{p}_{\alpha}(t_0) \rangle,
\end{eqnarray}
where $(KN)^{(I)}_{\alpha}({\bf r}, t+t_{0})$ is the $KN$ sink operator defined as
\begin{eqnarray}
(KN)^{(I)}_{\alpha}({\bf r}, t) =  \sum_{\bf{x}}\frac{1}{\sqrt{2}}[K^{0}(\vb{r+x},t)p_{\alpha}(\vb{x},t) +(-1)^{I+1} K^{+}(\vb{r+x},t)n_{\alpha}(\vb{x},t)],\label{eq:sinkop}
\end{eqnarray}
and $\bar{K}^{0}(t_0)$ and $\bar{p}_{\alpha}(t_0)$ are the wall source operators, where local quark source operators in Eqs.~(\ref{eq:def_kaonop}) and (\ref{eq:def_nucleonop}) are replaced with the wall source operators.  
Note that the baryon operators at the sink and source in Eq.~\eqref{eq:4ptfunc} share the same spin indices,
since the spin flipping does not occur in the S-wave scatterings.

\subsection{\label{subsec:details} Simulation details}

We use $(2+1)$-flavor gauge configurations,``HAL-conf-2023'', generated with the Iwasaki gauge action and the non-perturbatively $\order{a}$-improved Wilson quark action with stout smearing at $\beta = 1.82$ and $c_{SW} = 1.11$ on a $96^4$ lattice~\cite{Aoyama:2024cko}. The corresponding lattice spacing and size are $a\approx  0.084~\textrm{fm}$ and $L\approx 8.1~\textrm{fm}$, respectively. 
The hopping parameters of the ensemble in our calculation are $\kappa_{ud}=0.126117$ and $\kappa_{s}=0.124902$, which corresponds to $m_{\pi}\approx 137~\textrm{MeV}$.
The periodic boundary condition is imposed in all four directions. 
We use 360 gauge configurations which are picked up one per five trajectories across three independent runs.
On each configuration, 
we perform measurements for 96 temporal source locations together with averages over forward-backward propagations, rotating the temporal direction 4 times by hypercubic symmetry. 
Therefore, the total number of the measurements is $360\times 96\times 2 \times 4 =276,480$.
The statistical errors are estimated by 
the jackknife method with a bin size of 40 configurations.
By comparing with a bin size of 20 configurations, we confirm that the bin-size dependence is small.

We also average the four-point correlation functions over spin indices to increase statistics, as the LO potential is spin-independent. 
In order to obtain the S-wave components, we employ the partial wave decomposition by the Misner method~\cite{Miyamoto:2019jjc} with $n_{\textrm{max}} = 4$, $l_{\textrm{max}} = 4$, and $\Delta/a = 1.0$, after projecting the four-point correlation functions onto the A$^{+}_{1}$ representation of the cubic group. 
However, at the three shortest points on the lattice, $r=0$, $a$, and $\sqrt2 a$, the LO potentials are computed using the correlation functions without the Misner method.
 
The kaon and nucleon masses are determined from the corresponding two-point correlation functions. 
We fit a cosh function in a temporal range $25\leq t/a \leq 39$ for kaon and a single exponential function in $14 \leq t/a \leq 20$ for nucleon, leading to $m_{K}= 502.0(2)~\textrm{MeV}$ and $m_{N} =  942.3(3.4)~\textrm{MeV}$, respectively, which are used in subsequent analysis for the LO potentials and scattering observables.

\section{\label{sec:results} Numerical results}
\subsection{\label{subsec:results_pot} $KN$ potentials}

\begin{figure}[htbp]
    \begin{center}
        \includegraphics[width=0.49\textwidth]{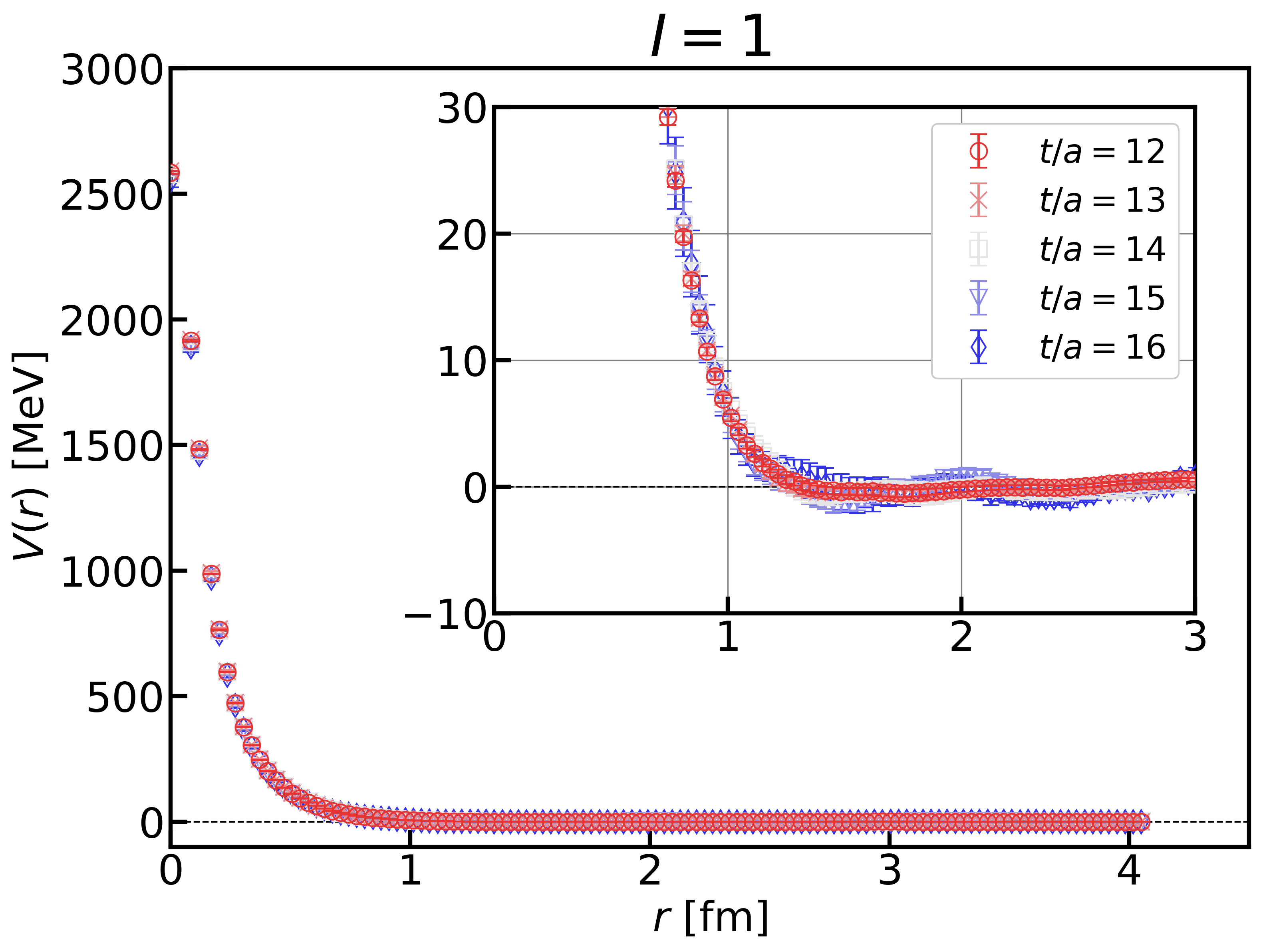}
        \includegraphics[width=0.49\textwidth]{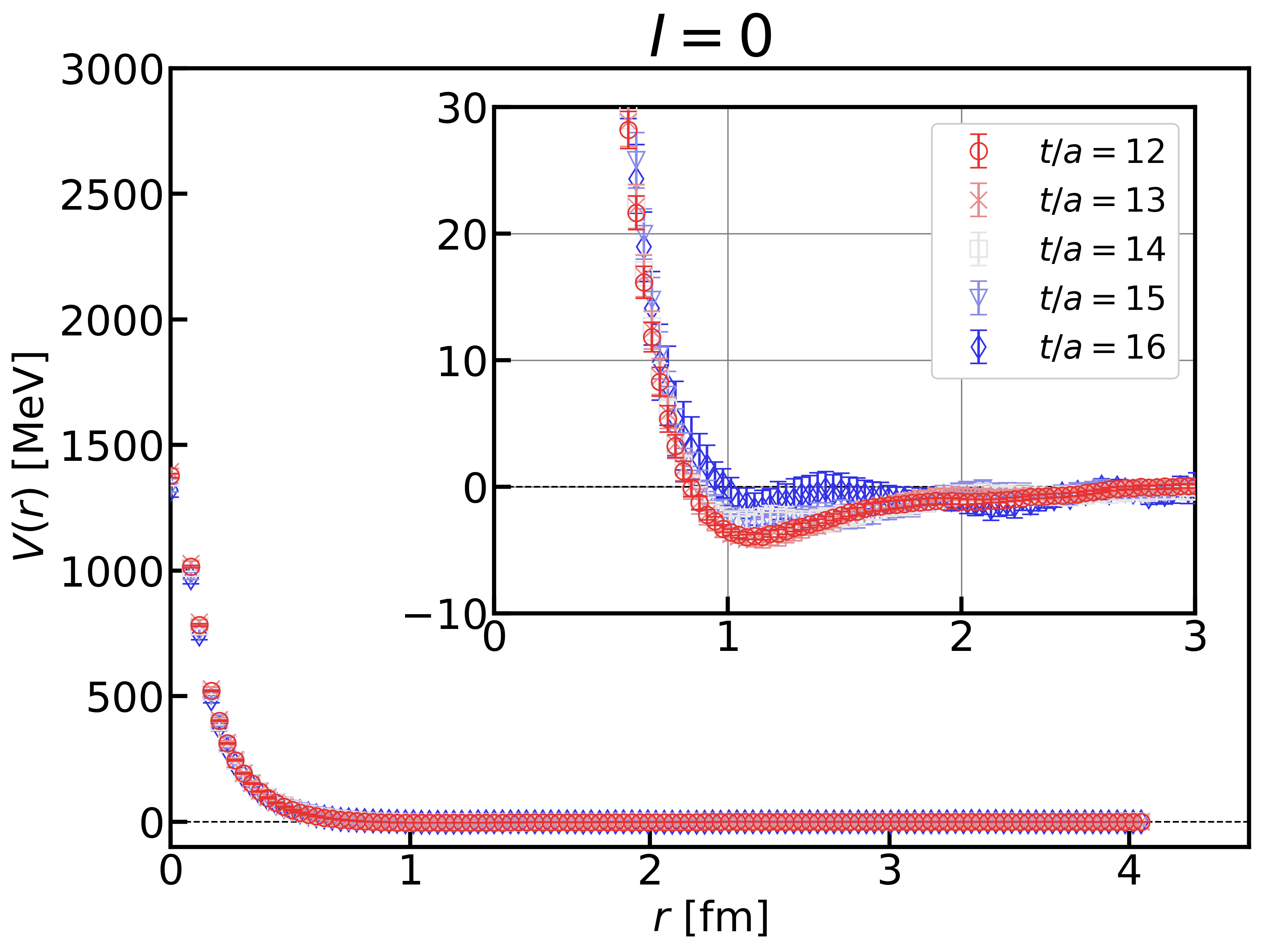}
	    \caption{Leading-order potentials in $I=1$ (Left panel) and $I=0$ channels (Right panel) at $t/a=12\textrm{--}16$.
        }
	\label{fig:results_pot_orig}    
    \end{center}
\end{figure}

Figure~\ref{fig:results_pot_orig} presents the LO potentials extracted at $t/a=12\textrm{--}16$, which do not suffer from large statistical uncertainties.
No significant time dependence is observed in the potentials, which indicates that the LO term in the derivative expansion is dominant and that contamination from the inelastic scattering states is suppressed.

The potential in both $I=1$ and $0$ channels has a repulsive core with a range of about $1~\textrm{fm}$. 
Between two channels, the range of the core is larger for $I=1$.  
In addition, the $I=0$ potential possesses an attractive pocket of a few MeV at intermediate distances. 

We make a comparison of the previous HAL QCD results at heavier pion masses,
both isospin channels at $m_{\pi}\approx 570~\textrm{MeV}$~\cite{Murakami:2020yzt} and only the $I=1$ channel at $m_{\pi}\approx 705~\textrm{MeV}$~\cite{Ikeda:2011qm}~\footnote{There is another result in the HAL QCD method at $m_{\pi}\approx 870~\textrm{MeV}$~\cite{Ikeda:2010sg}. We exclude this work in the comparison, however, since the potentials suffer from a large finite-volume effect.}. 
A visual comparison of the results in the present study with those calculated at $m_{\pi}\approx 570~\textrm{MeV}$ is provided in Appendix~\ref{appendix:comp_pot_570MeV}. 
The potentials in previous studies share the same qualitative feature as the present ones, while the range of the repulsive core is smaller in the previous studies. 
This indicates that the repulsion is weaker at heavier quark masses, suggesting that the color-magnetic interaction between quarks, that scales inversely with respect to the quark mass, is responsible for the repulsion at short distances. 
On the other hand, no significant difference is found at long distances.

\begin{figure}[htbp]
    \begin{center}
        \includegraphics[width=0.49\textwidth]{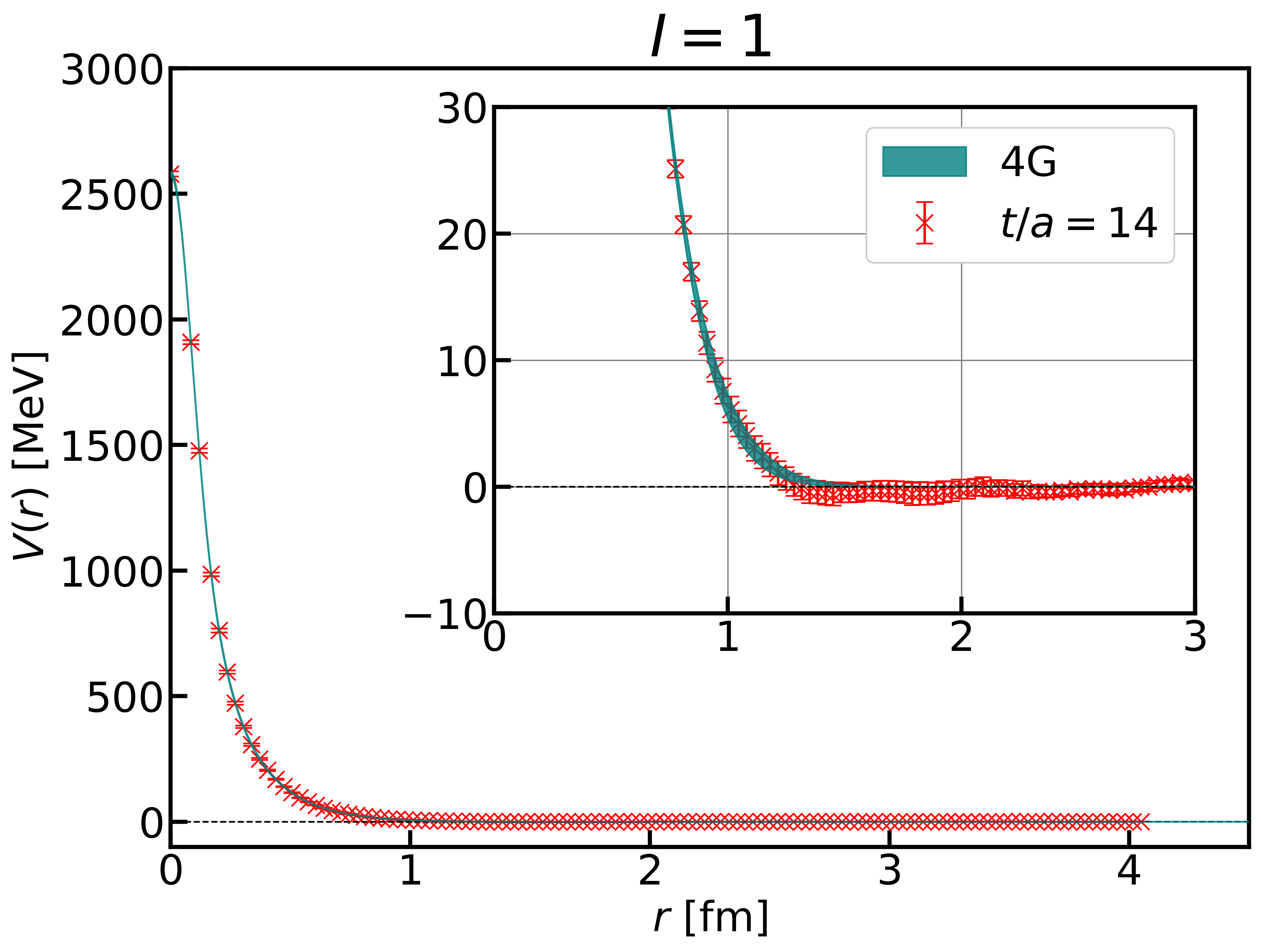}
        \includegraphics[width=0.49\textwidth]{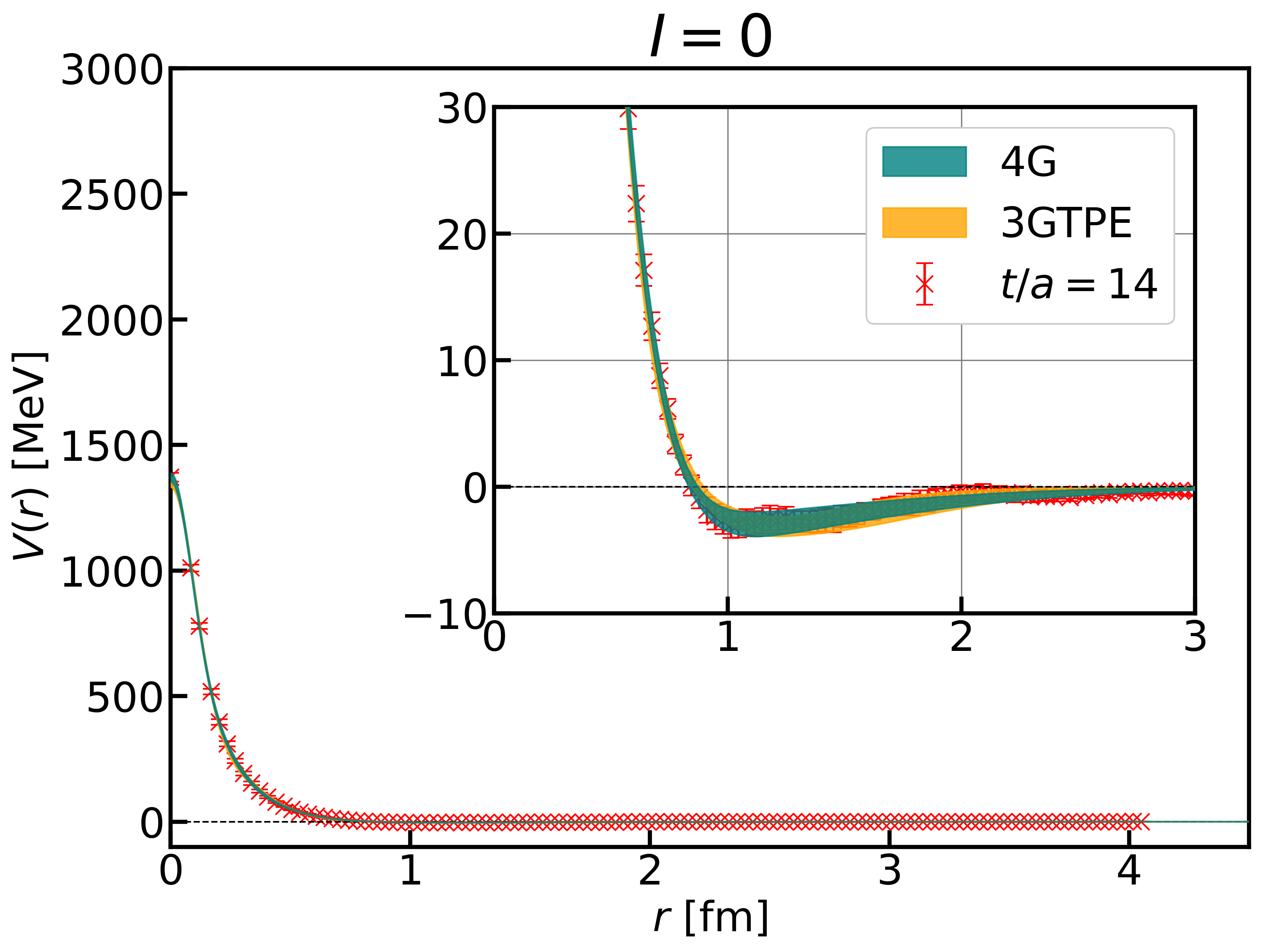}
	    \caption{Fit results for the $I=1$ (Left panel)  and $I=0$ channels (Right panel) at $t/a=14$. 
        Teal and orange bands correspond to the fit results using $V^{\textrm{4G}}_{I}(r)$ and $V^{\textrm{3GTPE}}_{I}(r)$, respectively.
        The red cross plots are the original potential data (same as those in Fig.~\ref{fig:results_pot_orig}).
        }
	\label{fig:results_pot_fit}    
    \end{center}
\end{figure}

We then perform the uncorrelated fit to the obtained potentials. 
We employ the following two types of fit functions: a four-range Gaussian as
\begin{eqnarray}
V^{\textrm{4G}}_{I}(r) &=& \sum_{i=1}^{4} a^{I}_{i}e^{-\frac{r^2}{(b^{I}_{i})^2}}, \label{eq:fitfunc_4G} 
\end{eqnarray}
and a three-range Gaussian plus a two-pion-exchange (TPE) term
\begin{eqnarray}
V^{\textrm{3GTPE}}_{I}(r) = \sum_{i=1}^{3} c^{I}_{i}e^{-\frac{r^2}{(d^{I}_{i})^2}}+\alpha^{I}\left[(1-e^{- \beta r^2})\frac{e^{-m_{\pi} r}}{r}\right]^2, \label{eq:fitfunc_3GTPE}
\end{eqnarray}
where $b^{I}_{1}<b^{I}_{2}<b^{I}_{3}<b^{I}_{4}$ and $d^{I}_{1}<d^{I}_{2}<d^{I}_{3}$, and $\beta$ is fixed to $2.0~\textrm{fm}^{-2}$, a value used in, e.g.,  Ref.~\cite{Wiringa:1984tg}~\footnote
{We cannot take $\beta$ as a fit parameter since the fit results strongly depend on the initial value of $\beta$.
}. 
The TPE term is associated with a phenomenological two-pion exchange process via $K^{*}N$  intermediate states~\cite{Aaron:1971ka}~\footnote{The vector-meson exchange term is not considered in this analysis since it is short-ranged and cannot be distinguished with the interaction process at the level of quarks.}.
The fit using $V^{\textrm{4G}}_{I}(r)$ works well for both isospin channels. 
The fit parameters are listed in Table~\ref{tab:fitparams_4G} and the corresponding fit curves are shown by teal bands in Fig.~\ref{fig:results_pot_fit}. 
For $V^{\textrm{3GTPE}}_{I}(r)$ in the $I=1$ channel, no reasonable fit results can be obtained for some jackknife samples, while the resulting coefficient $\alpha^{I=1}$ is negligibly small for the others, which suggests that the TPE term is not suitable to describe its long-range part. 
For $I=0$, on the other hand, the function $V^{\textrm{3GTPE}}_{I=0}(r)$ produces good fit results. 
The parameters are shown in Table~\ref{tab:fitparams_3GTPE}, and the corresponding curve is depicted by orange bands in Fig.~\ref{fig:results_pot_fit}. However, the range of the third Gaussian term ($d^{I=0}_{3}$) is comparable with that of the TPE term, $(2m_{\pi})^{-1}\approx 0.72~\textrm{fm}$, which indicates that the TPE term is not enough to reproduce the long-range part of the $I=0$ potential. 
These results suggest that the coupling of the TPE is likely to be very small 
since it is suppressed by $(m_{K^{\ast}}m_{N})^{-1}$ with the $K^{*}$ meson mass $m_{K^{\ast}}$. 
\begin{table}[h]
\centering
\caption{
Fit parameters of the $KN$ potential at $t/a=14$ for $V^{\textrm{4G}}_{I}(r)$. 
The values of $\chi^2/\textrm{d.o.f.}=0.71(40)$ ($0.64(49)$) for $I=1$ ($I=0$).
}
\label{tab:fitparams_4G}
\begin{tabular}{c||c|c|c|c|c|c|c|c}
 \hline \hline
 $I$ & $a^{I}_{1}$ [MeV] & $b^{I}_{1}$ [fm] & $a^{I}_{2}$ [MeV] & $b^{I}_{2}$ [fm] & $a^{I}_{3}$ [MeV] & $b^{I}_{3}$ [fm] & $a^{I}_{4}$ [MeV] & $b^{I}_{4}$ [fm]  \\
\hline
 $1$ & $1172.2(93.6)$ & $0.113(2)$ & $782.0(47.2)$ & $0.191(16)$ & $446.2(84.1)$ & $0.324(36)$ & $178.6(60.5)$ & $0.548(38)$   \\
\hline
 $0$ & $726.9(53.8)$ & $0.117(3)$ & $443.5(42.5)$ & $0.228(16)$ & $205.0(28.4)$ & $0.427(20)$ & $-5.1(1.7)$ & $1.629(118)$   \\
\hline \hline
\end{tabular}
\end{table}
\begin{table}[htbp]
\centering
\caption{
Fit parameters of the $I=0$ $KN$ potential at $t/a=14$ for $V^{\textrm{3GTPE}}_{I}(r)$. 
The value of $\chi^2/\textrm{d.o.f.}=0.91(49)$.
}
\label{tab:fitparams_3GTPE}
\begin{tabular}{c|c|c|c|c|c|c}
\hline  \hline
 $c^{I=0}_{1}$ [MeV] & $d^{I=0}_{1}$ [fm] & $c^{I=0}_{2}$ [MeV] & $d^{I=0}_{2}$ [fm] & $c^{I=0}_{3}$ [MeV] & $d^{I=0}_{3}$ [fm] & $\alpha^{I=0}$ [MeV fm$^2$] \\
 \hline
 $943.0(58.7)$ & $0.135(7)$ & $343.5(68.8)$ & $0.327(25)$ & $65.0(21.2)$ & $0.750(167)$ & $-71.7(54.2)$ \\
\hline \hline
\end{tabular}
\end{table}

\subsection{\label{subsec:results_obs} $KN$ scattering observables}

We solve the Schr\"{o}dinger equations with the fitted potentials to extract phase shifts of the S-wave $KN$ scatterings. 
We first notice that two fits for $I=0$ data using $V^{\textrm{4G}}_{I=0}(r)$ and $V^{\textrm{3GTPE}}_{I=0}(r)$ lead to almost identical results on the phase shifts, of which the details will be discussed in the next subsection. 
We therefore present results obtained from $V^{\textrm{4G}}_{I}(r)$ in both $I=1$ and $0$ channels hereafter. 
In addition, for visual clarity, we show the scattering results only for $t/a=12$, $14$, and $16$ unless otherwise stated, although we have confirmed that the potentials for the other time slices yield scattering results consistent within statistical uncertainties.

\begin{figure}[htbp]
    \begin{center}
        \includegraphics[width=0.49\textwidth]{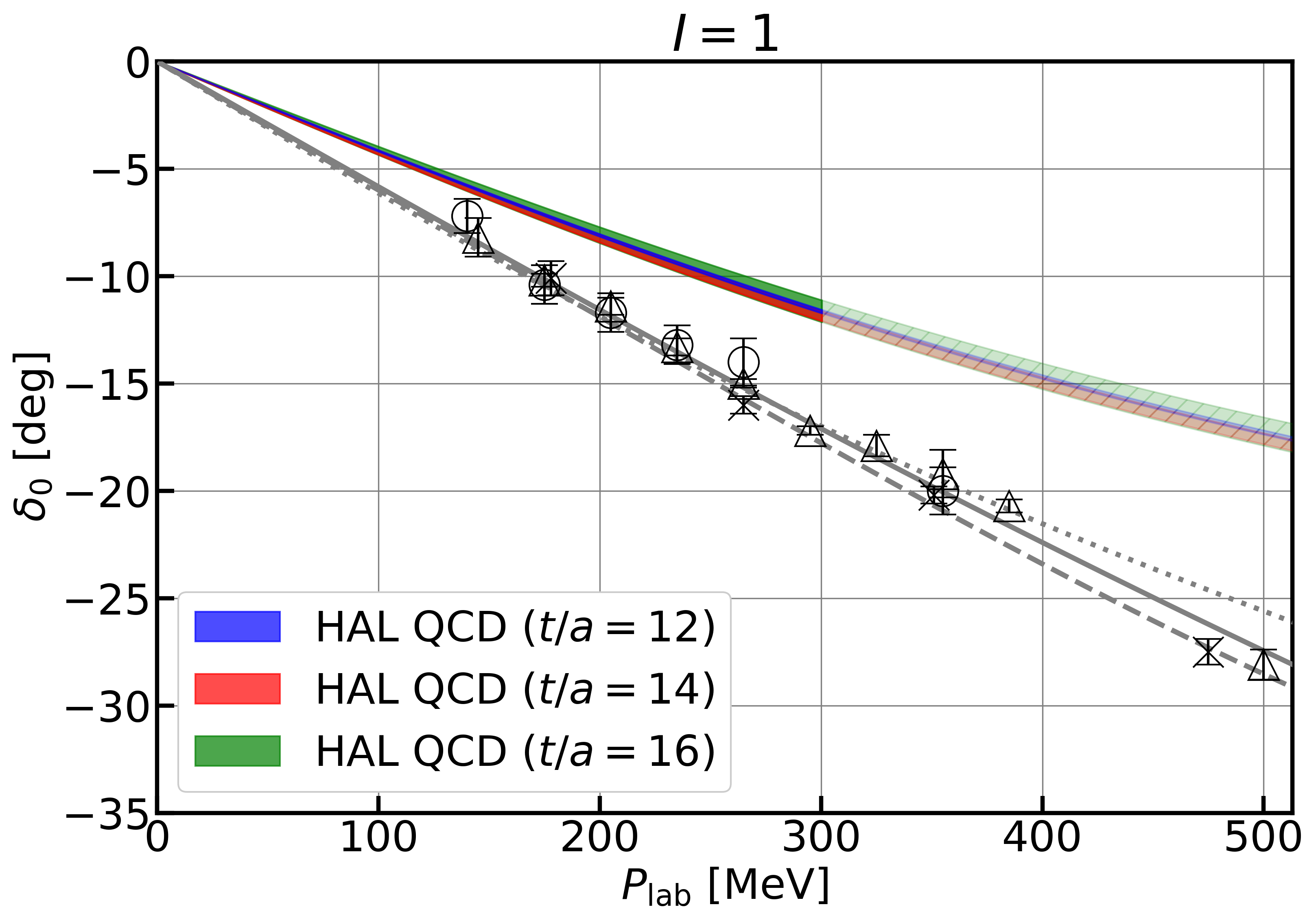}
        \includegraphics[width=0.49\textwidth]{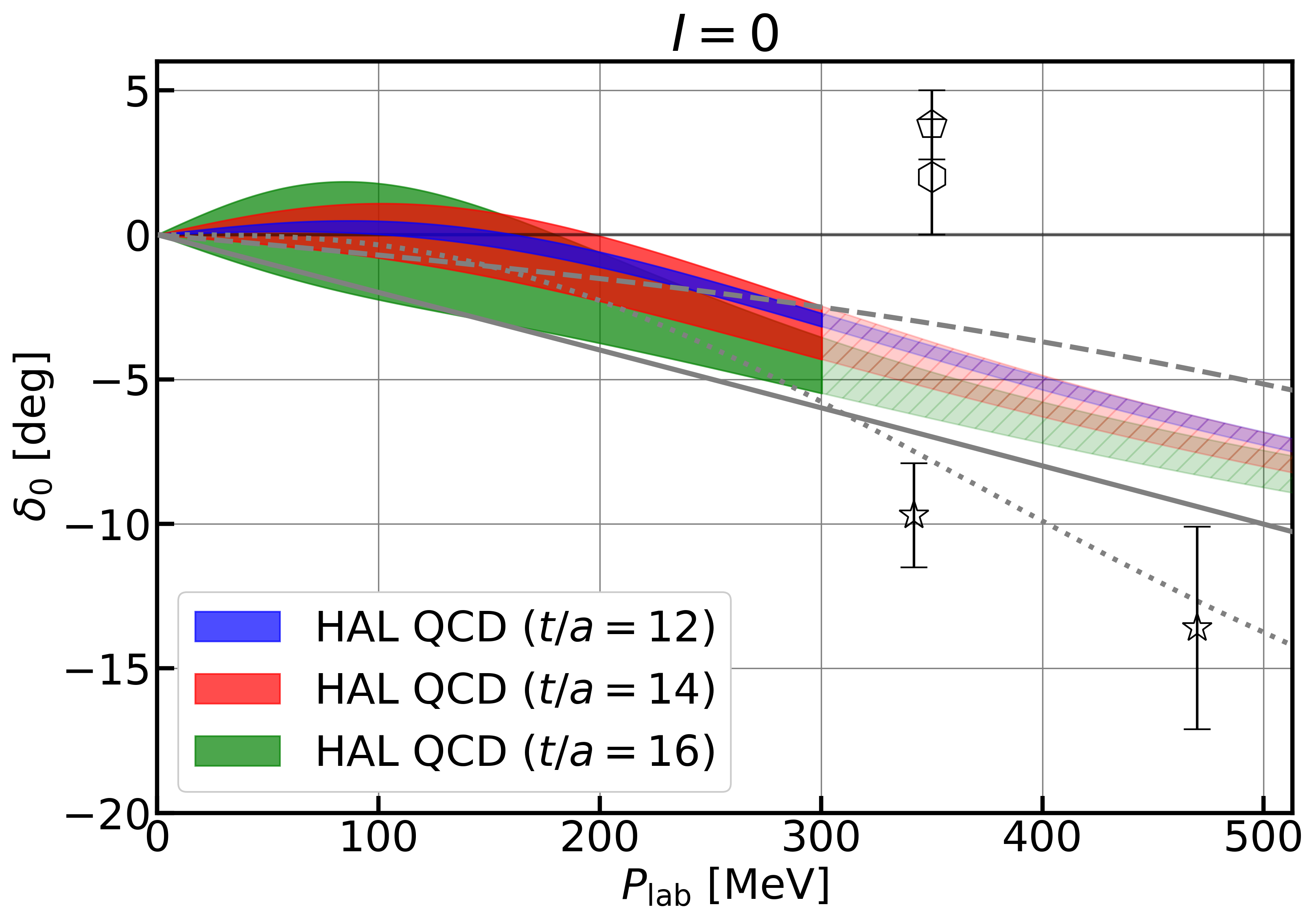}
	    \caption{
        Phase shifts for the S-wave $KN$ scatterings in $I=1$ (Left panel) and $I=0$ channels (Right panel) at $t/a=12$ (Blue), $14$ (Red), and  $16$ (Green).
        Experimental results are taken from Refs.~\cite{Goldhaber:1962zz} (Circles), \cite{Burnstein:1974ax} (Crosses), and~\cite{Cameron:1974xx} (Triangles) for $I=1$ channel, and from Refs.~\cite{PhysRev.134.B1111} (Pentagon and Hexagon) and~\cite{PhysRevD.15.1200} (Stars) for $I=0$ channel.
        Dashed, dotted and solid lines depict the results of the energy-dependent partial-wave analysis in Refs.~\cite{Martin:1975gs},~\cite{Hyslop:1992cs} and~\cite{Gibbs:2006ab}, respectively. 
        }
	\label{fig:results_phase}    
    \end{center}
\end{figure}

Fig.~\ref{fig:results_phase} shows results of phase shifts as a function of the kaon momentum in the laboratory frame $P_{\textrm{lab}}$. 
In the figure, we show data at $P_{\rm lab}\le 300$ MeV by bands with dark colors, while
those at $P_{\rm lab} >  300$ MeV with light colors and stripe pattern, in order to remind readers of a fact that results in the higher momentum region contain larger systematics due to non-locality of the potentials neglected in our LO analysis.
The value $P_{\textrm{lab}}=300~\textrm{MeV}$ lies between the first and second excited states in a non-interacting $KN$ system in our finite spatial volume of $(8.1 {\rm fm})^3$.
For comparisons, experimental results (in Refs.~\cite{Goldhaber:1962zz,Burnstein:1974ax,Cameron:1974xx} for $I=1$ and Refs.~\cite{PhysRev.134.B1111,PhysRevD.15.1200} for $I=0$) and results from the partial wave analysis in the energy-dependent approach~\cite{Martin:1975gs,Hyslop:1992cs,Gibbs:2006ab} are also shown in the figure. 

Our results of the phase shifts monotonically decrease for $I=1$, due to the pure repulsive interaction.
Such behaviors are also seen in all experimental results as well as the partial-wave analysis, though they are larger in magnitude than ours. 
Possible systematic uncertainties in our calculation that might cause this difference are discussed in the next subsection. 

For $I=0$, phase shifts from the potential are nearly zero within errors below $P_{\textrm{lab}}\sim 180~\textrm{MeV}$ and start decreasing above it, which is probably implied by a combination of the repulsion with the small attractive pocket. 
Two experimental results~\cite{PhysRev.134.B1111,PhysRevD.15.1200} disagree with each other in both sign and magnitude, and three partial-wave analyses show quantitatively different $P_{\rm lab}$ dependence among them.
The lack of low energy experimental data for the $I=0$ phase shifts seems to be a reason for these disagreements. 
Our results lie between two experimental results.
Moreover, the phase shifts at $t/a=14$ are consistent with the two of partial-wave analysis~\cite{Martin:1975gs,Hyslop:1992cs} near the threshold.

In both isospin channels, phase shifts start from 0 degrees at the threshold (with the convention that phase shifts approach zero at infinitely large momentum), indicating an absence of bound states according to Levinson's theorem.
Above the threshold, they show neither sudden rise nor crossing $90$ degrees, which is a typical behavior of resonances. 
These observations lead to the conclusion that the $\Theta^{+}(1540)$ pentaquark does not exist in S-wave $KN$ systems.

\begin{figure}[htbp]
    \begin{center}
        \includegraphics[width=0.49\textwidth]{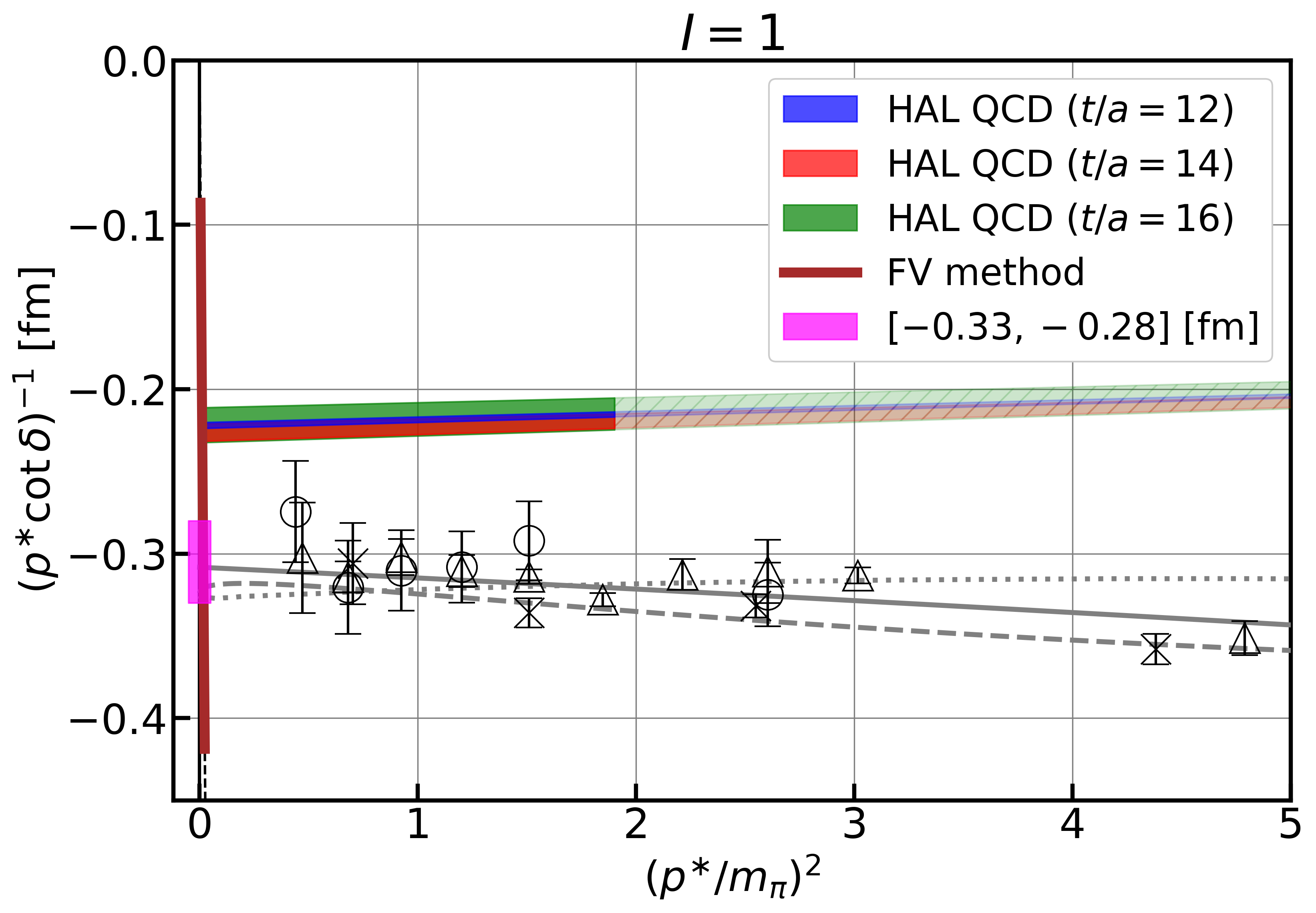}
        \includegraphics[width=0.49\textwidth]{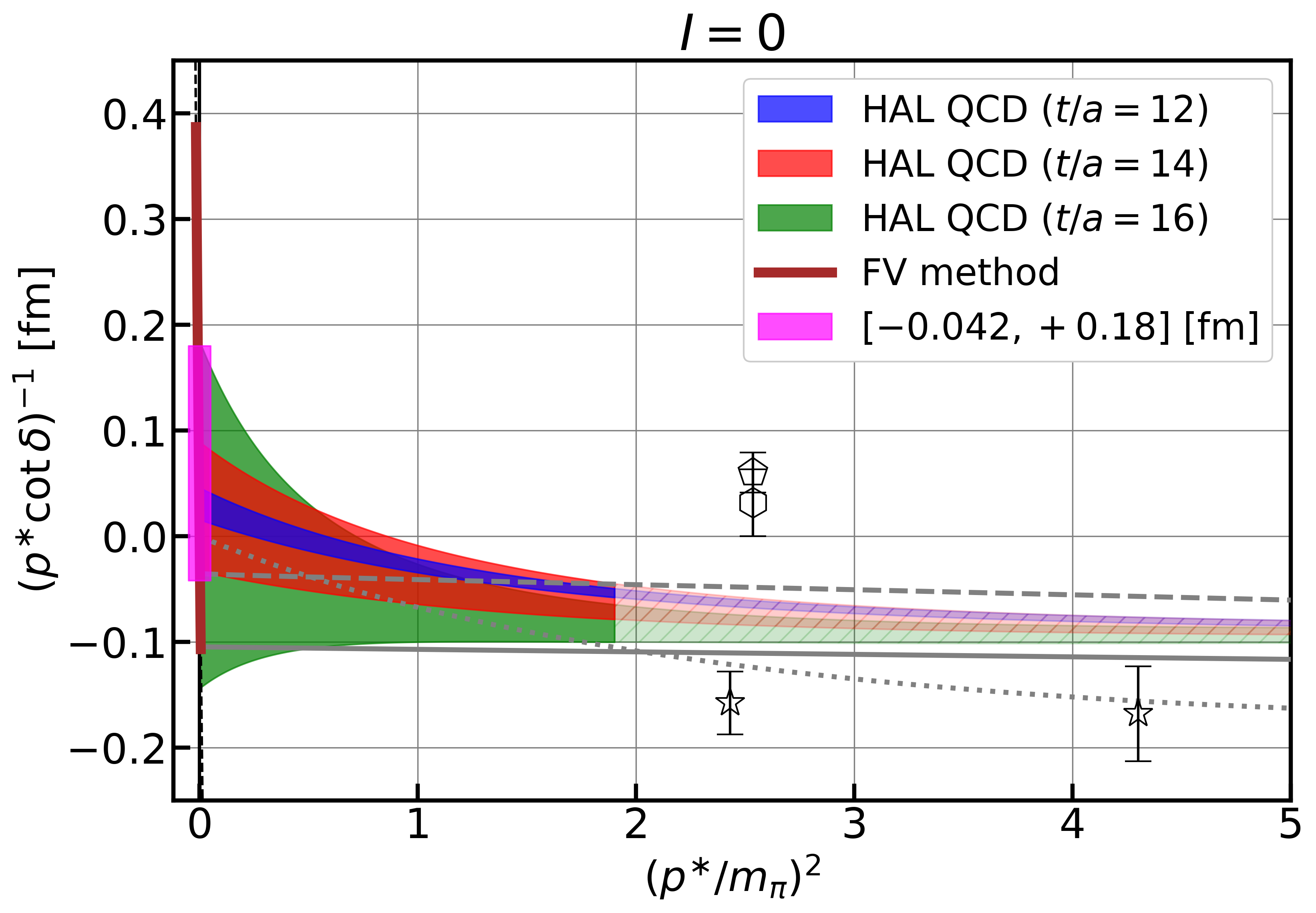}
	    \caption{
        $(p^{\ast}\cot{\delta_{0}(p^{\ast})})^{-1}$ for the S-wave $KN$ scatterings in $I=1$ (Left panel) and $I=0$ channels (Right panel) at $t/a=12$ (Blue), $14$ (Red), and $16$ (Green). 
        Experimental results are taken from Refs.~\cite{Goldhaber:1962zz} (Circles), \cite{Burnstein:1974ax} (Crosses), and~\cite{Cameron:1974xx} (Triangles) for $I=1$ channel, and from Refs.~\cite{PhysRev.134.B1111} (Pentagon and Hexagon) and~\cite{PhysRevD.15.1200} (Stars) for $I=0$ channel.
        Dashed, dotted and solid lines depict the results of the energy-dependent partial-wave analysis in Refs.~\cite{Martin:1975gs},~\cite{Hyslop:1992cs} and~\cite{Gibbs:2006ab}, respectively.         
        Magenta band located at $(p^{\ast}/m_{\pi})^2=0$ represents the scattering length taken from~\cite{Dover:1982zh}. 
        The phase shift at the ground-state energy on the finite volume is shown as a brown curve. 
        }
	\label{fig:results_kcotdinv}    
    \end{center}
\end{figure}
To evaluate scattering parameters, we calculate $p^{\ast}\cot{\delta_{0}}$ with $p^{\ast}$ being the relative momentum in the center-of-mass frame, of which the inversions, $(p^{\ast}\cot{\delta_{0}})^{-1}$, are shown in Fig.~\ref{fig:results_kcotdinv}. 
The experimental results~\cite{Goldhaber:1962zz,Burnstein:1974ax,Cameron:1974xx,PhysRev.134.B1111,PhysRevD.15.1200} and the results from the partial-wave analysis~\cite{Martin:1975gs,Hyslop:1992cs,Gibbs:2006ab} are also depicted in the figure. 
The scattering length $a_0$ and effective range $r_{\mathrm{eff}}$ defined as
\begin{eqnarray}
p^{\ast} \cot \delta_{0}(p^{\ast}) =  \frac{1}{a_{0}} + \frac{1}{2}  r_{\mathrm{eff}}(p^{\ast})^2 + \order{(p^{\ast})^4} \label{eq:def_of_ERE},
\end{eqnarray}
are given by
\begin{eqnarray}
\begin{aligned}
a^{I=1}_{0} &= -0.226(5)(^{+5}_{-0})~\textrm{fm}, \quad 
r^{I=1}_{\textrm{eff}} = -0.294(30)(^{+21}_{-7})~\textrm{fm}, \\
a^{I=0}_{0} &= +0.028(61)(^{+3}_{-26})~\textrm{fm},
\end{aligned}\label{eq:results_a0_reff}
\end{eqnarray}
where numbers in the first (second) parentheses are statistical (systematic) errors. 
Systematic errors are estimated from differences between the central value at $t/a=14$ and that within $t/a=12\textrm{--}16$. 
The effective range for $I=0$ is not shown here since statistical errors are too large relative to the small central value.
We notice that the scattering length corresponds to the $y$-intercept in Fig.~\ref{fig:results_kcotdinv}. 
In addition, since $(p^{\ast} \cot \delta_{0}(p^{\ast}))^{-1}=a_{0} - \frac{1}{2} a_0^2 r_{\mathrm{eff}}(p^{\ast})^2 + \order{(p^{\ast})^4}$, positive (negative) sign of the slope at the threshold in the figure indicates negative (positive) sign of the effective range. 
As a comparison of the scattering lengths, we show in Fig.~\ref{fig:results_kcotdinv} the summary results of various analyses based on the experimental data~\cite{Dover:1982zh}~\footnote{The summary results include those in Ref.~\cite{Martin:1975gs}, in which the results of $(p^{\ast} \cot \delta_{0}(p^{\ast}))^{-1}$ are shown as dashed lines in Fig.~\ref{fig:results_kcotdinv}.}, that $a_{0,\textrm{exp}}^{I=1}\in [-0.33,-0.28]~\textrm{fm}$ for $I=1$ and $a_{0,\textrm{exp}}^{I=0}\in [-0.042,0.18]~\textrm{fm}$ for $I=0$.

The scattering length is negative for $I=1$ and nearly zero within errors for $I=0$.
Both are qualitatively consistent with the summary results~\cite{Dover:1982zh} and the results in the partial-wave analysis~\cite{Martin:1975gs,Hyslop:1992cs,Gibbs:2006ab}, though the magnitude of our result for $I=1$ is a little smaller. 
On the other hand, the effective range for $I=1$ is negative, indicating a positive slope in Fig.~\ref{fig:results_kcotdinv}, which is opposite to the summary results $r_{\textrm{eff},\textrm{exp}}^{I=1}\in [0.32,0.5]~\textrm{fm}$ but consistent with the partial-wave analyses that exhibit positive slopes around the threshold in the figure~\cite{Martin:1975gs,Hyslop:1992cs}.
A discrepancy in $(p^{\ast} \cot \delta_{0}(p^{\ast}))^{-1}$ for $I=1$ between our results and experimental data, existing only above $P_{\textrm{lab}}=140~\textrm{MeV}$ ($(p^{\ast}/m_{\pi})^2=0.44$), as seen in Fig.~\ref{fig:results_kcotdinv}, seems to cause a difference in the scattering length.

\begin{figure}[htbp]
    \begin{center}
        \includegraphics[width=0.49\textwidth]{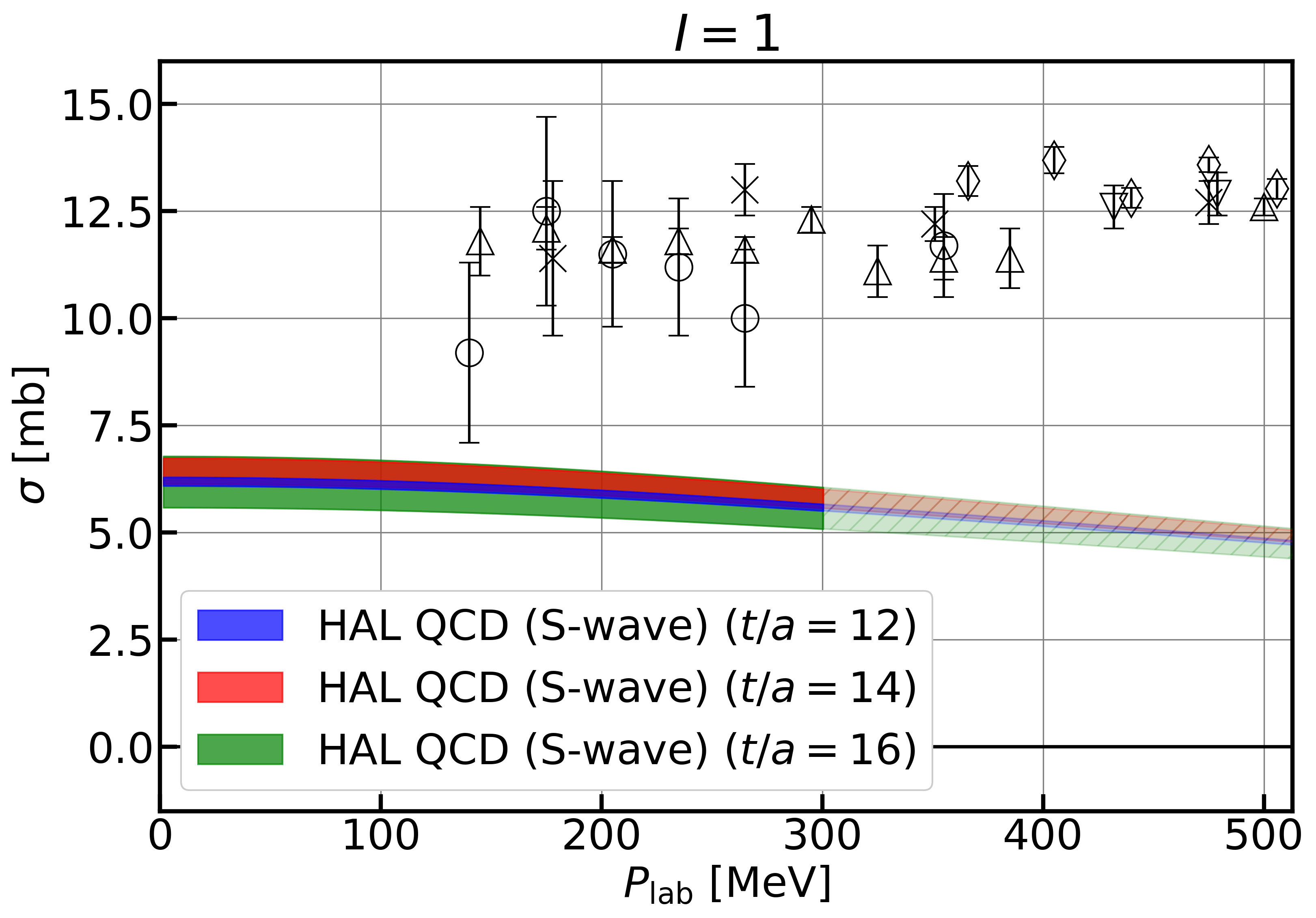}
        \includegraphics[width=0.49\textwidth]{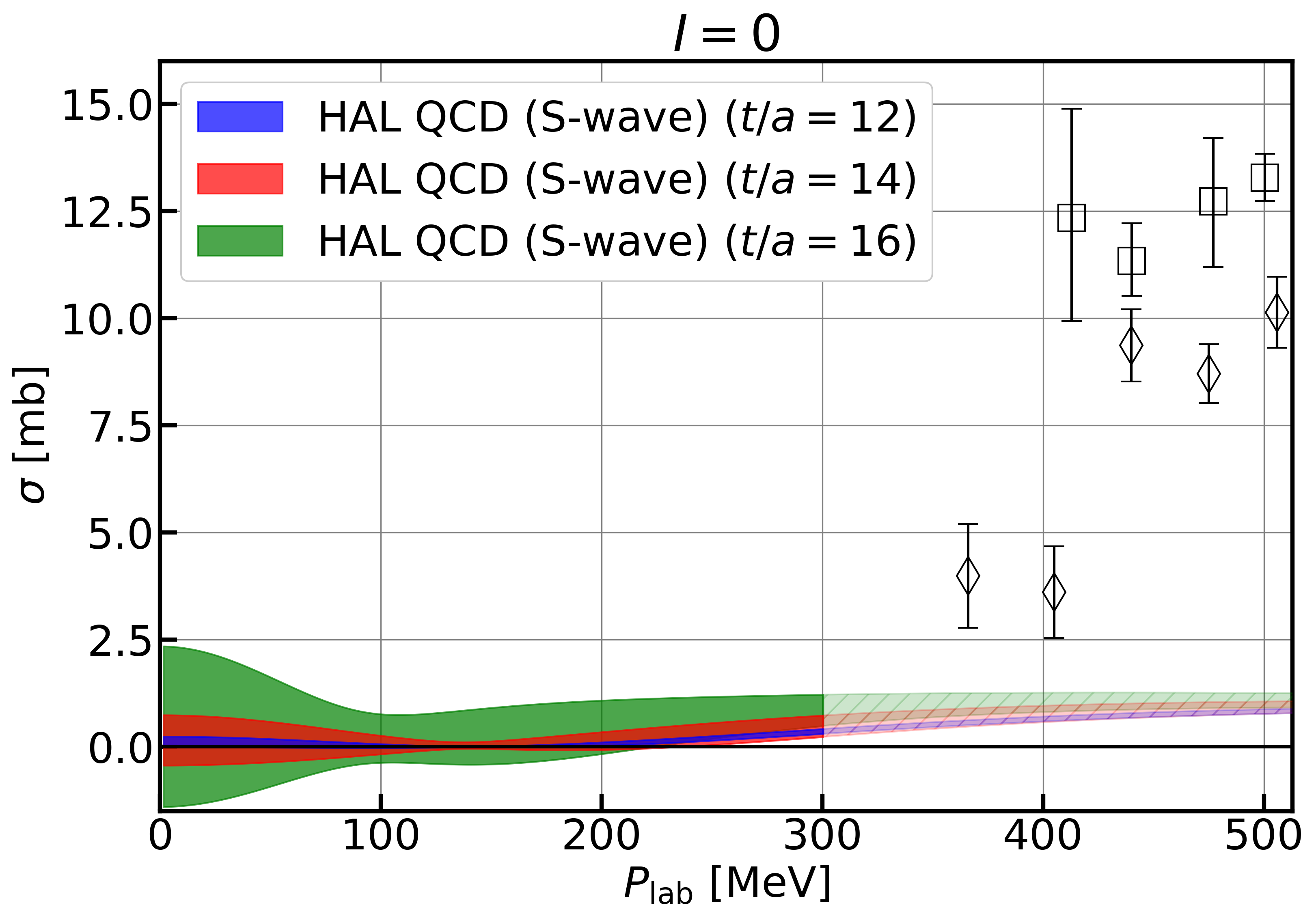}
	    \caption{S-wave total cross sections for $KN$ scatterings in $I=1$ (Left panel) and  $I=0$ channels (Right panel) at $t/a=12$ (Blue), $14$ (Red), and  $16$ (Green). 
        Experimental results are taken from Refs.~\cite{Goldhaber:1962zz} (Circles),~\cite{Burnstein:1974ax} (Crosses),~\cite{Cameron:1974xx} (Triangles),~\cite{Bowen:1970azd} (Diamonds) and~\cite{Adams:1973elr} (Inverted triangles) for the $I=1$ channel, and from Refs.~\cite{Bowen:1970azd} (Diamonds) and ~\cite{Carroll:1973ux} (Squares) for the $I=0$ channel.
        }
	\label{fig:results_crosssec}    
    \end{center}
\end{figure}

To make a more direct comparison of our lattice results and experimental data, we plot in Fig.~\ref{fig:results_crosssec} the S-wave component of the total $KN$ cross section obtained from lattice QCD, together with experimental results at $P_{\textrm{lab}} \leq 506~\textrm{MeV}$~\cite{Goldhaber:1962zz,Bowen:1970azd,Carroll:1973ux,Burnstein:1974ax,Cameron:1974xx}.
While the total cross section obtained in experiments contains all partial waves,
the one for $I=1$ is supposed to be dominated by the S-wave component at low $P_{\textrm{lab}}$, since the differential cross section for $K^{+}p\to K^{+}p$ scatterings has a small dependence on the angle except for forward scatterings, which diverges due to the Coulomb repulsion (see, e.g., Fig.1 in Ref.~\cite{Cameron:1974xx}). 

Our result on the $I=1$ S-wave cross section, which is $6.3\pm 0.3~\textrm{mb}$ at the threshold and slowly decreases with respect to $P_{\textrm{lab}}$, deviates from the experimental total cross section, which is larger than $10~\textrm{mb}$ at $P_{\textrm{lab}}<300~\textrm{MeV}$, though some data in Refs.~\cite{Goldhaber:1962zz,Burnstein:1974ax} are still consistent with ours within $2$--$3$ $\sigma$.
This deviation is also related to discrepancies in the $I=1$ phase shifts. 
In the next subsection, we will discuss possible sources for this difference.

The $I=0$ S-wave cross section, on the other hand, is nearly zero within errors below $P_{\textrm{lab}}\sim 180~\textrm{MeV}$, and 
becomes nonzero but less than $1~\textrm{mb}$ above $P_{\textrm{lab}}\sim 180~\textrm{MeV}$. 
This behavior is consistent with the recent (unitarized) chiral perturbation analysis~\cite{Aoki:2018wug,Iizawa:2023xsi}, 
which claims that experimental data above $P_{\textrm{lab}}=300~\textrm{MeV}$~\cite{Bowen:1970azd,Carroll:1973ux} are mainly explained by the P-wave rather than the S-wave component. 
The P-wave dominance has also been proposed by the analysis to experimental data~\cite{Slater:1961zz,PhysRev.134.B1111,Giacomelli:1974az,PhysRevD.15.1200,PhysRevD.12.3386,PhysRevD.15.1846,Martin:1975gs}. 
While it is likely that the $I=0$ scattering amplitudes are dominated by the P-wave rather than the S-wave, 
it is necessary to calculate the P-wave $KN$ interactions in lattice QCD for the definite conclusion.
As this calculation is challenging on the physical point due to considerable statistical and systematic errors, it is worth performing it even with heavier quark masses in future.

\subsection{\label{subsec:results_FV} Discussions on systematic uncertainties}

In this subsection, we discuss possible systematic uncertainties in our calculation that may lead to discrepancies observed between experimental results and ours, particularly in the $I=1$ channel.

We begin with the finite-volume effect on the interacting potential. 
The half of the spatial extension, $L/2\approx 4.03~\textrm{fm}$,
is much larger than the interaction range, which is less than $2.0~\textrm{fm}$ as estimated from Fig.~\ref{fig:results_pot_orig}.
Therefore, the finite-volume effect to our potentials is negligible.

We next examine the systematic uncertainty associated with the fit ansatz to potential. 
In Sec.~\ref{subsec:results_pot}, we introduced the four-range Gaussian Eq.~\eqref{eq:fitfunc_4G}, denoted by $V^{\textrm{4G}}_{I}(r)$, and three-range Gaussian plus two-pion exchange term Eq.~\eqref{eq:fitfunc_3GTPE}, denoted by $V^{\textrm{3GTPE}}_{I}(r)$.
To examine the fit function dependence, we additionally employ a three-range Gaussian plus a pion-exchange term with an arbitrary power 
\begin{eqnarray}
V^{\textrm{3GGPE}}_{I,n}(r) = \sum_{i=1}^{3} c^{I}_{i}e^{-\frac{r^2}{(d^{I}_{i})^2}}+\alpha^{I}\left[(1-e^{- \beta r^2})\frac{e^{-m_{\pi} r}}{r}\right]^n. \label{eq:fitfunc_3GGPE}
\end{eqnarray}
Note that the case $n=2$ is equivalent to $V^{\textrm{3GTPE}}_{I}(r)$.
For $I=0$, in addition to $V^{\textrm{4G}}_{I}(r)$ and $V^{\textrm{3GTPE}}_{I}(r)$, we find that $V^{\textrm{3GGPE}}_{I,n}(r)$ with $n=3$ provides good fit parameters. 
For $I=1$, as discussed in Sec.~\ref{subsec:results_pot}, we were able to obtain reasonable fit results for $V^{\textrm{4G}}_{I}(r)$ while we were not for $V^{\textrm{3GTPE}}_{I}(r)$. 
In addition to $V^{\textrm{4G}}_{I}(r)$, we identify two further fit functions that works well: $V^{\textrm{3GGPE}}_{I,n}(r)$ with $n=4$ and a three-range Gaussian without the pion-exchange term. 
Hereafter, the latter is referred to as $V^{\textrm{3G}}_{I}(r)$.
\begin{figure}[htbp]
    \begin{center}
        \includegraphics[width=0.49\textwidth]{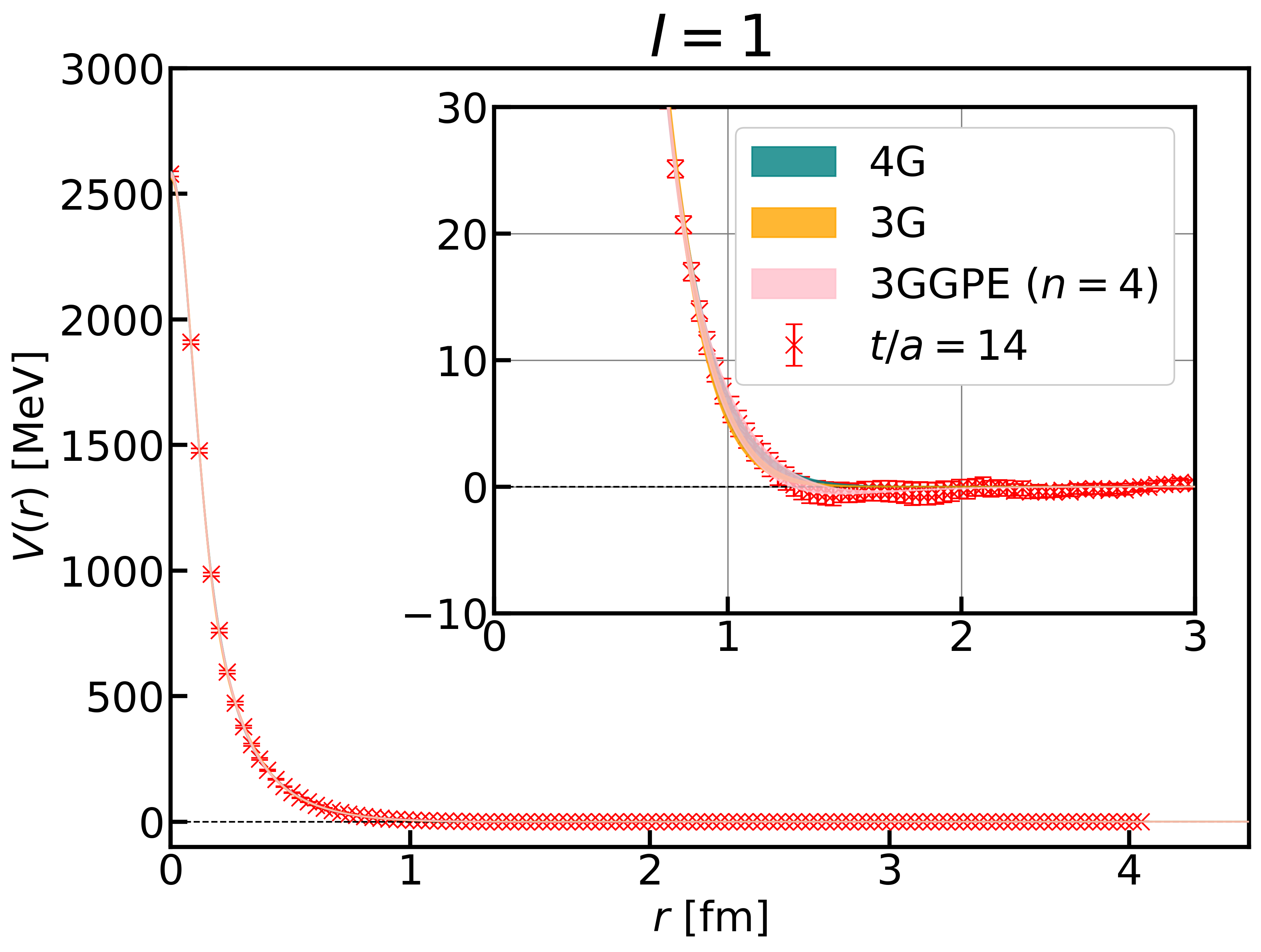}
        \includegraphics[width=0.49\textwidth]{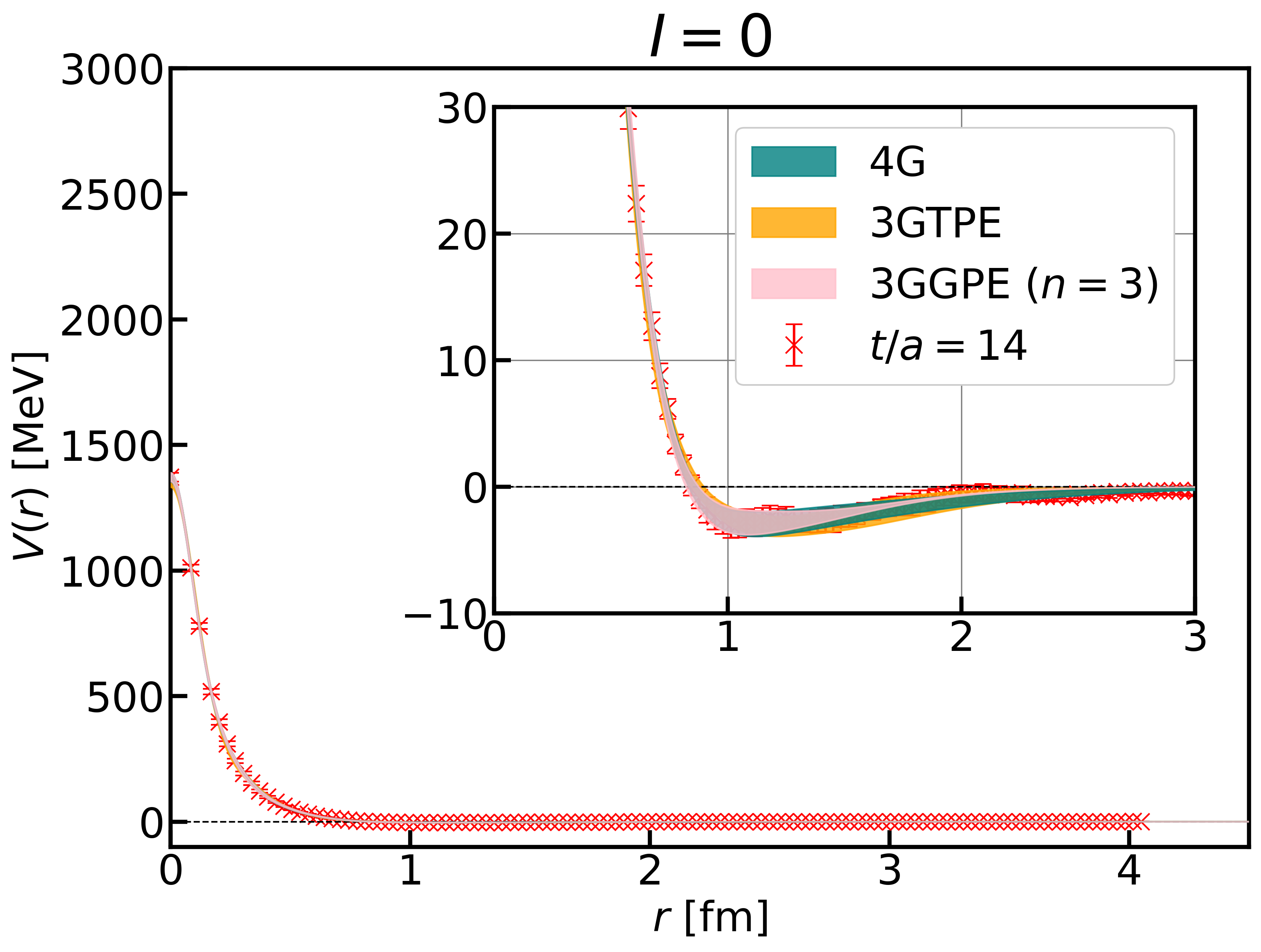}
	    \caption{
        Fit results at $t/a=14$ as well as the original potential data (Red crosses). 
        In the left panel, the results for $I=1$ are shown using $V^{\textrm{4G}}_{I}(r)$ (Teal), $V^{\textrm{3G}}_{I}(r)$ (Orange), and $V^{\textrm{3GGPE}}_{I,n=4}(r)$ (Pink), whereas in the right panel, the results for $I=0$ are presented using $V^{\textrm{4G}}_{I}(r)$ (Teal), $V^{\textrm{3GTPE}}_{I}(r)$ (Orange), and $V^{\textrm{3GGPE}}_{I,n=3}(r)$ (Pink). 
        }
	\label{fig:results_pot_fit_comp}    
    \end{center}
\end{figure}
\begin{figure}[htbp]
    \begin{center}
        \includegraphics[width=0.49\textwidth]{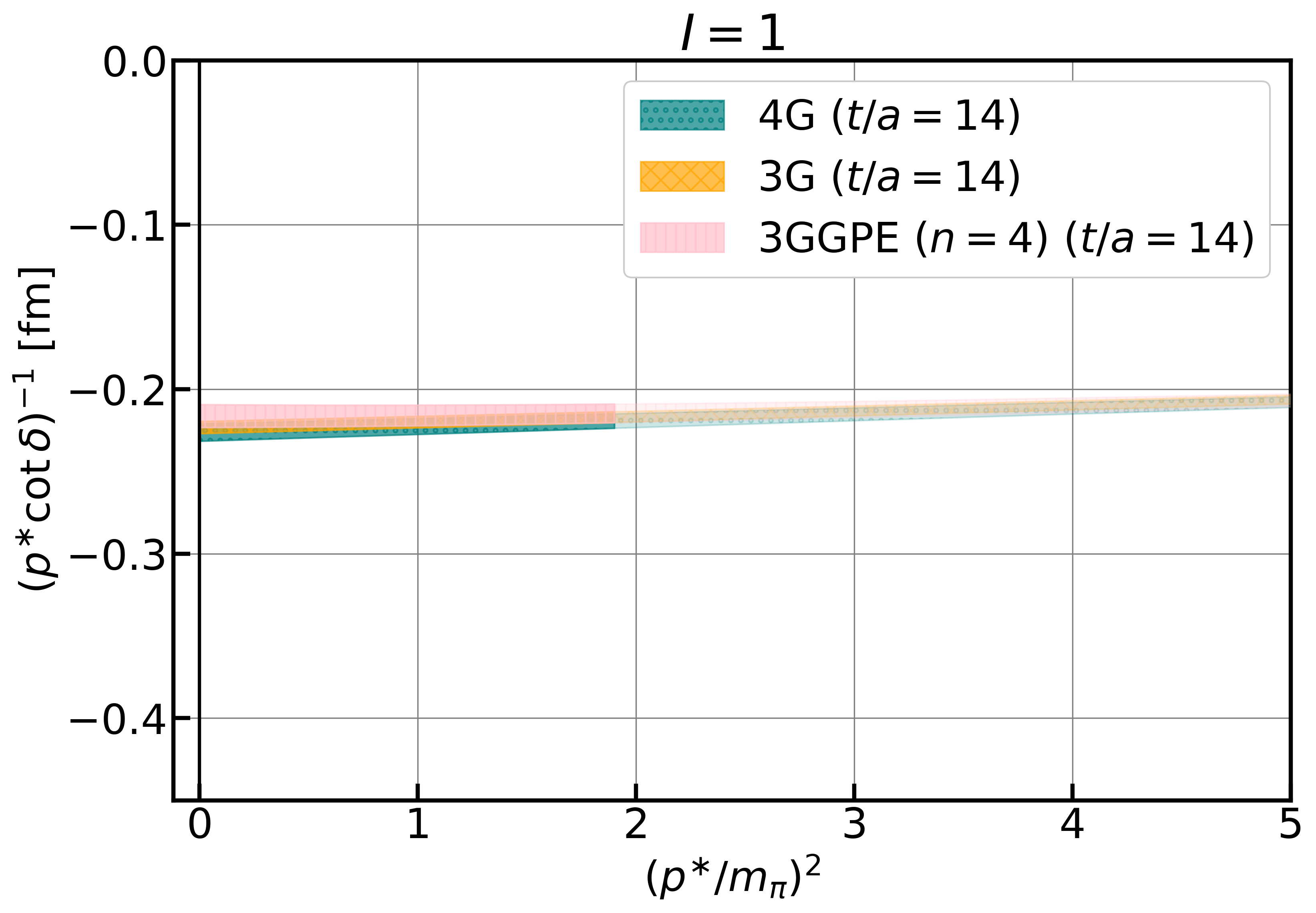}
        \includegraphics[width=0.49\textwidth]{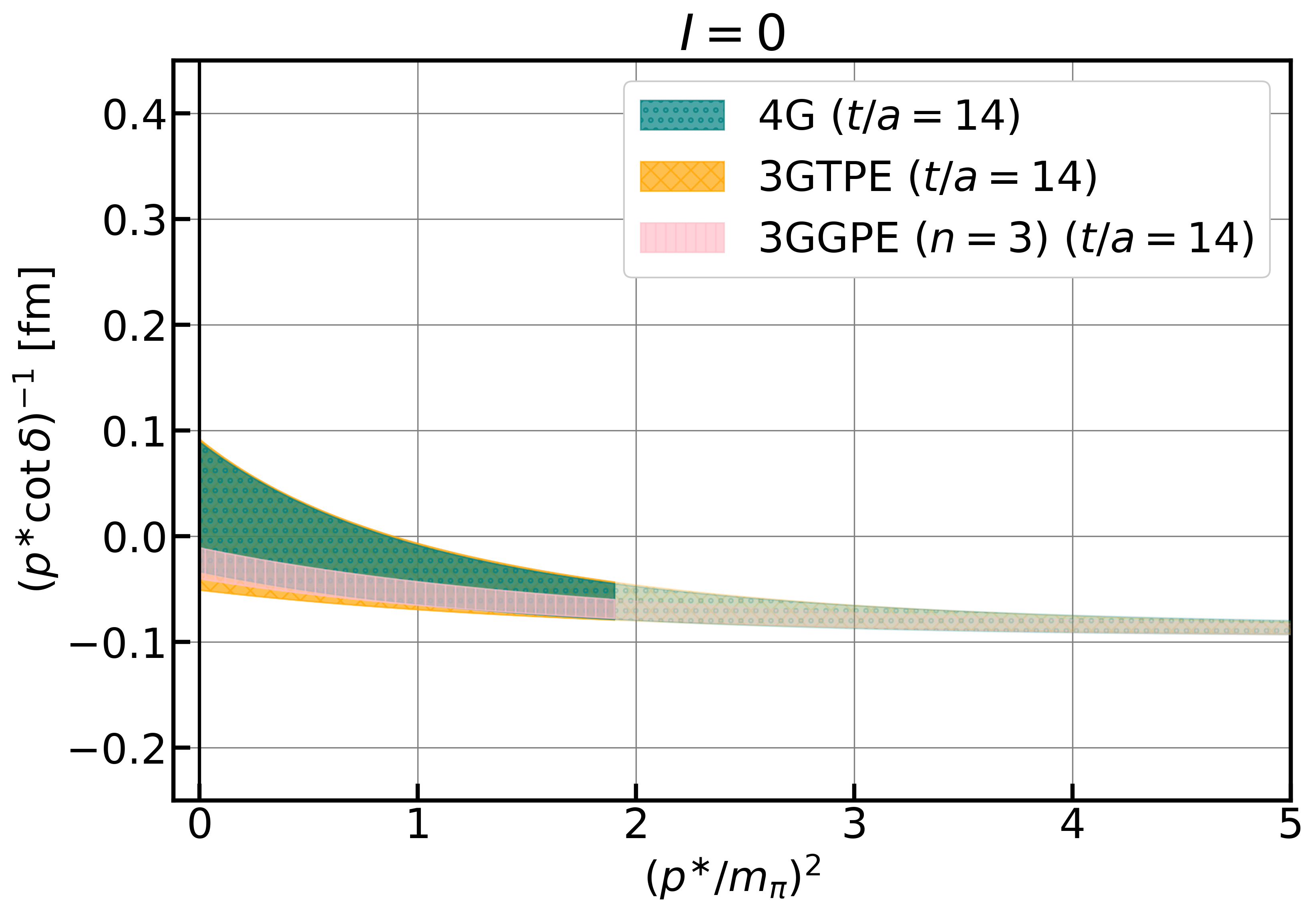}
	    \caption{
        $(p^{\ast}\cot{\delta_{0}(p^{\ast})})^{-1}$ at $t/a=14$. 
        For $I=1$ (Left panel), the results are shown for $V^{\textrm{4G}}_{I}(r)$ (Teal), $V^{\textrm{3G}}_{I}(r)$ (Orange), and $V^{\textrm{3GGPE}}_{I,n=4}(r)$ (Pink). 
        For $I=0$ (Right panel), the results are presented for $V^{\textrm{4G}}_{I}(r)$ (Teal), $V^{\textrm{3GTPE}}_{I}(r)$ (Orange), and $V^{\textrm{3GGPE}}_{I,n=3}(r)$ (Pink). 
        }
	\label{fig:results_kcotdinv_comp}    
    \end{center}
\end{figure}
The fit results for each ansatz are shown in Fig.~\ref{fig:results_pot_fit_comp}, and the corresponding $(p^{\ast}\cot{\delta_{0}(p^{\ast})})^{-1}$ obtained from the fitted potential are displayed in Fig~\ref{fig:results_kcotdinv_comp}. 
For both isospin channels, while the fitted potentials exhibit slight differences at middle and long distances, $0.8 \lesssim r \lesssim 2.5 ~\textrm{fm}$, the resulting phase shifts are consistent within statistical uncertainties. 
This indicates that the choice of fit ansatz does not significantly affect scattering observables in both channels. 
However, we note that there remains room for improvement in the examination of the model dependence; to quantify such an effect in a more rigorous manner, it is necessary to take an approach that does not rely on specific functional assumptions, such as one based on Bayesian inference. 
Establishing such a framework is left in future.

We then investigate systematics associated with contamination of inelastic states and the LO approximation to potentials. 
These effects typically appear as the time dependence of potentials. 
As seen in Fig.~\ref{fig:results_pot_orig}, however, such dependence is negligibly small. 
Furthermore, as a consistency check, we compute the phase shifts at the ground-state energies on a finite volume using the L\"{u}scher's formula~\cite{Luscher:1990ux}, which is shown as brown curves in Fig.~\ref{fig:results_kcotdinv}. 
The ground-state energy is evaluated from the $KN$ temporal correlation functions with optimized operators.
For details, see Appendix~\ref{appendix:FVanalysis}. 
We also evaluate the scattering lengths from the ground-state energies on a finite volume via the $1/L$ expansion~\cite{Luscher:1986pf} as
\begin{eqnarray}
E^{I}_{\textrm{gs}}-m_{N}-m_{K} = -\frac{2\pi a^{I}_{0}}{\mu L^3} \left[ 1+c_{1}\frac{a^{I}_{0}}{L}+c_{2}\left(\frac{a^{I}_{0}}{L}\right)^2\right] + \order{L^{-6}},
\end{eqnarray}
where $c_{1}=-2.837297$ and $c_{2}=6.375183$. 
We obtain $a^{I=1}_{0,\textrm{FV}}=-0.261 (164)~\textrm{fm}$ and $a^{I=0}_{0,\textrm{FV}}=+0.122 (243)~\textrm{fm}$, which should be compared with $a^{I=1}_{0} = -0.226(5)(^{+5}_{-0})~\textrm{fm}$ and $a^{I=0}_{0} = +0.028(61)(^{+3}_{-26})~\textrm{fm}$, respectively, obtained from the potentials.
The FV analysis and the potential method give consistent results on the phase shifts, while errors in the FV analysis are too large to establish the absence of inelastic contributions or non-locality effects.
In future studies, it is important to explicitly quantify the non-locality effect for the high momentum region shown by light color and stripes in Figs.~\ref{fig:results_phase},~\ref{fig:results_kcotdinv}, and~\ref{fig:results_crosssec}, by performing the next-to-leading-order analysis in the derivative expansion for the potentials.

We also investigate an effect due to a small difference in the quark masses between our setup and nature.
Since an extra simulation with different quark masses is beyond the scope of this paper, we have performed the following analysis. 
When we derive the LO potential and solve the Schr\"{o}dinger equation, we use the same R-correlator but with the isospin-averaged values of kaon and nucleon masses in nature, $m_{K}=495.64~\textrm{MeV}$ and $m_{N} =938.92~\textrm{MeV}$, instead of $m_K=502.0(2)$ MeV and $m_N=942.3(3.4)$ MeV in our setup. 
We find no considerable change in the phase shifts, which suggests that the phase shifts are insensitive to the small differences of quark masses in kinematics.

Lasts but not least, we consider the lattice discretization effect~\cite{Green:2021qol}. 
Again, an extra simulation with a different lattice spacing is beyond the scope of this paper.
To estimate how the lattice artifact at short distances affects scattering observables for $I=1$, we artificially enhanced the coefficient of the Gaussian term with the shortest range, namely $a^{I=1}_{1}$ in Eq.~\eqref{eq:fitfunc_4G}, by five times larger than the original value.
The modified potential provides almost the same phase shifts as the original ones, indicating that the discretization error at short distances does not affect our results significantly. 
While there still remains a possibility that the lattice artifact non-trivially affects potentials at intermediate distances, such an examination of the lattice discretization effect is left to future studies.

\section{\label{sec:conclusion} Conclusion}

In this work, we have investigated the S-wave kaon-nucleon ($KN$) interactions with strangeness $S=+1$ by (2+1)-flavor lattice QCD with the time-dependent HAL QCD method. 
By employing the ``HAL-conf-2023" gauge configurations,
we have performed the first physical point simulation of the $KN$ interactions
with $m_{\pi} \approx 137~\textrm{MeV}$, $m_{K} \approx 502~\textrm{MeV}$ and $m_{N} \approx 942~\textrm{MeV}$.
The kaon and nucleon operators with the wall quark sources are used to calculate the four-point correlation functions, and the Misner method~\cite{Miyamoto:2019jjc} has been applied after the projection onto the $A_1^{+}$ representation to obtain the S-wave component.

Our numerical results of the $KN$ potentials have repulsive cores with a size of around $1~\textrm{fm}$ in both $I=1$ and $0$ channels, while the $I=0$ potential has a small attractive pocket of a few MeV at intermediate distances. 
Compared with the previous results at heavier pion masses, we have observed a tendency that the repulsive core increases as the quark masses decrease, while an overall shape looks similar. 
The fit results for $I=0$ by the function including the two-pion exchange term indicate that the two-pion exchange processes seem negligible in the $KN$ interactions.

The S-wave phase shifts in the $I=1$ channel show monotonically decreasing behaviors with increasing the energies, which are also observed in experimental data and the partial-wave analysis.
The S-wave phase shifts in the $I=0$ channel nearly vanish below $P_{\textrm{lab}}\sim 180~\textrm{MeV}$ and decrease above it.
These behaviors are in agreement with some of the partial-wave analyses. 
Our phase shift analysis shows no signal for resonances or bound states in both isospin channels, which indicates the absence of the $\Theta^{+}(1540)$ pentaquark in the S-wave $KN$ systems.

Scattering parameters from the phase shifts become $a^{I=1}_{0} = -0.226(5)(^{+5}_{-0})~\textrm{fm}$ for the $I=1$ scattering length, 
which is qualitatively consistent with the summary results in various analyses based on the experimental data, 
and $r^{I=1}_{\textrm{eff}} = -0.294(30)(^{+21}_{-7})~\textrm{fm}$ for the effective range, which shares the same sign with certain results in the partial-wave analyses.
The scattering length for $I=0$ is $a^{I=0}_{0} = +0.028(61)(^{+3}_{-26})~\textrm{fm}$, which is in agreement with the experimental summary results within errors. 

We have also computed the S-wave component of the total cross section. 
Our result in the $I=1$ channel is $6.3\pm 0.3~\textrm{mb}$ at the threshold, decreasing slowly with respect to the momentum.
These values are smaller than most experimental data, except for a few cases.
In the $I=0$ channel, the cross sections are nearly zero around the threshold and are consistent with those in the recent (unitarized) chiral perturbation analysis, which suggests that the scattering amplitudes for $I=0$ are dominated by the P-wave components rather than the S-wave. 
Explicit examination of the P-wave dominance by lattice QCD simulations of the P-wave interaction is left for future studies.

Finally, we discussed possible sources of systematic uncertainties. 
Our analysis indicates that the finite-volume effect, inelastic contamination, and non-locality effect around the threshold are under control. 
It was also found that the dependence on the fit function for the potentials, the difference in the kaon and nucleon masses between our setup and in nature, and the lattice discretization artifact of the potentials at short distances do not significantly affect the observables. 
However, there remain possible effects: the non-locality effect in the momentum region far from the threshold and lattice discretization artifacts to potentials at middle distances. 
The first one is likely to affect our results in a high-momentum region and can be controlled by employing the next-to-leading-order analysis to potentials using multiple $KN$ correlation functions with different source operators. 
On the other hand, the second effects may affect our results at low momenta, causing the deviation between our results and the experimental values in the $I=1$ channel. 
Controlling the lattice artifacts and taking the continuum limit are mandatory for the definite prediction of the S-wave $KN$ interaction in lattice QCD.

Despite these systematic uncertainties mentioned above, our first-principles calculation provides quantitative understanding of the $KN$ interactions and the $KN$ scatterings near the threshold, and gives essential input for the investigation of the nuclear matter properties such as the in-medium effect of the strange-quark condensate and the exploration of exotic hadrons/nuclei.

\begin{acknowledgments}
We would like to thank the members of the HAL QCD Collaboration for fruitful discussions.
K.~M. appreciates Prof. D.~Jido and Prof. M.~Oka for their useful comments. 
The numerical lattice QCD simulations have been performed on 
the supercomputer Fugaku.
%
%
This work was partially supported by
RIKEN Incentive Research Project (``Unveiling pion-exchange interactions between hadrons from first-principles lattice QCD"), 
Adopting Sustainable Partnerships for Innovative Research Ecosystem (ASPIRE), Grant No. JPMJAP2318,
HPCI System Research Project
(hp200130, hp210165, hp220174, hp220066, hp230207, hp230075, hp240213, hp240157),
the JSPS
(Grants No. JP22H00129, JP22H04917, JP23H05439, JP25K17384)
``Priority Issue on Post-K computer'' (Elucidation of the Fundamental Laws and Evolution of the Universe), 
``Program for Promoting Researches on the Supercomputer Fugaku'' (Simulation for basic science: from fundamental laws of particles to creation of nuclei) 
and (Simulation for basic science: approaching the new quantum era)
(Grants No. JPMXP1020200105, JPMXP1020230411), 
and Joint Institute for Computational Fundamental Science (JICFuS).
\end{acknowledgments}

\appendix

\section{\label{appendix:comp_pot_570MeV} Plots of the comparison of the LO $KN$ potentials between different pion masses}

\begin{figure}[htbp]
    \begin{center}
        \includegraphics[width=0.49\textwidth]{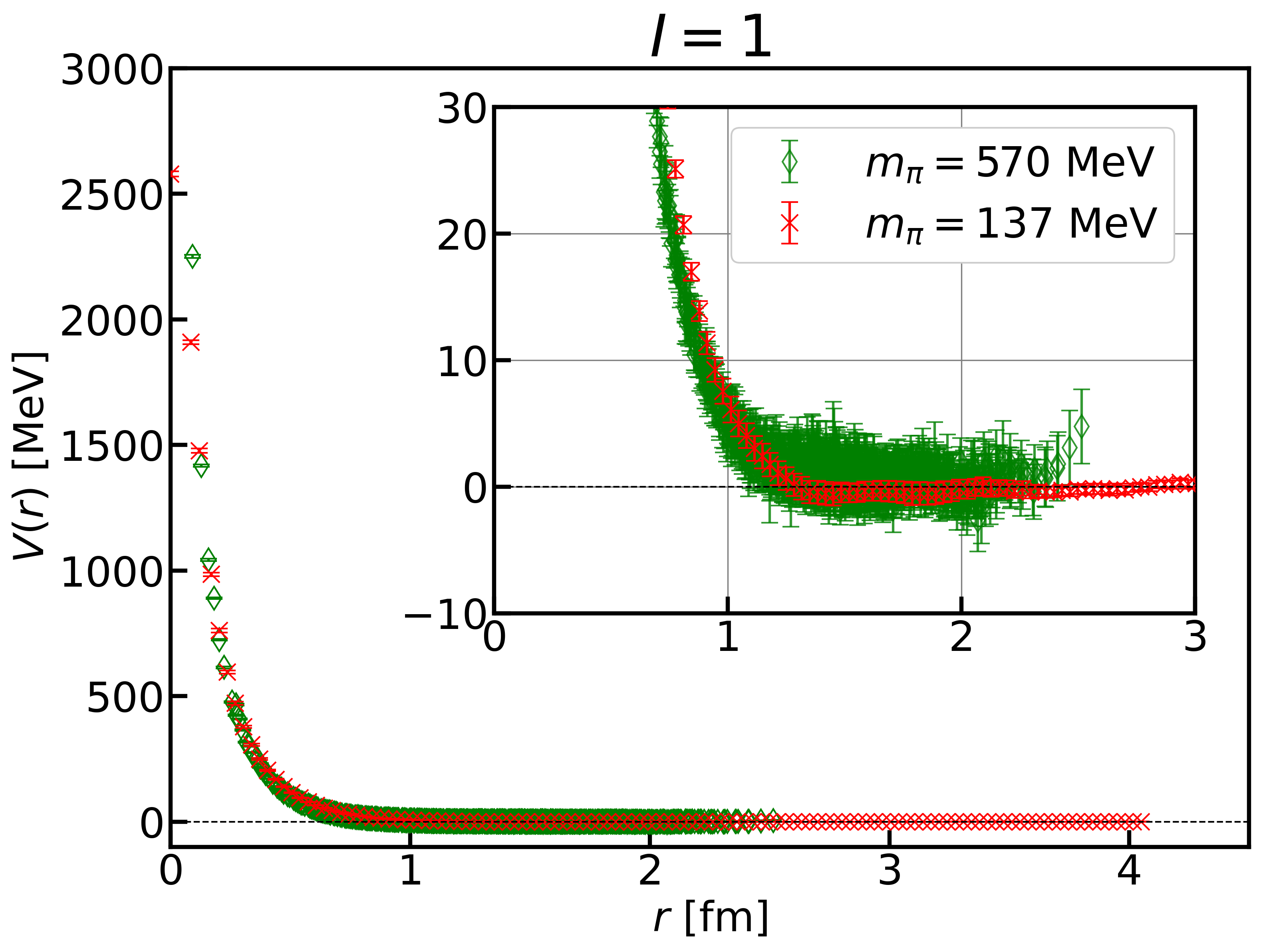}
        \includegraphics[width=0.49\textwidth]{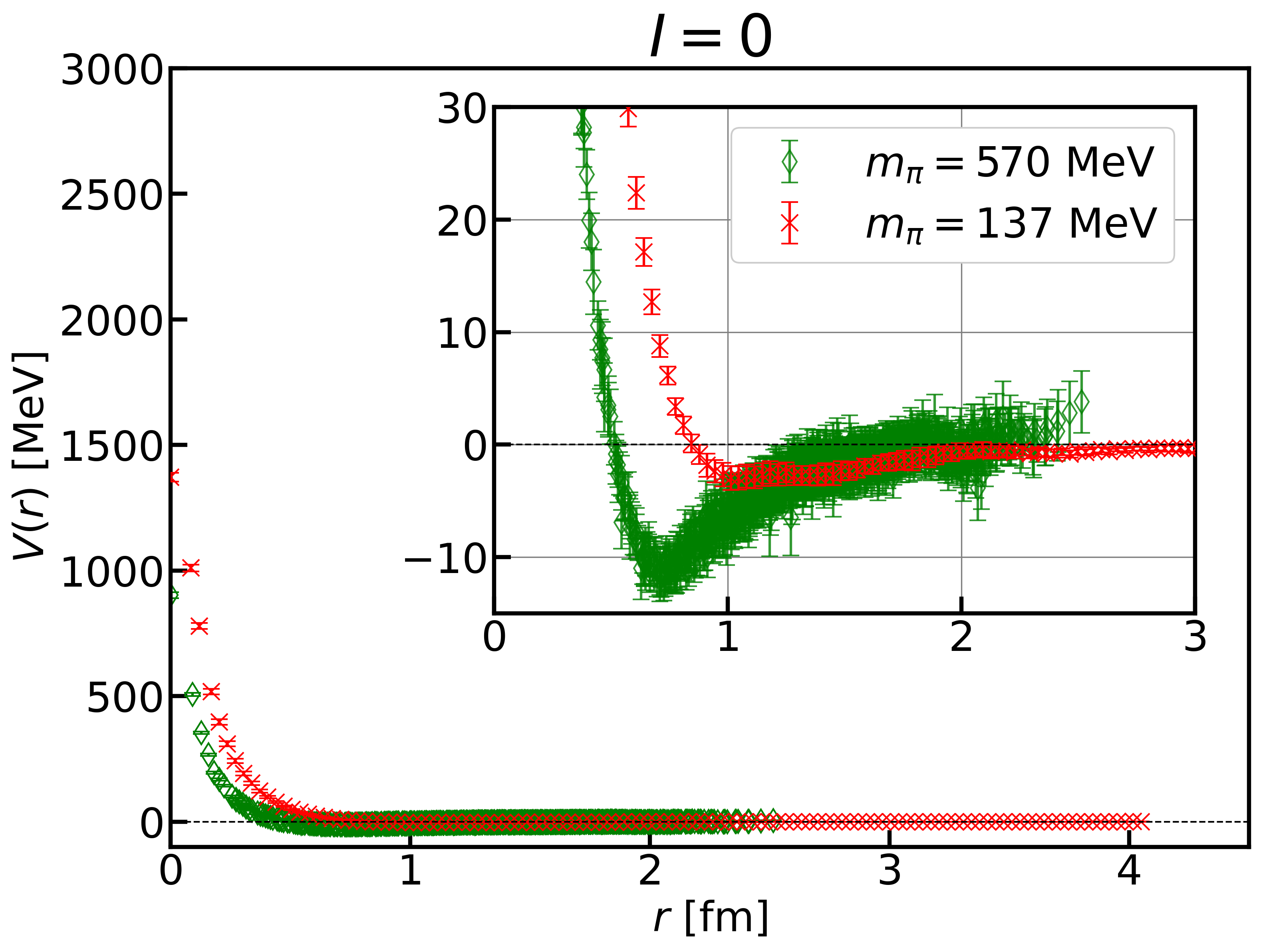}
	    \caption{LO potentials in our calculation 
        at $m_{\pi}\approx 137~\textrm{MeV}$
        ($t/a=14$) (Red crosses) and in the previous study at $m_{\pi}\approx 570~\textrm{MeV}$~\cite{Murakami:2020yzt} (Green diamonds) for $I=1$ (Left panel) and $I=0$ (Right panel).
        }
	\label{fig:results_pot_comp_570MeV}    
    \end{center}
\end{figure}
We show in Fig.~\ref{fig:results_pot_comp_570MeV} a comparison of the LO $KN$ potentials between those at the physical point in this work and those obtained in the previous HAL QCD study~\cite{Murakami:2020yzt}.
The range of the repulsive core for $I=0$ is smaller at $m_{\pi}\approx 570~\textrm{MeV}$ than at the physical point. 
For $I=1$, no significant difference is observed between $m_{\pi}\approx 137~\textrm{MeV}$ and $570~\textrm{MeV}$, whereas a decrease in the range of the core is clearly seen at $m_{\pi}\approx 705~\textrm{MeV}$~\cite{Ikeda:2011qm}.

\section{\label{appendix:FVanalysis} Energy shifts on finite volume from $KN$ temporal correlation functions with optimized operators}

In this appendix, we present the details of the calculation of the energy shift of the $KN$ system on finite volume.

\begin{figure}[htbp]
    \begin{center}
        \includegraphics[width=0.49\textwidth]{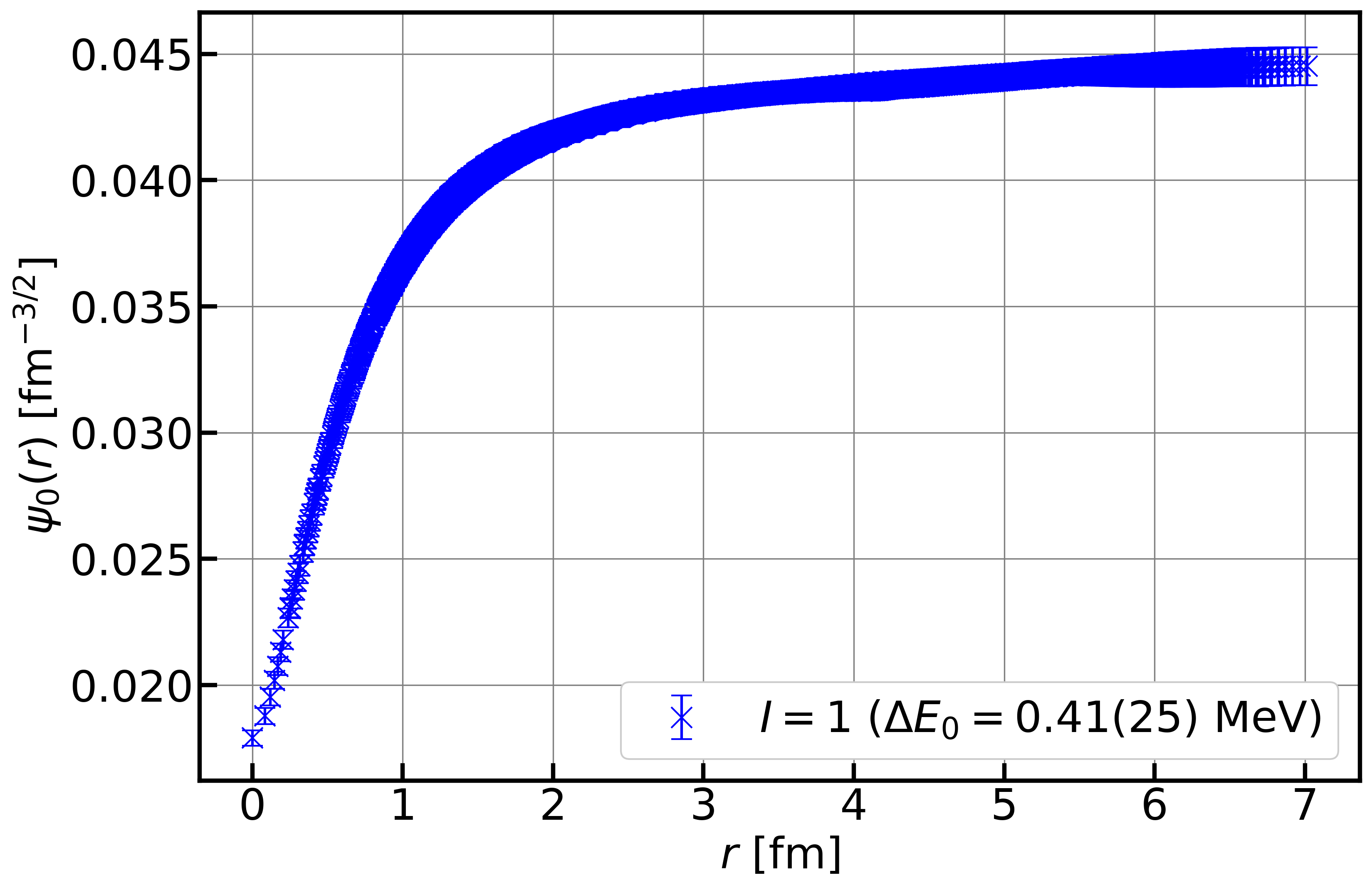}
        \includegraphics[width=0.49\textwidth]{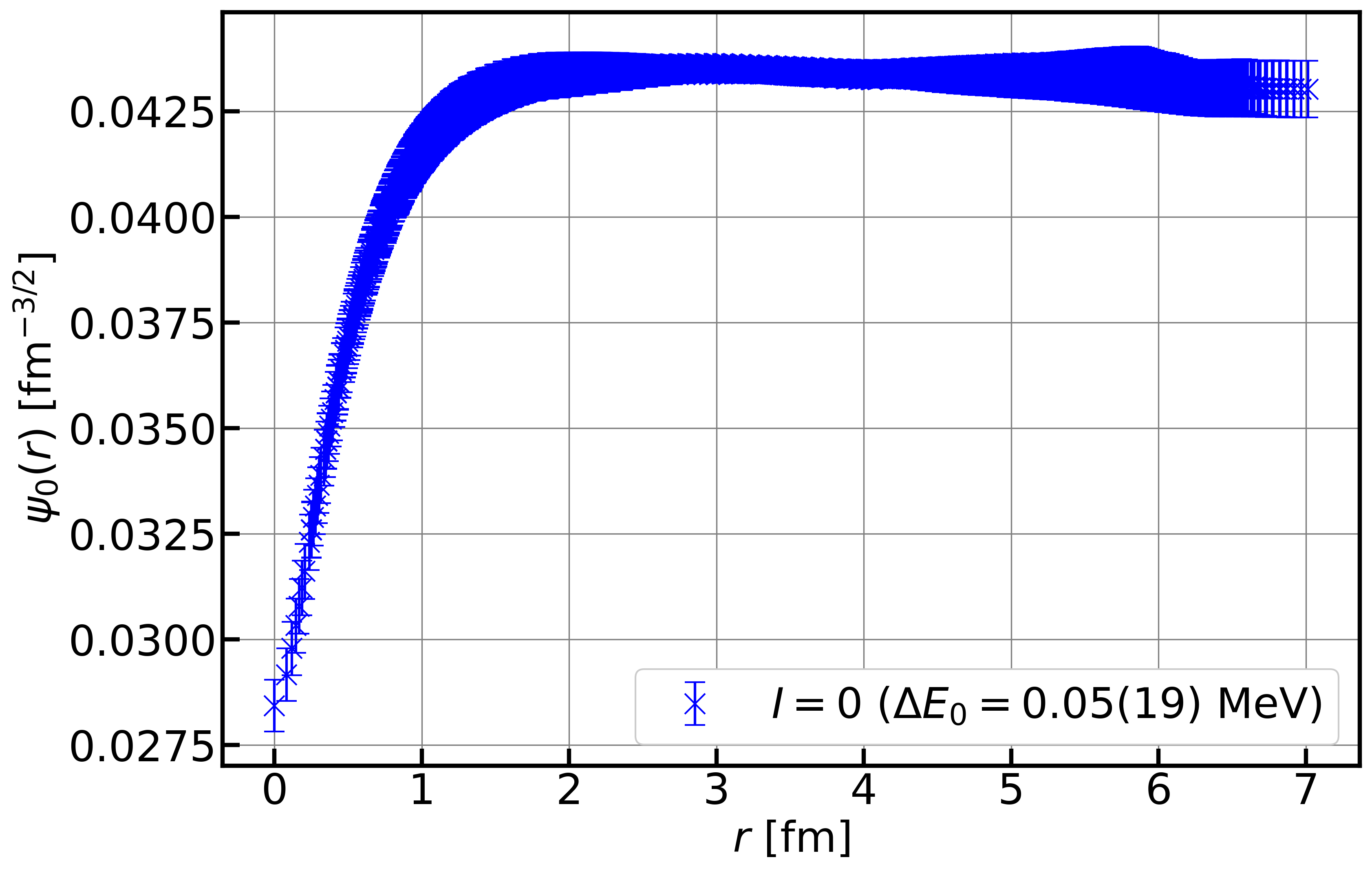}
	    \caption{Wave function of the ground state on the lattice in the finite volume
        with the potential in $I=1$ (Left panel)
        and $I=0$ channels (Right panel).
        }
	\label{fig:results_wavefunc_FV}    
    \end{center}
\end{figure}

We construct the temporal correlation functions following Refs.~\cite{Iritani:2018vfn,Lyu:2022tsd}.
We first solve the Schr\"{o}dinger equation with the obtained $KN$ potential on the same finite number of spatial points.
Here we do not employ the Misner method to extract the potential. 
Fig.~\ref{fig:results_wavefunc_FV} shows the wave functions with the lowest energy, which we denote $\Psi^{I}_{0}(\vb{r})$ hereafter. 
The $\Delta E_{0}$ shown in the legend on each panel of the figure is the corresponding eigen energy. 
Combining the wave functions and the R-correlators, Eq.~\eqref{eq:def_Rcorr}, for the $KN$ system, we then construct the temporal correlation functions with the optimized sink operators as 
\begin{eqnarray}\label{eq:2ptcorr_optimized}
R^{I}_{0}(t) = \sum_{\vb{r},\alpha}\Psi^{I}_{0}(\vb{r})R^{I}_{\alpha}(\vb{r},t)  =   \frac{\sum_{\alpha}\langle \left[\sum_{\vb{r}}\Psi^{I}_{0}(\vb{r})(KN)^{(I)}_{\alpha}({\bf r}, t+t_{0})\right] \bar{K}^{0}(t_0)\bar{p}_{\alpha}(t_0) \rangle}{C_{K}(t)C_{N}(t)}.
\end{eqnarray}

\begin{figure}[htbp]
    \begin{center}
        \includegraphics[width=0.75\textwidth]{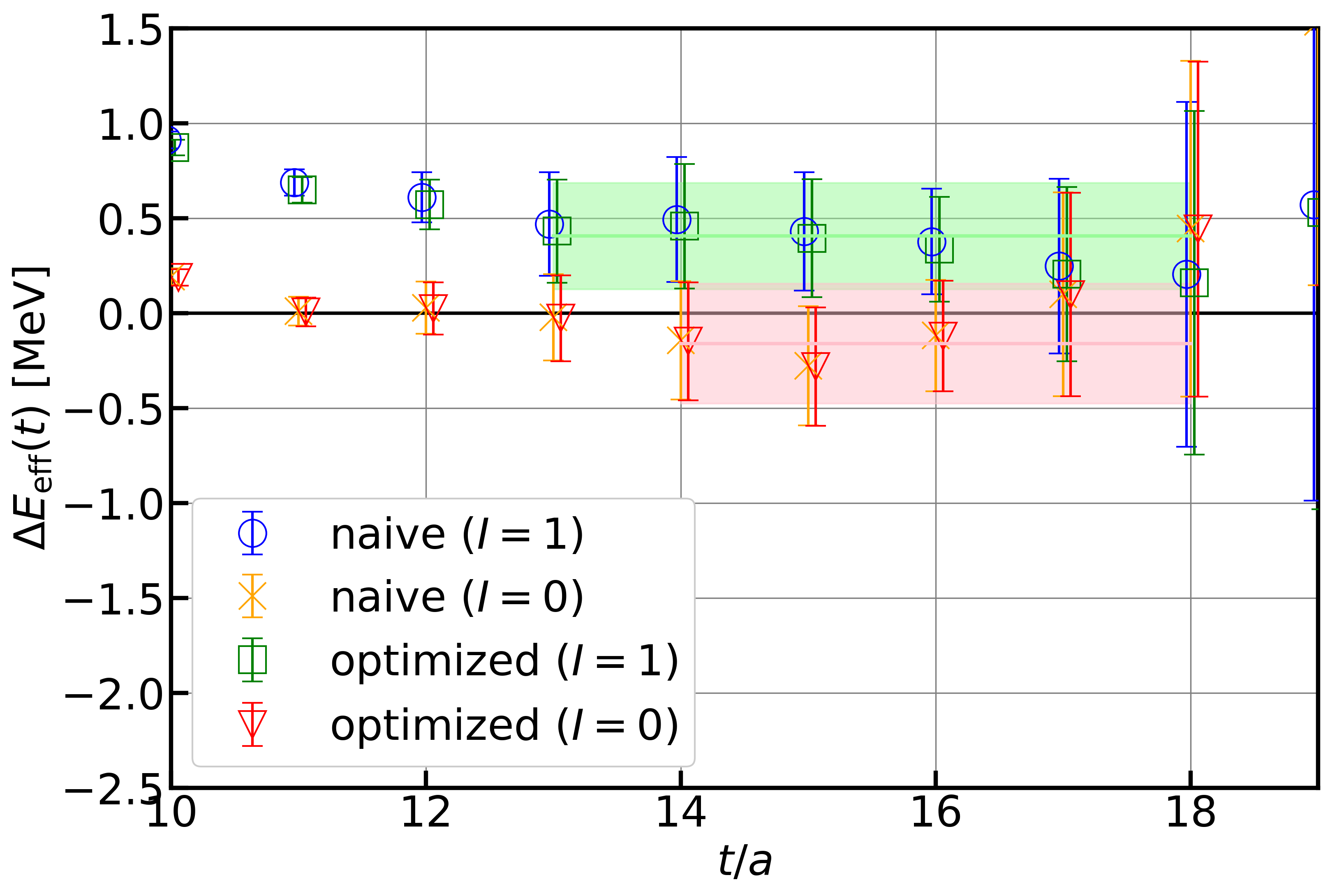}
	    \caption{Effective energy for the two-point correlation functions with the optimized sink operators for $I=1$ (green squares) and $0$ (red triangles), and the naive sink operators for $I=1$ (blue circles) and $0$ (orange crosses). 
        Light-green and pink lines with bands correspond to the energy shifts calculated by fitting to the two-point correlation functions with the optimized sink operators for $I=1$ and 0, respectively. 
        }
	\label{fig:results_eff_energy}    
    \end{center}
\end{figure}

Shown in Fig.~\ref{fig:results_eff_energy} are the effective energies for the temporal correlation functions, $\Delta E_{\textrm{eff}}(t)= \frac{1}{a}\ln\left(\frac{R^{I}_{0}(t)}{R^{I}_{0}(t+a)}\right)$. 
The value of the plateau in this figure corresponds to the energy shift from the threshold on the finite volume. 
For comparison, we show the results with the sink operators summed over spatial coordinate without any weight factor, which we call the naive sink operator here. 
We find that the results from the optimized and naive sink operators are almost consistent, which indicates that the R-correlators have negligible overlap with the excited elastic scattering states. 
Fitting the temporal correlation functions with the optimized sink operators using a single exponential function with the range $13 \leq t/a \leq 18$ for $I=1$ and $14 \leq t/a \leq 18$ for $I=0$, we obtain the energy shifts of the $KN$ ground states on the finite volume as
\begin{eqnarray}
\begin{aligned}
E^{I=1}_{\textrm{gs}}-m_{N}-m_{K} &= 0.405(280)~\textrm{MeV},\\
E^{I=0}_{\textrm{gs}}-m_{N}-m_{K} &= -0.161(317)~\textrm{MeV}. \label{eq:results_energy_shift}
\end{aligned}
\end{eqnarray}
These two results are consistent within the error with the eigen energies, $\Delta E_{0}$, shown in the legends of Fig.~\ref{fig:results_wavefunc_FV}.
These results are also used to obtain phase shifts using L\"uscher's finite volume formula,
shown in brown curves in Fig.~\ref{fig:results_kcotdinv}.

\bibliography{KN_Fconf}